%% file: stegomalware.tex
\documentclass[journal]{IEEEtran}
\usepackage{graphicx}
\usepackage{nomencl}
\usepackage{url}
\usepackage{multirow}
\usepackage{changes}
\usepackage{xcolor}
\usepackage{amsmath}
\usepackage{adjustbox}
\usepackage{supertabular}
\usepackage{booktabs}
\usepackage{rotating}
\usepackage{supertabular}
 \usepackage[bookmarks=false]{hyperref}

\DeclareUnicodeCharacter{2212}{-}

\usepackage{hyperref}
\usepackage{float}

\makenomenclature

\ifCLASSINFOpdf
\else
\fi
\begin{document}

\title{Stegomalware: A Systematic Survey of Malware Hiding and Detection in Images, Machine Learning Models and Research Challenges }

\author{Rajasekhar~Chaganti, Vinayakumar~Ravi, Mamoun~Alazab, Tuan D. Pham


\thanks{Rajasekhar~Chaganti was with Dept. of Computer Science, University of Texas at San Antonio, San Antonio, Texas 78249, USA. e-mail: (Raj.chaganti2@gmail.com)}

\thanks{Vinayakumar~Ravi was with Center for Artificial Intelligence, Prince Mohammad Bin Fahd University, Khobar, Saudi Arabia. e-mail: (vravi@pmu.edu.sa)}

\thanks{Mamoun~Alazab was with College of Engineering, IT and Environment, Charles Darwin University, Darwin, NT 0820 Australia (e-mail: alazab.m@ieee.org)}

\thanks{Tuan D. Pham was with Center for Artificial Intelligence, Prince Mohammad Bin Fahd University, Khobar 34754 Saudi Arabia (e-mail: tpham@pmu.edu.sa)}

}


\maketitle

\begin{abstract}
Malware distribution to the victim network is commonly performed through file attachments in phishing email or downloading illegitimate files from the internet, when the victim interacts with the source of infection. To detect and prevent the malware distribution in the victim machine, the existing end device security applications may leverage sophisticated techniques such as signature-based or anomaly-based, machine learning techniques. The well-known file formats Portable Executable (PE) for Windows and Executable and Linkable Format (ELF) for Linux based operating system are used for malware analysis and the malware detection capabilities of these files  has been well advanced for real time detection. But the malware payload hiding in multimedia like cover images using steganography detection has been a challenge for enterprises, as these are rarely seen and usually act as a stager in sophisticated attacks. In this article, to our knowledge, we are the first to try to address the knowledge gap between the current progress in image steganography and steganalysis academic research focusing on data hiding and the review of the stegomalware (malware payload hiding in images) targeting enterprises with cyberattacks current status. We present the stegomalware history, generation tools, file format specification description. Based on our findings, we perform the detail review of the image steganography techniques including the recent Generative Adversarial Networks (GAN) based models and the image steganalysis methods including the Deep Learning (DL) models for hiding data detection. Additionally, the stegomalware detection framework for enterprise is proposed for anomaly based stegomalware detection emphasizing the architecture details for different network environments. Finally, the research opportunities and challenges in stegomalware generation and detection are presented based on our findings. 
\end{abstract}


\IEEEpeerreviewmaketitle

\input{S1Introduction}
\input{S12AimandMotivation}

\input{S111Relatedwork}
\input{S11Stegohistory}
\input{S12stegotools}
\input{S2Fileformat}
\input{S3Steganography}

\input{S4Steganalysis}

\input{S5Stegoframework}

\input{S51Dataset}

\input{S52Performancemetric}

\input{S6ResearchChallenges}
\input{S7Conclusion}


\section*{Acknowledgment}
The authors would like to

{
\bibliographystyle{unsrt}
\bibliography{References}
}




\end{document}

%% file: S1introduction.tex
\section{Introduction}
\label{sec:intro}

\IEEEPARstart {W}{ith} the proliferation of the internet availability to everyone at affordable costs and users can communicate through Internet to even remote locations, enormous amount of data is being generated and exchanged between the public and private organizations, service providers, government, customers, small and medium businesses and other entities on a daily basis. The security and privacy of the user data has been a concern for a long time \cite{ElHaourani2020}. Cryptography is one of the way to improve the security and privacy when the user transmit the data through wired or wireless communication channels. The asymmetric and symmetric key combination techniques like SSL/TLS protocols developed to securely transmit the data between two endpoints. These protocols protect the original data being tampered and satisfies the security property "integrity". But, an adversary can still able to read the scrambled data in the communication channel if the man in the middle attack is performed and could apply the cryptanalysis to decipher the data \cite{Chris2019}. Additionally, the secret keys may be exposed to the public or compromised by the adversary with poor security practices and lack of security awareness. Information hiding is another way to securely share the information through communication channels to others with no exposure of the secret information. In contrast to cryptography, the data is concealed in multimedia files in information hiding and no normal user is able to know that the information is hidden in the multimedia files. The information hiding techniques further classified into two types such as watermarking and steganography \cite{Singh20208}. Watermarking is used for protecting and claiming the ownership of the digital assets, copyrights and intellectual property rights. It is most widely accepted to use for digital assets protection.

The another information hiding technique is the steganography. Steganography is the process of embedding the secret information into the cover medium for information hiding. The cover medium usually any multimedia file such as image, audio, video, text to hide the information. Anyone may use the steganography techniques to secretly share the data for legitimate purpose. For instance, government may share the confidential information with shareholders secretly using steganography. Despite steganography will improve the user data sharing privacy, the adversaries constantly look up new attacking ways and may be used steganography for illegitimate attack purpose. For example, an adversary may inject the malicious payload into the cover medium to evade the antimalware solutions detection. The malware hiding in the cover medium is termed as "stegomalware". We herein use the term stegomalware in the context of the malware hiding in multimedia files. The stegomalware can be incorporated in cyberattack life cycle to fool the security defender tools and perform the intended malicious activity. For instance, the lokibot malicious source code is concealed in the PNG image file for malware installation and avoid the detection of the malware by email security tools when distributing through phishing email campaigns \cite{Osborne2019}. The image files were used for malicious code obfuscation and sophisticated malware propagation in victim network. So, the stego malware detection technology is highly desired to identify the concealed data in cover medium for protecting the enterprise's assets and preventing the users for not becoming a victim of data breaches.

The multimedia file formats image, audio, video, and text are known to be considered for concealing the malicious content in enterprise attacks \cite{Davis2016} \cite{AnkitAnubhav2016}. The network protocols like TCP, UDP, and ICMP data format may be leveraged to hide the victim data and send over through  network communication medium for covertly exfiltrating the data in the text form. The detection of this covert communication requires the decryption of the data packets and monitor the packet header or data anomalies for identification. These are difficult to detect in enterprise environment, as the enormous amount of network outbound traffic generates and may arise performance issues of decrypting the network traffic. Another category of attacks may use the image, audio, and video files as cover media to deliver the malware payloads and artifacts in the enterprise networks. These stegomalware may not be easy to detect and identify using existing enterprise security tools due to the complex nature of malicious payload hidden in the multimedia files and the tools are not capable of handling image, audio files for malware detection. The proactive mitigation plan like blocking the delivery of the cover media file formats is not a viable solution, as these file formats are extensively used in enterprise network for business transactions and can cost halting the business operations. So, an efficient, robust and accurate detection methods are needed to effectively identify these threats. Our analysis on the stegomalware history showed that images are extensively used for malware hiding. So, we focus on handling stegomalware in images in this paper. 

Image steganography techniques are mainly categorized into spatial or transform domain based and spread spectrum techniques \cite{Hussain2018}. Recently, the GAN based image steganography solutions are popular and effective to hide the secret data \cite{Wang2020}. In spatial domain, the techniques applied directly on the images to manipulate the pixels and store the secret information. On the other hand, in transform domain, the image file is transformed into transform domains such as DCT or DWT, and then secret information can be embedded in the DCT or DWT coefficients of the image\cite{Hussain2018}. Some of the methods may use the combination of spatial and transform domain to hide the secrets. Additionally, spread spectrum technique can be used to embed the secret information in the noise and then add the noise to the original media. GAN based steganography typically includes the generator, discriminator and steganalysis modules to iteratively generate stego images such that minimize the distortion between stego and cover image. These techniques can also be used by adversaries to hide the malware payload in images. So, the steganalysis methods should be applied to detect the hidden malware payload in images.

There are five different ways to perform the steganalysis. Those are visual steganalysis, statistical steganalysis, signature based steganalysis, spread spectrum steganalysis, transform domain steganalysis and the universal or blind steganalysis \cite{Karampidis2018}. In visual steganalysis, the hidden information on the images can be determined using the naked eyes. There are no techniques need to be applied for deciphering the hidden secret in visual steganalysis. We may use statistical, and signature based steganalysis for detecting the hidden information and possibly revealing the malicious intent of the file if it is part of attack propagation. 

The signature based steganalysis focused on matching the known hidden secret technique patterns such as byte patterns, known text signatures for detection. The signature-based detection can be using the existing forensic tools. These techniques may identify the files, which used the existing steganography techniques. However, a small change in the cover image or using advanced steganography algorithms like UNIWARD \cite{Holub2014}, HILL \cite{Li2014}, WOW \cite{Holub2012} makes the signature-based techniques difficult to classify the stegomalware. Statistical steganalysis look for the statistical properties of the file such as entropy of Exif header, histogram of the pixel’s distribution in the images and other attributes information to find the anomalies in the files when compared with the original files \cite{Karampidis2018}. Spread spectrum steganalysis and transform domain steganalysis emphasize the importance of performing steganalysis indirectly on the image to improve the security and identify the classification even if the cover image is compressed, resized, and added with noise.  The universal steganalysis techniques such as ML or DL based solution techniques can be applied to learn the behavior of the image using statistical features or unique feature taken from the sample data \cite{Perez2016} \cite{Su2019}. But a proper selection of universal features for accurate classification is a challenge. The steganalysis classification is even harder when the dataset sample contain multiple image formats and want to obtain better detection performances.

Despite number of steganalysis techniques exist to detect the secret data, enterprises face challenges to identify the stegomalware because of lack of detection capabilities within the existing security tools deployed in the environment, the state-of-the-art detection accuracies are not enough to use in production environment \cite{Reinel2021}, \cite{Singh2021} lack of professional resource and difficulty of analyzing the large number of files traverse through enterprise. For instance, the malicious data embedded in an image may target the employees to compromise their credentials or install a backdoor code hidden in the image to establish a command-and-control (C\&C) server communication or send a suspicious attachment in the email to let the victim download into the machine and constantly communicate with C\&C for data exfiltration. 

Our state-of-the-art analysis shows that the research done so far considered the image steganography as an application to hide the data. Similarly, the image steganalysis is considered as a stego data detection tool. But, to our knowledge, there is limited work on exploring the malware hiding and detection in multimedia, specific to hiding in cover images, even though stegomalware is touted as one of sophisticated technique used by adversary in cyberattacks to evade the security detection \cite{Goodin2019} \cite{MatteoMauriIginoCorona2020} \cite{Votiro2020}. This gap could be because of Image steganography and steganalysis, and cybersecurity are considered to be two different technical fields. We wanted to bridge the knowledge gap between these fields by discussing the current status of the stegomalware and presenting the current state of the art in image steganography and steganalysis. The stegomalware discussion includes history of stegomalware used in cyberattacks, stegomalware generation tools, stegomalware hiding multimedia file format specifications and the proposed stegomalware detection framework deploying in datacenter, AWS cloud and multicloud environment for enterprises. Based on our stegomalware history study, we have determined that the image is actively used as a cover medium for hiding malware content. So, we have preformed a detailed review of the image steganography and steganalysis techniques. The image steganography review includes describing image domain feature based solutions and GAN based solutions for steganography covering major research contributions in the state of the art. The image steganalysis review includes statistical feature and ML based solutions in spatial and JPEG domain, Rich models with Ensemble classifier and Deep learning based steganalysis solutions. To the end, we discuss the research challenges and directions on stegomalware hiding and detection techniques in the future.
The main contributions of this work are as follows

\begin{itemize}
    \item We discuss the current state-of-the-art stegomalware emphasizing the evidence based approach (stegomalware history, tools and file formats) with focus on recent attack trends.
    
    \item We perform a detailed review of image steganography and steganalysis techniques used for data hiding in images highlighting the major contributions for the last decades in chronological order and investigating the stegomalware contributions towards either image steganography or steganalysis. 
    
    \item We provide a thorough review of GAN based image steganography for hiding the data securely and deep learning based image steganalysis for effective detection of the secret data, and also presented the performance comparison of the GAN based steganography and DL based steganalysis solutions.

    \item We present an enterprise stegomalware detection framework for identifying the stegomalware in multimedia, especially hiding malware in images and provided the framework implementation architecture for datacenter, AWS cloud and multicloud network environments.  
    
    \item We also describe the research opportunities and challenges in the stegomalware hiding and detection solutions space by analyzing the stegomalware trends and reviewing the existing image steganography and steganalysis solutions.

\end{itemize}

The rest of the paper is organized as follows. Section \ref{sec:motivation} describes our motivation and aim of this study. Section \ref{sec:relatedwork} compares the related work with our study. Section \ref{sec:history} discusses the stego malware history of using it for malicious purpose. Section \ref{sec:tools} illustrates the most familiar tools used for steganography to hide the content in multimedia. Section \ref{sec:fileformat} describes the various image and audio file formats used in malware activity. Section \ref{sec:steganography} includes the review of the image steganography algorithms and their performances. Section \ref{sec:steganalysis} illustrates the steganalysis techniques used for image steganography detection. Section \ref{sec:Stegoframework} present the proposed enterprise framework for stegomalware detection. Section \ref{sec:dataset} includes the datasets used for performing experiments in steganography and steganalysis space. Section \ref{sec:metrics} describes the various performance metrics used to assess the detection techniques and hiding algorithms in the prior art. Section \ref{sec:challenges} discusses the research opportunities and challenges based on our study. Section \ref{sec:conclusion} concludes the paper. 

%% file: S12AimandMotivation.tex
\section{Aim and Motivation}
\label{sec:motivation}

The traditional antivirus solutions existed for a long time to protect the users and enterprise customers from known malware by incorporating signature and anomaly based detection techniques in the tools. As the adversary look for new ways to easily evade the traditional antivirus solutions, enterprises started adapting ML/AI based next generation end point security solutions. These solutions possess advances data analytics capabilities to detect not only known malware but also sophisticated malware variants. However, the antivirus and next generation endpoint solutions still exhibits false negatives due to the adversary usage of advanced techniques for hiding the malware and evading the detection. In particular, the enterprise security solutions find it difficult to detect the malware hidden in multimedia format files such as image, audio file formats. The existing solutions are designed to mainly focus on the signature and behavior analytics of the executable files for the malware identification.  Additionally, the number of known malware samples hidden in multimedia file format are very less for applying the ML/DL techniques and predict the malware files. 

Security professionals may find it even difficult to perform stegomalware mitigation activities such as blocking file formats at Firewall, Intrusion Detection/Prevention System or endpoint security level, as the image, audio and video files formats are extensively used in the enterprise for business operations and transactions. So, the security workforce most likely may ignore to perform any proactive actions on stegomalware by considering the criticality of not allowing multimedia format in the enterprise for business continuation except reacting when the device compromise or data breach is identified. Additionally, the file signature update in security tools also may not be a viable option, as the malware constantly evolves to evade the detection. Furthermore, the malware hidden image, audio files are inherently complex to identify because of no suspicion, various file formats and structures to process and characterize the malware using the state-of-the-art security products. Based on these facts, it is clear that adversary using multimedia formats such as image, audio to hide the malicious executable, malicious IP or C\&C server domain name detection is challenging and there is no existing solution to claim to be effectively and accurately detecting these stegomalware files in enterprise organizations.

Our motivation of this study is to identify the technology gaps and reasons for not addressing the stegomalware problem actively so far, address the stegomalware by investigating the image steganography state-of-the art solutions since 2000, explore the cross discipline technical field dependencies to envision the effective solution space for stegomalware detection in enterprises. Interestingly, there is limited work in recent times exploring the stegomalware hiding in multimedia topic \cite{Cohen2020} \cite{Puchalski2020} \cite{MCaviglioneu2021}, even though great amount of research being done discussing the steganography and steganalysis techniques in multimedia files. The reasons could be the topic is multidisciplinary field focussing on the signal and image processing, cybersecurity and malware analysis. So, another reason of motivation to this study is performing a thorough study considering the industry needs and academic research current status.

The main aim of this study is multifold. Firstly, an adversary adapts stegomalware techniques nowadays for evading the detection. So, we wanted to present an overview of the stegomalware, including the stegomalware history, tools use for hiding malware, multimedia file for malware hiding format specifications, steganography and steganalysis techniques in data hiding. We believe that our study will enable the cybersecurity workforce to explore the prior art and build solutions to detect stegomalware. For instance, our study on the steganalysis using deep learning techniques in this paper can be used as a reference to train and test the models for image based stegomalware detection and implement the production level effective stegomalware solutions if optimal solution available in the state of the art. Secondly, our study is seen as a reference for bridging the knowledge gap between the signal and image processing researchers, deep learning/GAN specialists and the cybersecurity technologists especially malware researchers to continue performing research in unfathomed malware hiding in multimedia file formats. Furthermore, we wanted to explore and discuss the research challenges for detecting and mitigating the stegomalware.

%% file: S111Relatedwork.tex
\section{Related work}
\label{sec:relatedwork}
In this section, we have described the relevant prior art works in information hiding specific to image as a cover and their contributions. Researchers reviewed various aspects of the steganography and steganalysis mainly focused on the generic data hiding using multimedia file formats as a cover. But, as our study on the stegomalware history reveals that adversary mostly leverage images as a cover to hide the malware or malicious artifacts, We restrain our scope of study to image based steganography and steganalysis even for related work comparison. The stegomalware covert channel communications hiding in network protocol stack i.e. text based hiding stegomalware \cite{Repetto2021} \cite{Bj2021} \cite{Carrega2020} also not considered in our study because the enterprises could decrypt the network traffic and monitor for anomalies to identify cover channels and may prevent stegomalware spread. The image steganography is further divided based on the spatial or transform domain algorithms used for data hiding in images \cite{Hussain2018} and GAN models \cite{Subramanian2021} used to generate stego images. The steganalysis is further classified based on the combination of the rich models for feature extraction and ML models for classification \cite{Karampidis2018}, and the application of deep learning models for steganography detection \cite{Tabares-Soto2019}.  

Mazurczyk et al. \cite{Mazurczyk2015} performed a survey on the steganography techniques and mitigation solution on the smartphones in 2015. The authors emphasized the malware attacks leveraging steganography to hide malware on smartphone, discussed the steganography tools, steganography techniques in the context of feature domain and ML based solutions, mitigation solutions mainly focusing to cover channel communication mitigation. However, the paper did not cover the recent advancements of using GAN to generate stegomalware and deep learning models for efficient detection stego content instead of rich models with classification techniques. Furthermore, the paper survey is limited to smartphones, which may not cover the enterprise end point based study. We also described the datasets used in the state-of-art, performance metrics needed to evaluate new models if someone wanted to evaluate them and file format specification to explore the opportunities for analyzing the specifications either for steganalysis or for steganography for secured hiding. We believe that the paper \cite{Mazurczyk2015} is somewhat close to our study topic exploring the stegomalware in multimedia. 

Other review studies mainly focused on a particular aspect of image steganography either focusing on the prior art work review in particular domain features \cite{Kadhim2019} \cite{Abduallah2016} \cite{Hussain2018} \cite{Tabares-Soto2020} \cite{Evsutin2020} or review on the GAN models \cite{Subramanian2021} \cite{Liu2020} \cite{Qin2019}   . Some works reviewed the steganalysis detection techniques using rich models or domain based extractions along with Machine learning models  \cite{Karampidis2018} \cite{Evsutin2020} and other works mainly reviewed deep learning based steganalysis detection techniques \cite{Tabares-Soto2019} \cite{Ruan2020} \cite{Selvaraj2021}. None of these works provided a holistic overview of the information hiding and hidden data detection particularly in the context of the hiding malware in images. So, we could see that our work can be good reference to continue research either steganography or steganalysis particularly using deep learning or GAN models and most importantly in the context of enterprise security for advanced stegomalware detection.

\begin{sidewaystable}
\centering
\caption{Comparison of the State-of-the art Reviews in Multimedia Steganography and Steganalysis}\label{T:priorreview}
\resizebox{\textwidth}{!}{%
\begin{tabular}{|l|l|l|l|l|l|l|l|l|l|l|l|l|l|l|l|}
\hline
\multirow{2}{*}{\textbf{Authors}}                                     & \multirow{2}{*}{\textbf{Cover Image}} & \multicolumn{4}{l|}{\textbf{Steganography}}    & \multicolumn{5}{l|}{\textbf{Steganalysis}}      & \multirow{2}{*}{\textbf{Datasets}} &   \multirow{2}{*}{\textbf{Metrics}} & \multirow{2}{*}{\textbf{File Formats}} & \multirow{2}{*}{\textbf{Attack History}} & \multirow{2}{*}{\textbf{ Framework}} \\ \cline{3-11}
                                                              &  & Spatial & Transform  & DL  & GAN & Spatial & Frequency & ML  & DL  & Tools &                           &                                          &  &                               &                                         \\ \hline
Mazurczyk et al. \cite{Mazurczyk2015}       & Yes     & Yes     & Yes    &     &     & Yes     & Yes       & Yes &     & Yes   &             &               &                                          & Yes                             &                                         \\ \hline
Karampidis et al. \cite{Karampidis2018}     & Yes         &         &        &     &     & Yes     & Yes       & Yes & Yes &       & Yes             &            &                                          &                                 &                                         \\ \hline
Inas et al. \cite{Kadhim2019}               & Yes        & Yes     & Yes        & Yes &     &         &           &     &     &       &                  &         Yes  &                                          &                                 &                                         \\ \hline
Nandhini et al.  \cite{Subramanian2021}     & Yes         & Yes     &    & Yes & Yes &         &           &     &     &       & Yes               &          &                                          &                                 &                                         \\ \hline
Jia et al. \cite{Liu2020}                   & Yes       &         &           &      & Yes &         &           &     &     &       &                    &         &                                          &                                 &                                         \\ \hline
Abduallah et al. \cite{Abduallah2016}       & Yes         & Yes     & Yes       &       &     &         &           &     &     &       &                  &           &                                          &                                 &                                         \\ \hline

Mehdi et al. \cite{Hussain2018}             & Yes        & Yes     &         &     &     &         &           &     &     &       &                      &       &                                          &                                 &                                         \\ \hline
Tabares-soto \cite{Tabares-Soto2019}        & Yes       &         &           &     &     &         &           &     & Yes &       &                  &           &                                          &                                 &                                         \\ \hline
Evsutin et al. \cite{Evsutin2020}           & Yes        & Yes     & Yes     &     &     & Yes     & Yes       & Yes &     &       &            Yes     &            &                                          &                                 &                                         \\ \hline
Jiaohua \cite{Qin2019}                      & Yes     & Yes     & Yes  & Yes & Yes &         &           &     &     &       &                 &            &                                          &                                 &                                         \\ \hline

Tabares et al.\cite{Tabares-Soto2020}       & Yes    & Yes       & Yes  & Yes &     &           &           &     & Yes &       &                 &            &                                          &                                 &                                         \\ \hline
Ruan \cite{Ruan2020}                        & Yes         &           &           &     &     &           &           &     & Yes &       &                   &          &                                          &                                 &                                         \\ \hline
Selvaraj \cite{Selvaraj2021}                & Yes        &           &    &     &     &           &           & Yes & Yes &       &                   &          &                                          &                                 &                                         \\ \hline
Chutani \cite{Chutani2019}                  & Yes       &           &     &     &     &           &           & Yes & Yes &       &                  &         Yes  &                                          &   
        &                                         \\ \hline     
Our work                                                     & Yes    & Yes     & Yes        & Yes & Yes & Yes     & Yes       & Yes & Yes & Yes   & Yes                & Yes       & Yes                                      & Yes                             & Yes                                     \\ \hline
\end{tabular}%
}
\end{sidewaystable}

\begin{table}[]
\renewcommand{\arraystretch}{0.1}
\centering
\caption{Description of the Abbreviations}\label{T:Abbreviation}
\resizebox{\columnwidth}{!}{%
\begin{tabular}{|l|l|}
\hline
\textbf{Acronym} & \textbf{Description}                           \\ \hline
ACGAN   & Auxiliary Classifier GAN\\ \hline
ANOVA & analysis of variance \\ \hline
ASDL    &   Automatic steganographic distortion learning \\ \hline
APT     & Advanced Persistent Threat             \\ \hline
BSM & Binary Similarity Measures \\ \hline
BMP     & Bitmap                                 \\ \hline
BBC     & Block Boundary Continuity            \\ \hline
CNN     & Convolutional Neural Networks          \\ \hline
CRM & color rich model\\ \hline
DCT     & Discrete Cosine Transform              \\ \hline
CWT     & Complex Wavelet Transform              \\ \hline
DFT     & Discrete frequency transform          \\ \hline
DRI     & Define Restart Interval                \\ \hline
DTCWT   & Dual Tree Complex Wavelet Transform    \\ \hline
DWT     & Discrete Wavelet Transform             \\ \hline
DQT     & Define Quantization table              \\ \hline
DBN     & Deep Brief Network                     \\ \hline
DNS     & Domain Name Server                     \\ \hline
DCTR & Discrete Cosine Transform Residual \\ \hline
DAT     & Define Arithmetic Coding               \\ \hline
DCTR    & discrete cosine transform residual     \\ \hline
DHT     & Define Huffman Table                   \\ \hline
DLL     & Dynamic Link Library                   \\ \hline
ELF     & Executable and Linkable Format         \\ \hline
EC & Ensemble Classifier \\ \hline
EA      & Edge Adaptive                        \\ \hline
GIF     & Graphics Interchange Format            \\ \hline
GFR & Gabor filter Residual \\ \hline
GNCNN   &     Gaussian-Neuron CNN            \\ \hline

GRDH    & Generative reversible data hiding\\ \hline
GAN     & Generative Adversial Networks          \\ \hline
HUGO    & High Undetectable steGO                \\ \hline
Hidden  & Hiding data with deep networks \\ \hline
HIGAN & Hiding images GAN\\ \hline
HILL    & High-pass, Low-pass, and Low-pass      \\ \hline
IWT     & Integer Wavelet Transform              \\ \hline
ICMP    & Internet Control Message Protocol      \\ \hline
IUERD   & Improved UERD   \\ \hline
JPEG    & Joint Photographic Experts Group       \\ \hline
JRM & JPEG Rich model \\ \hline 
LSB     & Least Significant Bit                  \\ \hline
LSER & Local source enhanced residual \\ \hline
ML        &    Machine learning                  \\ \hline
MME     & Modified Matrix Encoding      \\ \hline
MiPOD   & Minimizing the Power of Optimal Detector  \\ \hline
MPEG    & Moving Picture Experts Group           \\ \hline
MSB     & Most significant bit                   \\ \hline
NPQ     & Normalized Perturbed Quantization     \\ \hline
PE      & Portable Executable                    \\ \hline
PCM     & pulse code modulation                  \\ \hline
PNG     & Portable Network Graphics              \\ \hline
PPM     & Pixel Pair Matching                    \\ \hline
PQt     & texture-adaptive Perturbed Quantization \\ \hline
PVD     & Pixel value differencing              \\ \hline
PQ      & Perturbed Quantization      \\ \hline
PHARM & PHase Aware pRrojection Model \\ \hline
PRSM & projection spatial rich model \\ \hline
QMF & quadrature mirror filters \\ \hline
RIFF    & Resource Interchange File Format       \\ \hline
RNN     & Recurrent Neural Network               \\ \hline
ReLU    & Rectified Linear Unit                \\ \hline
RBF     & Radial Basis Function                  \\ \hline
SVM     & Support Vector Machine                 \\ \hline
SPAM    & Second-order Markov chains             \\ \hline
SEGAN   & Steganographic Encryption GAN\\ \hline
SSGAN   & Secure Steganography Based on GAN \\ \hline
SPAM & Subtractive Pixel Adjacency Model \\ \hline
SGF & steerable Gaussian filter\\ \hline
SRM & Spatial-domain Rich Model  \\ \hline
SGAN    & Steganographic GAN \\ \hline
SCA & Selective Channel Aware \\ \hline
TIFF    & Tagged Image File Format               \\ \hline
TLU & truncated linear unit \\ \hline
TCP     & Transmission Control Protocol          \\ \hline
UDP     & User Datagram Protocol                 \\ \hline
WAV     & Waveform Audio File Format             \\ \hline
WGAN    &     Wassertian GAN                        \\ \hline
WOW     & Wavelet Obtained Weights               \\ \hline
UNIWARD & universal wavelet relative distortion  \\ \hline
UED     & Uniform Embedding Distortion           \\ \hline
UERD    & Uniform Embedding Revisited Distortion \\ \hline
\\ \hline

\end{tabular}%
}
\end{table}

%% file: S11Stegohistory.tex
\section{Stegomalware History}
\label{sec:history}

We discuss the existing malware families and their behavior when leveraging the steganography on multimedia files to hide malicious commands or source code. The steganography technique is rarely seen in malware generation. However, this technique is used to create sophisticated malware for organized crimes like operated by APT groups and hard to detect those malware attempts. So, it is indeed very important to detect the stegomalware targeting the enterprises and other sectors. The APT malware “Operation shady RAT” was first seen to be using digital images to hide the command C\&C server addresses in 2011 and use it to connect the C\&C servers from the compromised victim machines. The malware spread through targeted phishing email campaign and compromise the machine when the user downloaded the attached file from the email \cite{Higgins2011}. NSS lab researchers discovered that the duqu malware targeting the industrial manufacturers has steganography process to embed the victim data into an image file prior to sending to the attacker controller servers. This is a clear indication that the image file can be used to hide the secret content and send through the internet in image form so that network level detection mechanisms not able to detect the data exfiltration \cite{Goodin2011}.

Zeus variant Trojan used for stealing money from victim bank account incorporates the configuration file in the image for deceiving the enterprise security tools. The configuration file may include the list of bank details, malicious code to steal the online banking details and hijacking the login details to the attackers. To conceal the configuration file, the encoding and symmetric encryption techniques applied on the cover image \cite{Paganini2014}.  In \cite{Mosuela2016}, the authors describe that the Gatek/stegoloader encrypts the shellcode and hidden in the digital image. The shellcode contains bot command to retrieve web browsing history, victim system information, executing payload and retrieving the software installed in the machine. When the malware installed in the victim machine, it connects to the C\&C server to download the stegomalware into the machine. Then, decrypt the shellcode from the image to run bot commands and exfiltrate the victim data. TeslaCrypt is a ransomware Trojan to encrypt the victim data once the targeted machine is compromised and expect ransom to receive the decryption key. Steganography is used to conceal the C\&C server commands in the cover image \cite{Teslacrypt2015}.  The stegomalware images displayed on the website ads click using the web browser may be enough to infect the victim machine. Stegosploit malware exploit can leverage the HTML5 $<$canvas$>$ tag to make the browser read the pixel data as JavaScript and execute the malware functionality. This technique can be used to exfiltrate the machine data to the attacker controlled server \cite{Miao2016}. Cerber Ransomware uses the steganography image to store the malicious executable code. So, when the victim clicks on the phishing email delivered with malicious word document, the downloaded document runs the macro code. This code loads a steganographic image to run the malicious executable code and perform the encryption of all the data on the machine \cite{AnkitAnubhav2016}. 

\begin{table*}[h!]
\centering
\caption{Historical malware variants employing steganography for exploitation}\label{T:history}
\resizebox{\textwidth}{!}{%
\begin{tabular}{|l|l|l|l|l|l|}
\hline
\textbf{Malware}                            & \textbf{Technique}                                & \textbf{Exploitation Stage}                & \textbf{Targeted Industry}                            & \textbf{File Type}    & \textbf{File Format}                 \\ \hline
Operation Shady RAT  \cite{Higgins2011} & Phishing email                           & C\& C server connection   domains & Government, International Corp, Nonprofit    & Image        & Not known                   \\ \hline
Duqu \cite{Goodin2011}                & Installing Rootkit                       & Data exfiltration                 & Industries                                   & Image        & Jpg                         \\ \hline
ZeusVM \cite{Paganini2014}            & Maladvertising campaign                  & Exfiltration                      & Banking sector                               & Image        & Jpeg                        \\ \hline
Gatak/ Stegoloader \cite{Mosuela2016}& Hosting malicious image in legit website & Download malware                  & -                                            & Image        & Png                         \\ \hline
teslacrypt \cite{Teslacrypt2015}      & Browsing malicious Page                  & Download C\&C commands             & Generic Internet users                       & Image        & Jpeg                        \\ \hline
Stegosploit \cite{Miao2016}           & Leverage HTML5 canvas tag                & Download malicious code           & Internet users                               & Image        & Not known                   \\ \hline
Cerber \cite{AnkitAnubhav2016}       & Phishing email                           & Malware Delivery                  & Various sectors                              & image        & Jpeg  \\ \hline
DNSchanger \cite{Khandelwal2016}       & Advertisements                           & Malware Delivery                  & Internet Users running vulnerable Routers    & Image        & Png                         \\ \hline
Vawtrak \cite{Davis2016}          & Hide in Favicon Icon                     & Download malware                  & Internet users                               & Image        & Favicon icon                \\ \hline
AdGholas \cite{Kafeine2017}            & Maladvertising Campaign                  & Exploitation                      & Education, Travel                            & image        & jpeg                        \\ \hline
Sundown \cite{Olenick2017}             & Malvertising campaign                    & Exfiltration                      & Internet users                               & Image        & Png                         \\ \hline
Synccrypt \cite{Socprime2017}          & Click Malicious URL                      & Install malware                   & Generic Internet users                       & Image        & Jpeg                        \\ \hline
ZeroT \cite{NJCCIC2017}               & Phishing Campaign                        & Command and Control               & Not known the target                         & Image        & Bmp                         \\ \hline
Verymal \cite{Goodin2019}              & Maladvertising                           & Downloading Shlayer Malware       & Internet users                               & Image        & Jpeg                        \\ \hline
Waterbug \cite{Cimpanu2019}           & Legitimate application vulnerabilities   & Downloading DLL                   & Government, Education, IT  & Audio        & Wav                         \\ \hline
Loki Bot \cite{Osborne2019}          & Phishing emails                          & Install malware                   & Internet users                               & Image, Video & Jpg, Video formats          \\ \hline
\end{tabular}%
}
\end{table*}

The stegomalware is also extensively being used recently to target the users visiting websites. For instance, the “stegano” exploit conceal the malicious code in the pixel of banner advertisements on websites. When a user clicks on the advertisements, the malicious code run on the machine and may install the DNSchanger to change the local DNS IP address. From then onwards, the victim machine directs the DNS requests to the attacker-controlled DNS server. An attacker may redirect the victim to the phishing website to install drive-by download malware and more, when the victim perform a legitimate request to trusted sites \cite{Khandelwal2016}.  AdGholas group targeted UK universities infecting with ransomware in 2017. Further investigation on the infection revealed that the maladvertising campaign involves stegano exploit for the widespread infection of the ransomware \cite{Kafeine2017}. These real time attack instances portray the stegomalware infection through maladvertisements has a large scale widespread across the nations.  Sundown pirate exploit kit also use stegomalware hiding the vulnerabilities of popular application in the cover image to test against the victim machine and execute the vulnerabilities code to infect the victim machine. When the user visits a website, the maladvertisements has iframe links to redirect them into malicious URL hosting a white image and download the image into the victim machine. The white image is a stegomalware, which can infect the machine with known vulnerabilities in popular application \cite{Olenick2017}.  

A researcher from Emmisoft discovered that Synccrypt ransomware distributing the malicious code zip file using steganographic JPEG images. The attack vector consists of sending phishing emails with attached windows script file (WSF); when the user run the script, it downloads the JPG stegomalware malware with payload embedded in the zip format and extract the ransomware components from the zip file so that it can create windows scheduled task to run the ransomware for encrypting the victim data \cite{Socprime2017}. ZeroT Trojan malware mainly used by China espionage group reportedly using PlugX RAT module to infect the targeted industries. The infection process involves sending a phishing email with malicious doc file as an attachment. If the victim user opens the document and agrees to run the hidden malicious application, the malware application is installed on the victim machine. Then, the malware connects to C\&C server to exfiltrate the user data and downloads the stegomalware image containing the PlugX RAT payload for performing malicious activities on the victim machine. 

Recently, shlayer adware is extensively used by adversary to infect the Mac machines. An adversary usually tricks the user to install fake adobe player for downloading the shlayer adware. A recent maladvertisements campaign noted to be downloaded by 5 million users in a day, which contains the stegomalware as part of the attack lifecycle. The stegomalware image hide the malicious URL used for downloading the fake adobe player. An HTML5 object run through the image when user click on the advertisements and in each loop, it extracts a malicious URL character. Once the loop is completed, all the extracted characters clubbed together to form the malicious URL and then redirects to the malicious website hosting fake player \cite{Goodin2019}. The audio files are also considered as a cover media for performing malicious activity. Waterbug cyber espionage group utilizing the audio file format WAV to conceal the malicious DLL files for installing on the victim machine during the attack lifecycle \cite{Cimpanu2019}.

Table \ref{T:history} illustrates the list of malware seen in the history using stegomalware to hide their activity or download the malicious code without being detected by anti-malware tools. Based on the description of the role of stegomalware in famous malicious campaigns and the collection of attacks in Table \ref{T:history}, we can observe that most of the malware leverage the image files formats for carrying the payload and occasionally seen video and audio formats as a payload carrier. The malware infection in the history shows that these stegomalware are targeting multiple industries and particularly used to evade the malware detection by Deep Packet Inspection (DPI) or endpoint security tool detection capabilities. Furthermore, we identify that the stegomalware is mostly seen to conceal the malware payload, exfiltrate the data in the form of steganographic multimedia files and store the C\&C IP address, domain name and the commands to communicate with the attacker server. Overall, it is evident that the stegomalware is popularly used by cyber espionage groups as part of attack propagation, even though it is not frequently touted in malware industry.

\textbf{Lesson Learned}:
Our stegomalware history review clearly show that the stegomalware is used in many cyberattack campaigns to hide the artifacts or malware payloads and evade the security detection. We are convinced that cyberattacks using stegomalware to hide the activity will keep growing, and the organizations should monitor and adapt image file stegomalware detection solutions.

%% file: S12stegotools.tex
\section{Stegomalware Creation tools}
\label{sec:tools}

There are various techniques used to conceal the payload in the multimedia files based on the cover media type and payload type. Over the years, there were a number of tools developed to perform the compression, encryption and encode the payload so that the scrambled data can be inserted into the multimedia files. Table \ref{T:tools} depicts the popular tool set used for performing steganography on the audio, image and video file formats but not limited to. As the image form as cover media is most used for hiding the content and deceiving the users, most of the existing tool’s support to perform steganography in images. The most popular steganography tools steghide, openstego, Hide’N’Send, SSuite Piscel, Camouflage, Xiao, Openpuff are used for performing steganography and most of them supports JPEG, BMP file formats, as shown in Table \ref{T:tools}. Some of the tools Xiao, Steghide, Deepsound, Openpuff and Silenteye support for hiding the content in audio files like wav, mp3. But few of the tools openpuff and camouflage supports for hiding the content in video file formats. We can observe that most of the tools supports windows operating system to hide content. The popular tool steghide has open source code available for analysis written in C++ and using advanced LSB techniques for hiding the content. Openpuff offering the subscription-based solutions for steganography and provides advanced capabilities like multi-layered data obfuscation. 

\begin{table*}[]
\centering
\caption{Steganography tools for multimedia files}\label{T:tools}
\resizebox{\textwidth}{!}{%
\begin{tabular}{|l|l|l|l|l|l|l|l|}
\hline
\textbf{Tool}                             & \textbf{Year} & \textbf{Last Update} & \textbf{OS}                  & \textbf{Image} & \textbf{Audio} & \textbf{Video} & \textbf{Supported Format}                                                     \\ \hline
Steghide\cite{StefanoDeVuono2003}    & 2003 & 2013        & Windows,   Linux    & Yes   & Yes   & No    & Jpeg, Bmp, Wav and Au                                                \\ \hline
Openpuff \cite{OpenPuff2004}         & 2004 & -           & All Platforms       & Yes   & Yes   & Yes   & Bmp, Jpg, Png and more \\ \hline
Xiao \cite{Steganography2006}       & 2006 & 2007        & Windows             & Yes   & Yes   & No    & Bmp, Wav                                                             \\ \hline
MP3stego \cite{MPStego2006}       & 2006 & 2015        & Windows,   Linux    & No    & Yes   & No    & Mp3                                                                  \\ \hline
Crypture  \cite{Crypture2007}        & 2007 & 2007        & Windows,   Linux    & Yes   & No    & No    & Bmp                                                                  \\ \hline
Our Secret \cite{Oursecret2008}      & 2008 & 2013        & Windows             & Yes   & No    & No    & Generic image   file format                                          \\ \hline
SteganographX Plus \cite{Plus2010}   & 2010 & 2010        & Windows             & Yes   & No    & No    & Bmp                                                                  \\ \hline
Hide’N’Send \cite{HideNSend2012}  & 2012 & 2012        & Windows             & Yes   & No    & No    & Jpeg                                                                 \\ \hline
SSuite Piscel \cite{Piscel2014}       & 2014 & 2018        & Windows             & Yes   & No    & No    & Bmp, Png                                                             \\ \hline
SteganPEG \cite{SteganPEG2014}       & 2014 & 2014        & Windows             & Yes   & No    & No    & Jpeg                                                                 \\ \hline
Silenteye \cite{Silenteye2014}      & 2014 & 2020        & All Platforms       & Yes   & Yes   & No    & Generic image   and audio                                            \\ \hline
rSteg \cite{Rsteg2015}             & 2015 & 2015        & All Platforms       & Yes   & No    & No    & Png                                                                  \\ \hline
Openstego \cite{Openstego2015}      & 2015 & 2020        & Windows and   Linux & Yes   & No    & No    & Bmp, Gif,   Jpeg, Jpg, Png and Wbmp                                  \\ \hline
Outguess for mac \cite{Rbcafe2007}   & 2017 & 2018        & Mac                 & Yes   & No    & No    & Jpeg                                                                 \\ \hline
Deepsound 2.0 \cite{DeepSound2.02018} & 2018 & 2018        & Windows             & No    & Yes   & No    & Wave,   Flac                                                         \\ \hline
\end{tabular}%
}
\end{table*}

\textbf{Lessons Learned}:
There are number of public tools available for performing steganography and hiding the malware payload or data. However, the advanced steganography solutions like UERD, UNIWARD, Mi-POD, HILL may not be available in in-built tools for generating stego images. Overall, the stegomalware generation using the existing tools is much easier and definitely a concern for organizations to defend stegomalware. 

%% file: S2fileformat.tex
\section{Stegomalware multimedia file formats}
\label{sec:fileformat}
\begin{figure}[H]	
\centering
\includegraphics[width=8.3cm,height=6cm]{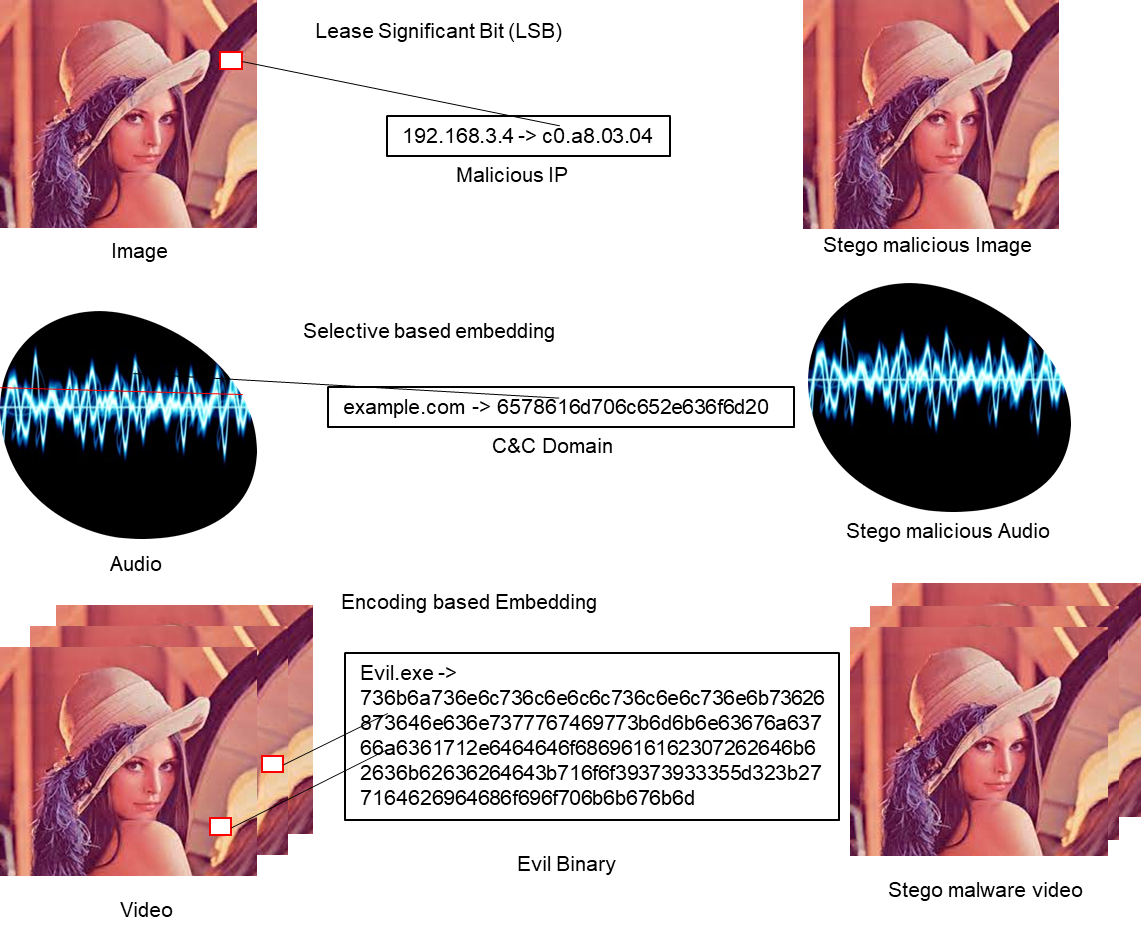}
\caption{Multimedia Stegomalware} 
\label{f.mmstego}
\end{figure}

The stegomalware payload embedding in different multimedia file formats such as image, audio or video is illustrated in the Figure \ref{f.mmstego}. The cover image in the Figure \ref{f.mmstego} is embedded with malicious command and control server IP address payload using lease significant bit algorithm. During the attack progress, the malware triggers to retrieve the malicious IP address from the cover image and initiate outbound connection to that IP address for data filtration or other malicious activity. In cover audio file scenario, the command and control server domain name is embedded with audio using selective based embedding steganography algorithm. The same malware outbound communication process seen in cover image  applicable for this scenario as well. In video cover steganography, the malware binary "evil.exe" is converted into hexadecimal representation and encoding based steganography is used to embed the malware binary in consequent frames of a video. As we have seen in Section \ref{sec:history}, most of the stegomalware used image as a cover for hiding the malicious artifacts. So, we have decided to give a detailed description of image file format specifications here.

The multimedia file formats used to hide the malicious payloads in the real time attacks seen over the last decade are JPG, BMP, GIF, PNG and WAV formats. The detail description of these file format structure is needed to understand where the attackers hide the malicious content and see the possibilities identifying the file features to accurately detect the stegomalware in each of these file formats. 

JPEG stands for Joint Photography Experts Group and the JPEG standard created in 1992. It represents the digital image in compressed form using the DCT algorithm and widely used standard for viewing the images in the digital devices. The JPEG images ends with the .jpg or .jpeg format \cite{Langjpeg}. 

\begin{figure}[H]	
\centering
\includegraphics[width=8.3cm,height=6cm]{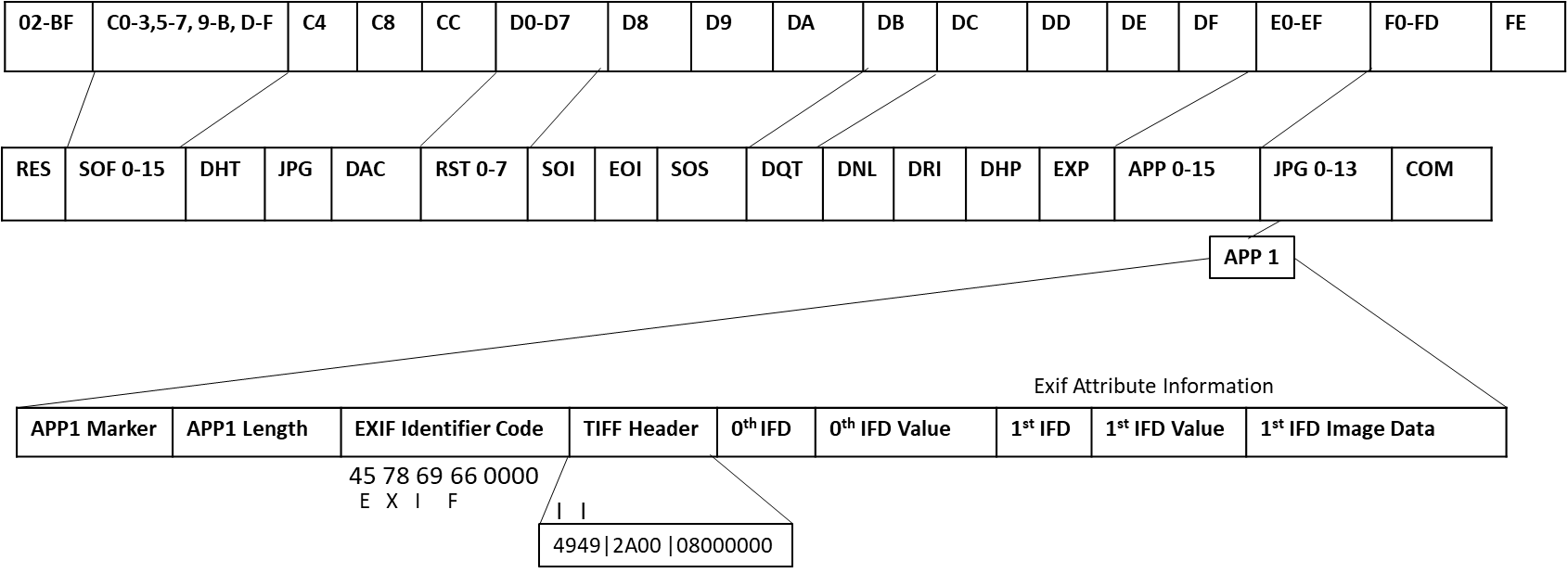}
\caption{JPEG and Exif file structure format} 
\label{f.s2jpeg}
\end{figure}

The Figure \ref{f.s2jpeg} illustrates the different markers used to represent the image data in the JPEG structure format; the markers hexadecimal representation to identify the markers in the given image \cite{Steenmarkers} and the Exif file format structure to store the image captures or processed device attributes and other information \cite{exif}. The JPEG markers categorization start with the value “0x FF02” and ends with 0x FFFE. The byte “FF” is not shown in the Figure 2.1 for simplification and good presentation purpose. The JPEG image usually starts with marker Start of the Image (SOI) data “0x FFD8” and ends with End of Image “0x FFD9”. The encoded content in the image is stored inside the frames. Each frame typically contains more than one scan units to store the image content. A JPEG may contain up to 15 frames starting with the segment “Start of Frame” and having a hexadecimal values in between the range from 0xFFC0- 0xFFCF.  The Start of Scan (SOS) is used to store the encoded image components and can be identified with the hexadecimal value 0x FFDA in the image. The JPEG images may also contain “Reserved” marker segment starts with 0xFF02-0xFFBF to add additional information to the JPEG image. The Define Huffman Table (DHT), Define Quantization table (DQT), and the Define Arithmetic Coding (DAT) store the Huffman, Quantization and Arithmetic coding data used during the lossless compression of the JPEG image. These markers start with 0x FFC4, 0x FFDB and 0x FFCC are used to parse the image and extract the relevant data for decompressed image presentation. Some of the markers can be repeated based on the Define Restart Interval (DRI) in the JPEG image representation and these can be started with 0xFFD0-0xFFD7 having a naming convention Restart Marker 1 to 7. The Define Hierarchical Progression (DHP) and Expand Reference Component (ERC) are not commonly used markers and assigned the hexadecimal values 0xFFDE and 0xFFDF respectively to identify them in the images. There are various JPEG extensions 0-13 options defined in the JPEG specifications having a marker hex ranges 0xFFF0-0xFFFD. JPEG extension 7 is commonly used for lossless JPEG image representation. The JPEG specification also contain the application segments 0-15 to add the application content in the image and can be identified with hex values 0xFFE0-0xFFEF.

The App1 segment 0xFFE1 is seen to be used for representing the Exif metadata, JPEG thumbnail and TIFF IFD format, when the image is taken from digital camera or the creation of image using the Adobe extensible metadata platform. The App1 data comprises the APP1 data size represented in bytes length, Exif Identification code, Tiff header, 0th Image File Descriptor (IFD), 1st IFD value and thumbnail image data for representing the Exif data in Tiff format. The data size is not a fixed value, and it depends on the other fields in the Exif format.  To identify the Exif format, the Exif identified value is assigned as hex value “Exif”. The tiff header contains 2bytes of byte align definition of the data, 2 bytes of Tag Mark “0x2A00” and the offset value to first IFD “0x00000008”.  The byte align values can be “I I” or “M M”, which represents either Intel or Motorola byte align representation of the data. The 0th IFD represents the main information image data and the stored in the 0th file directory, and it links to the 1st IFD; the 1st IFD represents the IFD of thumbnail image data and the link is terminated. IFD0 may also contain the Exif SubIFD, which stores the digital cameras information. Exif format support three formats JPEG, RGB TIFF and YCbCr TIFF format to represent the thumbnail images. If the thumbnail image is saved in JPEG, then the structure of the JPEG follows the same structure format as shown in Figure \ref{f.s2jpeg}.  Having seen the typical JPEG file format specifications so far, there are many ways an adversary hides the malicious payload in JPEG images. In particular, the Exif data and thumbnail images storage give more space to embed the malware content.

\begin{figure}[H]	
\centering
\includegraphics[width=8.3cm,height=6.5cm]{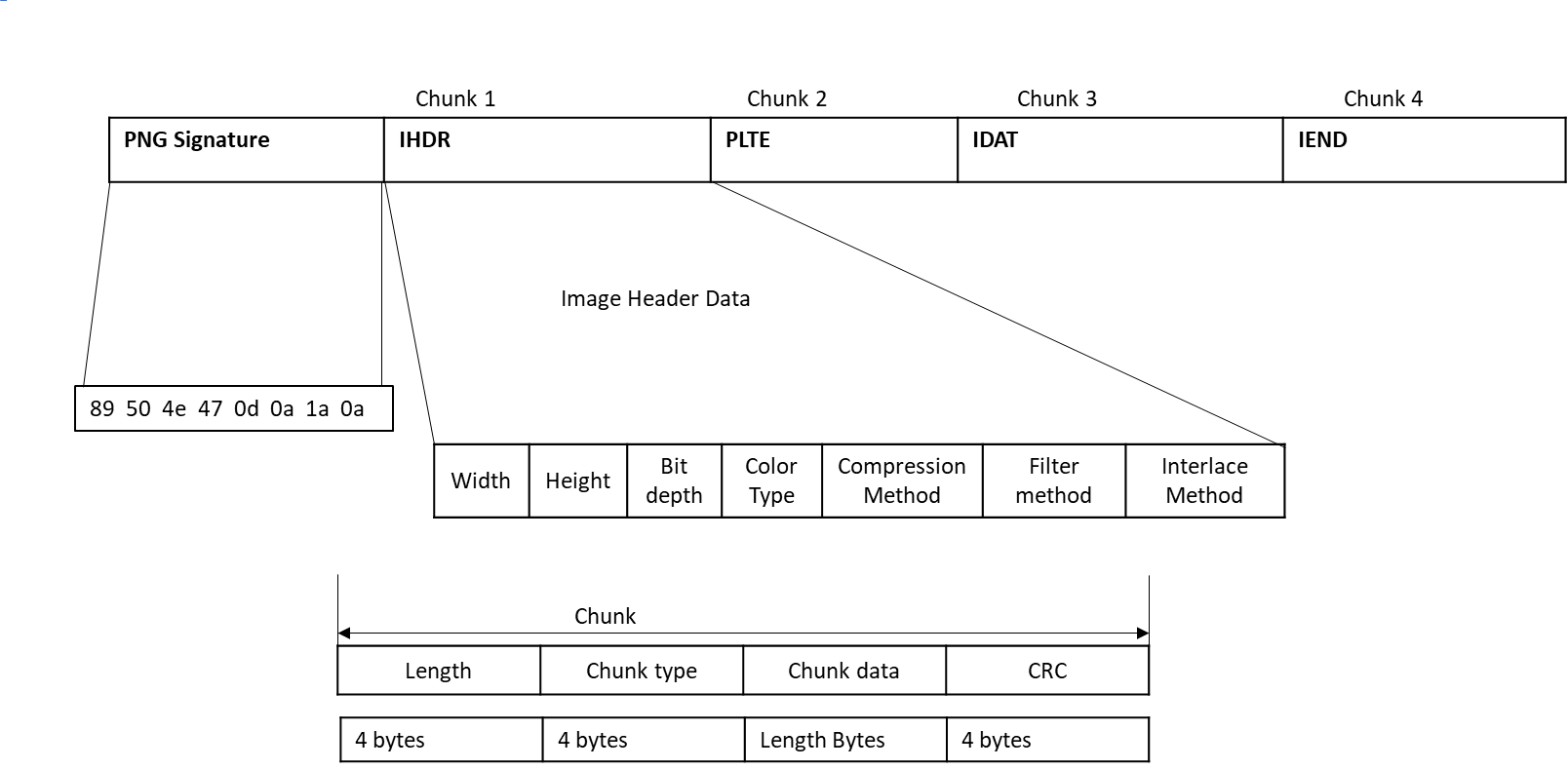}
\caption{PNG file structure format} 
\label{f.s2png}
\end{figure}

The PNG image file start with PNG hex decimal signature “89 50 4e 47 0d 0a 1a 0a” and the byte sequence ”0x50 0x4e 0x47” represents in ASCII letters as “PNG” is used to identify the given image is PNG image or not. The value “89” is used to detect the transmission systems that do not support the 8-bit data and eliminating the chances of mistakenly interpreting the text file as PNG files. The last 4 bytes in the signature is used to detect the DOS-Unix line ending conversation of the data \cite{PNG}. 
The PNG data is divided into chunks in the image. Each chunk has four fields as shown in the Figure \ref{f.s2png}. The chunk data size is determined by the first 4 bytes assigned for representing the data length in bytes. The chunk types can be classified as Critical chunks and Ancillary chunks. The critical chunks must be presented in the image to read by the image readers. There are four chunks comes under critical chunks. Those are Image Header (IHDR), Palette (PLTE), Image data (IDAT), Image End (IEND). The Cyclic Redundancy check (CRC) is used to perform the integrity check on the chunk data for errors and data corruption. The four critical chunks are shown in the PNG specification format Figure \ref{f.s2png}. The image header chunk type has a length of 13 bytes and contains the image metadata. The PNG image width and height values consumes the first 8 bytes of the image data field. The number of bits per palette or index represented by the one-byte bit depth. The one-byte color type signifies the grey scale or color of the images. The two bytes used for compression and filter method enables the image compression and filtering the image before compression if needed. The interlace method can have two values: no interlace and Adam 7 interlace. When no interlace is selected, the image is scanned from the top to bottom and the pixel scan starts from left to right sequentially. 
The Palette chunk may contain 1 to 256 palette entries and each entry can be represented in RGB image form. The number of entries can be selected based on the chunk length. The Image Data (IDAT) contains the image data processed using the meta data assigned in the Image header, which includes finding the size of the data and performing filtering and compressing the data.  The Image End (IEND) chunk contains a single empty byte to mark the end of the PNG data stream. In addition to the critical chunks, the optional ancillary chunks like bKGD, sBIT, tIME can be embedded in the PNG image for conveying other information about the image. The detail description of the ancillary chunks can be found \cite{pngchunk}.

\begin{figure}[H]	
\centering
\includegraphics[width=7cm,height=7cm]{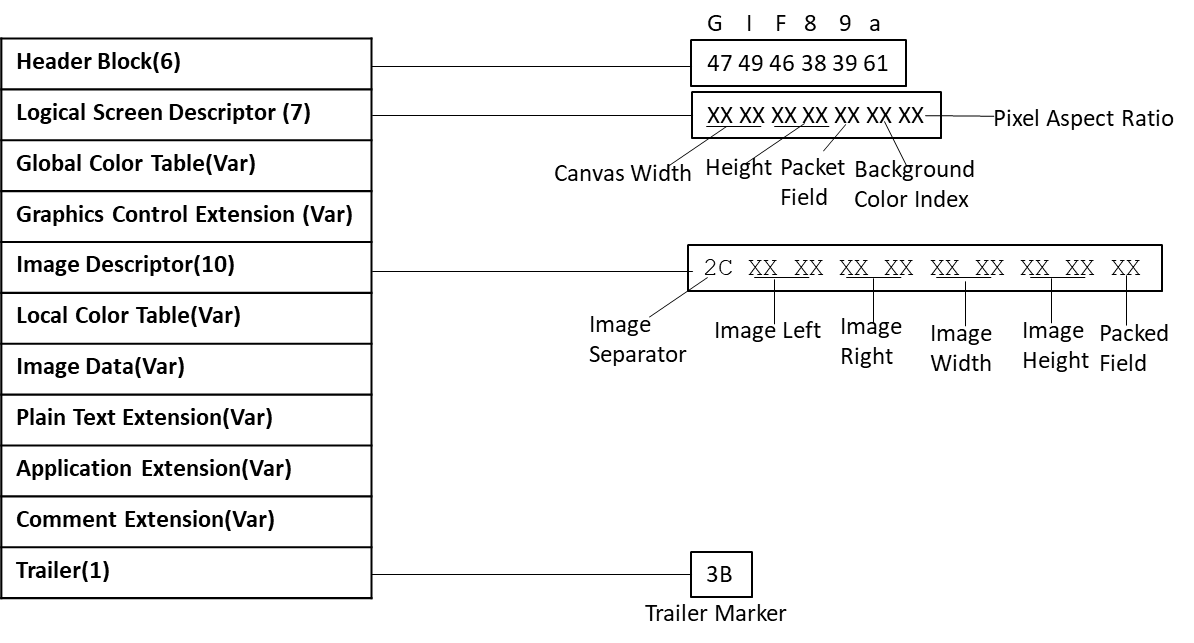}
\caption{GIF file structure format} 
\label{f.gif}
\end{figure}

The GIF image format ads can be used by adversary to host fake ads with embedded malicious URLs so that the victims are redirected to the malicious web pages when the user click the ad page. The first three bytes representing as “GIF” in hex decimal can be used to determine if the image is GIF or not. The next three bytes signifies the GIF version. As shown in Figure \ref{f.gif}, the GIF image has version 89a in the image. These first 6 bytes forms the GIF image header block.  The next block “Logical screen descriptor” is used for knowing the amount of space needed for the GIF image. The canvas width and height determine the image size and the pixel aspect ratio to set the proportions of the width and height.  The packet field does contain multiple color flags like Sort flag, color resolution, Global color table flag and size global color table, which are used to set the global color aspects of the GIF image. If the global color table flag set in the logical screen descriptor, the global color table should be followed to specify the global color image configurations. The color table consists of the RGB color intensities represented color values in between 0 and 255. The GIF may use the global color table for all the sub images or local color table for each sub images in the GIF. Based on the color depth value mentioned in the logical screen descriptor, the global color table is determined and used for coloring the GIF canvas image. The graphical control extension is optional and varied in length. This block specifies the transparency settings and control the image animations. The first byte for extensions always starts with “21”, which is also called extension introducer and ends with block terminator “00”. The image descriptor block stores the local or sub images data and number of these blocks used if the number of sub images present in the GIF image. The first byte is an image separator and denoted by the hex value “2C”. The next 8 bytes represents the position and size of the local image in the GIF image. The packet field contains multiple color related flags including the local color table flag. If the local color table flag is enabled, the next block can be local color table for representing the color aspects of the local image. If the local color table flag is not enabled, the global color table is used for the local image color. Then, the image data block stores the data in data sub-blocks. The image data block starts with LZW minimum code size “02” and the sub block first bytes signifies the length of the sub block. So, the next number of the bytes in the sub block are the real data to read until reaches the length of the bytes and then the next subblock length can be seen in the first byte and continue this procedure to read all data until the last sub block reaches to “00” block terminator. There are few extensions such as plaint text, application and comment extensions are used for specifying the text captions in the GIF image, embedding the application specific information in the GIF image and adding the ASCII text information to GIF image, respectively. These extension blocks are all optional to the GIF image. The image ends with trailer marker “3B” to confirm that the image is ended. However, an adversary may add the malicious payload after the trailer marker to hide the content and infect the victim machines when triggered in the scheduled action of items. 

The BMP file format is also called as device independent bitmap (DIB) image file format used to store the two-dimensional images both in gray scale and color \cite{bmp}. Microsoft defined the image representations with color resolutions and can still map the image colors with internal representations even when move the bitmap move from one device to another device. The BMP file starts with 14 bytes file header. The BMP signature hex value “42 4D” is used to determine the BMP images and the next four bytes contains the file size. The following next 4 bytes are reserved and application specific. These fields usually unused and contain zero-byte values. The last 4 bytes in the file header signifies the starting address of the BMP image data. The next DIB header represents the image information such as size, compression, color of the BMP image and the length of the header based on the version used for representing the image. The first 4 bytes are used for storing the DIB header size value. The next following fields are as follows 4 bytes for image width, 4 bytes for image height, 2 bytes for number of planes, 2 bytes for bits for pixel, 4 bytes for compression, 4 bytes for image size, 4 bytes for horizontal pixels per meter, 4bytes for vertical pixel per meter and 4 bytes for colors used. The color palette typically depends on the number of colors mentioned in the DIB header. The Gap1 and Gap2 are usually optional for the BMP image. Depends on the image size described in the DIB header, the image data is presented in the pixel array header. The International color consortium (ICC) color profile header only exist when the Bitmap version 5 header is used as a DIB header to link the color profile data to the image. Due to the historical changes of the BMP headers, an adversary can leverage the different versions and can still be able to hide the content so that the BMP image can be used as a stegomalware carrier.

\begin{figure}[H]	
\centering
\includegraphics[width=8.5cm,height=6cm]{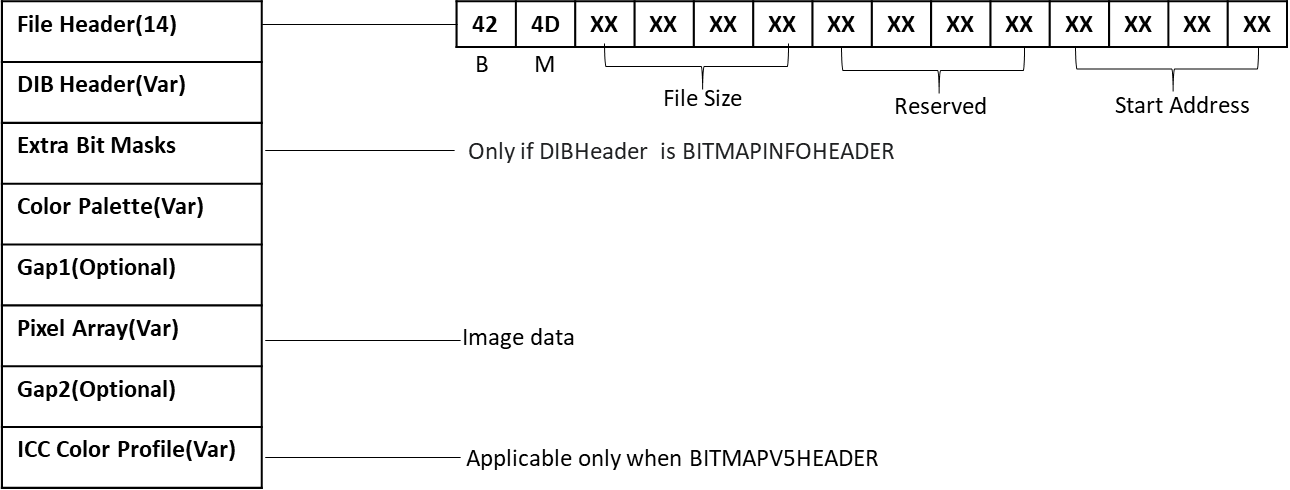}
\caption{BMP file structure format} 
\label{f.bmp}
\end{figure}

We have also seen that the audio format WAVE is used as a malware stager in the advanced persistent threat attack life cycle. So, we are discussing the wave file structure format for understanding and analyzing the possible hidden areas in the audio file. Microsoft Resource Interchange File Format (RIFF) specification is commonly used for storing and managing the multimedia applications, which include image, audio, and video files. Some of the file formats that RIFF support are AVI, WAV, RDI, ANI, BND and RMI. The RIFF specification starts with chunk ID “RIFF” and the chunk size 4 bytes to represent the size of the chunk \cite{wav}. The Wave file may contain one chunk with RIFF specification. The wave chunk may further consist of two sub chunks. Those are “fmt” and “data”. The format header represents the file format that the RIFF specification of the file referring, and, in this case, it is “wave” with hex value “0x57 0x41 0x56 0x45”. The sub chunk 1 representing the audio format consists of the fields such as sub chunk 1 file size, audio format, number of channels, sample rate, byte rate and block align with each field varying the byte sizes from 2 to 4. The audio format by default can be pulse code modulation (PCM) with the number of channels can be either mono (1) or stereo (2). The sample rate, byte rate and bits per sample can be selected based on the audio requirements. The audio data is stored in the sub chunk 2. The data include the right channel and left channel samples representing the audio data. The data may be embedded with malicious payload to hide the stegomalware and use audio format WAV as a carrier to deliver it to the victim device.

\begin{figure}[H]	
\centering
\includegraphics[width=7cm,height=7cm]{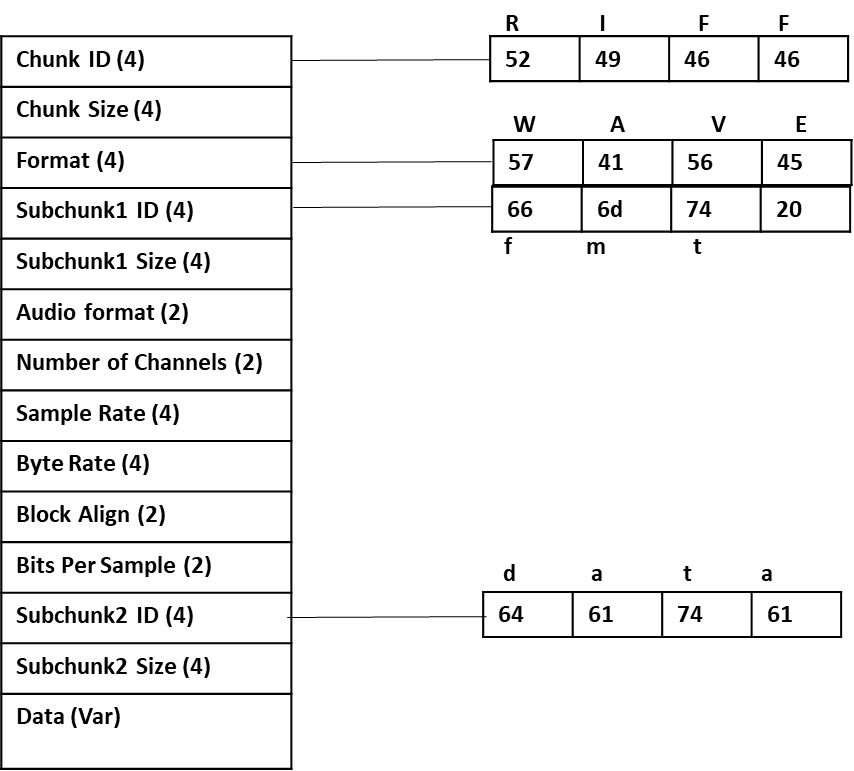}
\caption{WAV file structure format} 
\label{f.wav}
\end{figure}

\textbf{Lessons Learned}:
The stegomalware used file formats JPEG, BMP, PNG, GIF, WAV are suitable for hiding the data or malware payload because these file format specifications poses new avenues for hiding the data. The files also contain unused space, which is used for storing the intended malware payload or other content. The file metadata also can be considered for hiding the data. Overall, it is evident that the mentioned file formats are perfect cover medium image file formats for hiding the malware. The proposal of universal stego detection solution for detecting the hidden malware in any of these file formats seems to be challenging.

%% file: S3steganography.tex
\section{Steganography}
\label{sec:steganography}

Steganography process involves the cover medium, secret data and the technique/algorithm used to conceal the secret data in the cover medium. The proper selection of the steganography algorithm is needed to effectively hide the content in the cover medium and defend the steganalysis attacks. As the computational, processing capabilities on data increased over the period of time, advanced techniques are being proposed in the prior art to improve the maximum embedded content capabilities and resisting the steganalysis methods. In this section, various steganography algorithms are listed and discussed based on the chronological advancement of technology used for embedding the data in the cover image. The cover image steganography is considered for review because the images are extensively used for hiding the malware in the reported attacks historically as described in section \ref{sec:history}, academic research contributions to image steganography are enormous, and image steganography is still considered as a very active research area. Additionally, the image steganography techniques are classified further based on algorithms applied on the spatial or transform domains with machine learning techniques. Furthermore, the GAN based for data hiding in cover images are also discussed along with their performances. 
\subsection{Domain feature based Image Steganography}
The image steganography techniques are mainly categorized into spatial domain and transform domain techniques. In spatial domain, the steganography algorithm/technique is applied directly on the pixel values of the cover image to embed the intended secret data. For instance, in the LSB method, the least bit of every pixel in the image is leveraged to perform the Exclusive "OR" operation with secret data so that the least bit values of the pixels store the secret data. There are various spatial domain steganography techniques such as LSB, PVD, histogram shifting, difference expansion and multiple bit planes-based technique are proposed in the literature \cite{Kadhim2019}. These methods include utilizing the gray images and color images as cover medium for data concealing purpose. As the steganographic operations performed directly on the pixel intensity values in the spatial domain for concealing the content, the decoding process may be easier with the existing tools and may detect the hidden content in the cover image. On the other hand, the transform domain converts the cover images into frequency or wavelet domains for data compression, fine detail separation in the data or localization purpose.  In frequency domain, the DFT and DCT  are well known and effective methods for image transform representation. The algorithms like DWT, IWT, CWT, and DTCWT are some of the most familiar in wavelet domain \cite{Kadhim2019}. The transformed domain coefficients like DCT may be manipulated to hide the data and represent the images back in spatial domain to maintain the same quality. So, the transform domain based techniques can be more effective in steganography applications because decoding is complex process and steganalysis can be harder. Additionally, the operations like compression, scaling performed on the images may not have any impact on the quality of the images and hidden data. However, the transform domain techniques may require more computation operations and processing capacity than spatial domain techniques. Some of the secured steganography algorithms like HUGO \cite{Pevny2010}, WOW \cite{Holub2012}, HILL \cite{Li2014} and UNIWARD  \cite{Holub2014} have also been proposed in the literature to effectively embed the data in spatial domain by adding the distortion and resist the steganalysis attacks.

There are number of works proposed or improved the existing works in the image steganography (pre GAN models) for the last two decades \cite{Kadhim2019}. Unlike the existing studies performed detailed classification of image steganography algorithms \cite{Hussain2018}, we have performed the review for the most widely used and well known image steganography algorithms and the compiled list is illustrated in the Table \ref{T:domain}. The algorithms are arranged in chronological order for easy comparison and track the progress of the technological advancements. The proposed algorithm, their advantage and the generic comments are also included in the table \ref{T:domain} for each contribution.  However, we have not provided further more details on minor contributions and additional relevant works \cite{Hussain2018} due to the limited space and align with the aim of our work.

Westfeld et al. \cite{Westfeld2001} proposed JPEG domain based F5 steganography algorithm to address the issues present in the prior techniques such as JSteg, F3, and F4 around the year 2001. In general, trade-off exist between the maximum embedding capacity and the resistance to defend the steganalysis attacks. Most of the algorithms JSteg, F3, and F4 are vulnerable to visual and statistical attacks when the embedding capacity is fixed. In order to improve the data embedding capacity and resisting against the statistical and visual attacks, the authors proposed steganographic algorithm "F5" in \cite{Westfeld2001} considers two techniques such as permutative straddling and matrix encoding for effectively concealing the data in the image. In permutative straddling, F5 shuffles the DCT coefficients using permutations and then embeds the data in the permutation sequence. The matrix encoding helps to reduce the number of necessary changes needed in the cover image.

In \cite{Fridrich2005}, the authors proposed perturbed quantization based steganography to hide the data in frequency domain. Quantization is one of the essential step need to be performed in data compression or down sampling. In perturbed quantization technique, the data is embedded in the cover image while performing the image compression using standard JPEG domain techniques like DCT. When the pixel values are quantized to assign them as level based final integers, the data is embedded in the quantized process to adjust the final integers. The authors reported that their technique performed better than existing solutions in the prior art. Cancelli et al. \cite{Cancelli2006} proposed spatial domain based MpSteg algorithm for hiding the data in images. MPSteg performs steganography based on the redundant basis image decomposition and matching pursuit. Although the PSNR value for the MpSteg is high, the MpSteg steganography is less detectable in comparison with LSB $\pm$ embedded algorithm when applied sophisticated steganoalysis techniques. In 2007, Fredrich et al. \cite{Fridrich2007} evaluated the JPEG steganography techniques to determine the maximum embed capacity techniques while securing against the steganalysis detectors. They deduce that PQt can embed maximum data compared to other techniques in JPEG domain and recommended syndrome coding methods to decrease the impact of embedding further. Kim et al. \cite{Kim2007} proposed an efficient steganographic algorithm for hiding the data in LSBs of JPEG coefficients. To select the minimal embedding distortion coefficients,  modified matrix encoding scheme is applied on the image in JPEG domain. The distortion results produced in the paper matches with theoretical distribution to embed the data in cover image.

In the year 2010, Pevny et al. \cite{Pevny2010} proposed HUGO algorithm to perform the steganography and is still considered as one of the benchmark algorithm for assessing the steganography and steganalysis solutions. HUGO uses high dimensional images for performing undetectable steganography. The distortion function as the weighted difference of the extended features from the steganoalysis is given as the steganography input to perform HUGO steganography such that to minimize the distortion. The feature set may contain more than $10^{7}$ features for HUGO, which may help reduce the security weaknesses. The HUGO algorithm was able to embed 7 times the message length compared to the LSB techniques for offering the same security level. The authors in \cite{siva2010} demonstrated the edge adaptive based steganography to select the data hiding positions in the image. The LSB is used to hide the content in the pixels and the sharp edges in the images are considered as image positions for hiding the content. This may help to protect the smooth regions with data corruption and protect from visibility and statistical attacks. In \cite{Holub2012}, the authors proposed WOW steganography algorithm to apply in wavelet domain. The algorithm uses syndrome trellis codes to minimize the expected distortion for a given payload. The embedded changes are done on the highly textured or noisy regions and avoid the sharp edges in an image, in contrast to the embedding done in edge adaptive algorithm. The WOW method is shown to be resistant to rich model based steganalaysis. The authors in \cite{Huang2012} introduced a normalized perturbed quantization to select the DCT coefficients for embedding the data. The perturbation error, quantization step (QS), the modified magnitude of quantized DCT coefficient are considered for choosing the DCT coefficients. Additionally, the authors showed that normalized perturbed quantization improved security of the embedded data against steganalysis techniques.

Guo et al. \cite{Guo2012} described UED algorithm to perform the uniform embedding along  with syndrome trellis coding (STC). The uniform embedding distortion metric is measured based on the magnitude of the DCT coefficients and their intra and inter-block neighborhood coefficients. Uniform embedding makes less detectable with steganalysis techniques. The authors reported that UED obtained superior performance to defend against the steganalysis. In \cite{Holub2014} presented universal distortion design method UNIWARD, which can be applied for arbitrary domains. The authors showed UNIWARD outperformed the existing solutions in spatial, JPEG and side-informed JPEG domains. The embedded distortion in UNIWARD is calculated as the sum of the relative changes of coefficients in a directional filter bank decomposition of the cover image. The three algorithms such as S-UNIWARD, J-UNIWARD, and SI-UNIWARD are named based on the operating domains such as spatial, JPEG and side informed JPEG. Li et al. \cite{Li2014} defined a new cost function for minimizing the embedding distortion in steganography algorithm using high pass filter and two low pass filters. The  HILL method locates the less predictable parts with high pass filter and then uses two low pass filters for clustering the low cost values. For the prior art steganalysis methods, Hill performed better than standard steganography algorithms such as HUGO, WOW and S-UNIWARD to hide the data. \cite{Denemark2015} synchronizes the embedded region in the cover image to improve the security against the steganalysis detection. A non-additive distortion function formed after cost assignment to each pixel enables the adjacent embedded changes to synchronize. The tests performed on HILL reveals that Synch-HILL exhibits more probability of error compared to the HILL when applied rich model steganalysis method. 

Sedighi et al. \cite{Sedighi2016} used the locally-estimated multivariate Gaussian cover image model to capture even non-stationary character of natural images and coined MiPOD steganography method, which minimizes the power of the optimal detector. The authors \cite{Pan2016} improved the minimum distortion framework UERD by considering the mutual correlations among DCT blocks such that the less statistical detectability is achieved. The reported results dictate that the IUERD performed better than the UERD by a considerable margin and achieved comparable performable with J-UNIWARD. \cite{Li2018} proposed a joint distortion in JPEG using  BBC principle. The BBC restrain the blocking artifacts caused by the inter block adjacent modifications to preserve the spatial continuity at the block level. The BBC addition into the Decomposing Joint Distortion (Dejoin) results in improving the large payload embedding capacity compared to the modern JPEG steganalyzers. \cite{Meng201809} propose a fusion method to combine the HUGO, WOW, S-UNIWARD steganography techniques. Additionally, the object detection method in the image is also used to select the complex textual regions for embedding the data. The authors \cite{Taburet2021} extended MiPOD in to JPEG domain and named it "J-MiPOD". The MiPOD minimizes the statistical detectability and obtained comparable performance with prior art. Additionally, the authors also addresses the problem of data embedding in the JPEG domain color images and mentioned that JPEP color image steganography is a potential research direction to pursue. Cogranne et al. \cite{Cogranne2021} performed natural steganography in JPEG images captured at ISO sensitivity. Natural steganography is performed on the natural cover images obtained from the CMOS sensor with ISO sensitivity and add stego signal, which mimics the sensor photonic noise. So, the stego image may mimic the sensitivity of the cover image and hide the date. The authors showed that JPEG domain NS (J-Cov-NS) achieved high embed capacity and security when tested with DCTR and SRNet steganalysis.

\textbf{Lessons Learned}: There were number of works proposed towards image steganography techniques starting from simple LSB to recent natural steganography based J-CoV-NS. However, there are few techniques such as HILL, WOW, UNIWARD in different domains, UERD, MiPOD stood out to be more secured and benchmarks for testing or comparing the new steganography or steganalysis techniques. The steganography in JPEG domain color images and natural steganography based techniques are two active areas to be explored for advancing the image steganography techniques (not including DL and GAN) \cite{Taburet2021} \cite{Cogranne2021}. We also conclude that there is no prior art testing the security of the proposed steganography techniques over the years in stegomalware image datasets.

\begin{table*}[] 
\centering
\caption{Spatial, Transform domain Image Steganography techniques}\label{T:domain}
\resizebox{\textwidth}{!}{%
\begin{tabular}{|l|l|l|l|l|l|}
\hline
\textbf{Authors} & \textbf{Year} & \textbf{Algorithm/Technique} & \textbf{Domain} & \textbf{Advantages}  & \textbf{Comment} \\ \hline

Westfeld et al. \cite{Westfeld2001} & 2001 & F5 & JPEG & defend visual and statistical attacks   & used permutative straddling and matrix encoding \\ \hline
Fridrich et al.\cite{Fridrich2005} & 2005 & PQ & JPEG & PQ is the most secured technique compared to prior art till 2005 & the data is hidden in the Quantization fields\\ \hline
Cancelli et al. \cite{Cancelli2006} & 2006 & MPsteg & Spatial & MPSteg less detectable than LSB$\pm$1 embedding  &  Embedding data in lines, corners,and flat regions\\ \hline
Fridrich et al. \cite{Fridrich2007} & 2007 & PQt & JPEG  & large embedding capacity compared to state of the art till 2007 & syndrome coding methods could minimize the embedding impact \\ \hline
Kim et al. \cite{Kim2007} & 2007 & MME & JPEG & The actual embedded distortion closely resembles the theoretical prediction  & Embeds message in the LSB of JPEG coefficients \\ \hline

Pevny et al.\cite{Pevny2010} & 2010 & HUGO &  Spatial & can hide 7 times longer messages  compared to LSB for same security & minimizes the distortion caused by the embedding \\ \hline

siva et al.\cite{siva2010} & 2010 & EA &  Spatial & preserve the statistical and visual features in cover images  & Edge regions are used to hide the content. \\ \hline
Holub et al. \cite{Holub2012} & 2012 & WOW  & Wavelet & WOW outperforms HUGO for large embedding payloads  & The embedded changes are done in noisy or highly textured regions\\ \hline
Guo et al. \cite{Guo2012} & 2012 &  UED &  jpeg & less detectability using existed steganalysis techniques  & Syndrome trellis coding is used during the data embedding. \\ \hline
Huang et al.\cite{Huang2012} & 2012 & NPQ & JPEG & Improved the Security performance of steganography algorithm  & NPQ focuses on the selection of the optimal DCT coefficients for data hiding \\ \hline

Holub et al. \cite{Holub2014} & 2014 &  UNIWARD & Arbitrary &  UNIWARD  outperformed state-of-art algorithms  & the universal design of the distortion function can work in any domain \\ \hline

Li et al. \cite{Li2014} & 2014 & HILL & Spatial  & better performance than HUGO, WOW, and S-UNIWARD & High pass filter and two Low pass filters are used in HILL \\ \hline

Denemark et al. \cite{Denemark2015} & 2015 & Synch-Hill & Spatial & resists against the rich media steganalysis models   &  non-additive distortion function to force adjacent embedding change to synchronize \\ \hline

Sedighi et al. \cite{Sedighi2016} &  2016 & MiPOD & Spatial & MiPOD provided comparable security to WOW, S-UNIWARD, HUGO  & minimizes the power of the optimal detector \\ \hline

Pan et al. \cite{Pan2016} & 2016 &  IUERD & JPEG & Embedding capacity improved compared to UERD  & The mutual correlations among DCT blocks in JPEG are efficiently utilized \\ \hline

Li et al. \cite{Li2018} & 2018 & IUERD- UpDist-Dejoin2 & Spatial and JPEG & The BBC improves the prior art JPEG additive schemes  & Preserves spatial continuity at block boundaries \\ \hline

Meng et al. \cite{Meng201809} & 2018 & fusion of \cite{Pevny2010}, \cite{Holub2012} \cite{Holub2014} &  Spatial & The fusion of WOW, HILL and UNIWARD obtained better performance  & data is hidden in selected objection detection block region. \\ \hline
 
Cogranne et al.  \cite{Cogranne2021} & 2020 & J-Cov-NS & JPEG &  Achieved higher capacity and security when tested against SRNet & The security may decrease with non-linear developments \\ \hline

Taburet et al. \cite{Taburet2021} &  2020 & J-MiPOD & JPEG & J-MiPOD showed competitive performance compared to prior art algorithms & first step towards embedding data into JPEG color images\\ \hline
\end{tabular}%
}
\end{table*}


\subsection{Deep Adversial based Image Steganography}

The advancement in the hardware computation capabilities in recent times helped the rapid innovation of solving complex problems using deep learning in many application areas. Deep generative models are one of the deep learning models, which are received major attention due to the recent advancements. The deep generative models learn different data distributions in an unsupervised manner.  There are two types of deep generative models. Those are Variational Auto Encoders (VAE) and GAN. The generative models have some drawbacks need to addressed when initially researchers explored the deep generative models for solving the problems. In order to address the drawbacks in generative models, Goodfellow \cite{Goodfellow2014} proposed GAN architecture in 2014. GAN mainly has two components such as generator and discriminator. The generator can randomly generate the input data  and also considers the feedback to incorporate the error data. The discriminator network discriminate between the generated image and the expected image for the feedback. The combination of these two components can be used to produce fake multimedia content and data hiding as well. 

Encoder decoder architecture are initially known to be used for data compression. The encoder helps to represent the same input with fewer data points and the decoder try to reconstruct the original input accurately given the encode output as decode input. However, this architecture may not be fitted into the generative model due to the encoder output is not regularized. Variational autoencoders mainly contain the encoder and the decoder modules. The authors in \cite{Kingma2014} proposed variational autoencoders to alleviate the regulation problem in traditional encoder and decoder architectures. These modules resemble the Convolutional neural network (CNN) layers with modification to the penultimate layers. Both VAE and GAN can be used to generate the data such as the image, text. Later on, these architectures also been used for media steganography and creating fake content for fooling the people. 

The deep adversial model solutions proposed for image steganography are illustrated in the Table \ref{T:GAN1} and discussed in chronological order.

Denis et al. \cite{DenisVolkhonskiy2017} proposed a SGAN network to generate the steganographic images. SGAN contains a generator G to produce the realistic images from noise, discriminator D to evaluate the produced image is real or synthetic and discriminator S performs the steganalysis to determine the given image is stego image or not. Haichao et al. \cite{Shi2017} proposed SSGAN to generate secure cover image. The SSGAN uses WGAN \cite{Arjovsky2017} to generate the cover image in generator network, discriminator D to evaluate the visual quality of the generated images and GNCNN to perform the steganalysis in discriminator network S. The authors in \cite{Chu2017} introduced  CycleGan to perform image steganography in specific hiding one image in another image. The CycleGAN output nearly imperceptible, high frequency signal, which makes it is difficult to perform steganalysis. However, a careful design of the CycleGAN framework is needed to hide the data and achieve secure performance against steganalysis techniques.  Tang et al. \cite{Tang2017} introduced ASDL GAN framework to hide the secret data in images. The ASDL framework consist of a generator and discriminator networks. The generator network contains CNN layers with groups from 1 to 25 and the discriminator adapts the Xu's deep learning model \cite{Xu2016} for steganalysis. The groups 2 to 24 configurations are the same in the generator network. Additionally, new activation function Ternary embedded simulator (TES) is proposed to generate the stego images according to the generator provided pixel level change probabilities.

Volkhonskiy et al. \cite{Volkhonskiy2017} proposed SEGAN model to hide the data in images using GAN. The SEGAN model contains one generative network, one discriminative network, and one decryption network. Considering the classic alice, bob and eva encryption problem mapping, the generative network performs the alice role to generate steganographic image, the decryption network performs bobs role to extract the hidden message in the image and the discriminator to determine the given image is real or generated image. In 2017, \cite{Hayes2017} proposed encoder decoder based architecture HayesGAN for embedding the data in the images. The technique involve alice, bob and  eve as the game participants, whereas alice and bob acts as an encoder and decoder and the eve acts as a steganalyzer. Like any deep learning solution, the domain knowledge is not required to perform the image steganography using these models. \cite{Zhu20184} also proposed an encoder and decoder based deep network technique Hidden to hide the data in cover images. Additionally, an adversary network is used to discriminate the cover and stego image. An adversial loss is measured to correct the stego image generation with generator for each iteration until secured image is created by encoder. In the paper \cite{Zhang2019}, the authors proposed  GRDH scheme for image steganography. The generator network is used to generate the cover images; CycleGAN is applied to translate the image to image and then performed the mapping of noise to hiding data for concealing the data in images. The recovery process includes the same components used for image generation. 

Zhang et al. \cite{Zhang201998} proposed StegnoGAN to hide the binary data in images. Three different models such as basic, residual and dense encoder architectures has been proposed to generate images. The decoder reconstruct the hidden data from the steganography image and another component "critic" provide feedback on the performance of the encoder to improve the secured stego generated image. Additionally, Reed-Solomon bits-per-pixel (RS- BPP) is proposed to evaluate the capacity of embedded data when using GAN models for steganography. Zhang et al.\cite{Zhang20191} demonstrated an ISGAN model to hide the grey scale image in the color image Y channel values. The architecture contains an encoder to embed the gray scale image in the color image and a decoder to reconstruct the original data from the stego image and a steganalyser tries to  distinguish stego images from cover images to improve the overall security performance. All 3 networks in the ISGAN model contain CNN based layers. The layers may include the inception module in encoder, spatial pyramid pooling (SPP) blocks in steganalyser.  \cite{Zhang2020} proposed  AC\-GANs, in which the secret data transmission is done without modifying the cover images. A set of labels and the image database is considered for selecting the cover image. The intended sending message is mapped to the labels using dictionary in the sender end and use the label to pick the image from the database for data transmission. At the receiver side, the image is mapped with the corresponding label already stored in the database. The message related to the label mentioned in the dictionary is used for retrieving the message.
 
 Zhangjie et al. \cite{Fu2020} proposed a GAN based model HIGAN to hide the images in another image. The color image is hidden in another color image using HIGAN model. The HiGAN is composed of three subnetworks to perform the GAN based steganography. The encoder can hide the color secret image into a cover color image with the same size. The output of the encoder is fed into the decoder to extract the secret image. A discriminator is used to distinguish whether the input image contains secret image or not. The encoder contains 3 down sampling layers, 9 residual layers, and 3 up sampling layers to hide the image in another image. Wang et al. \cite{Wang2020} proposed GAN model, which consists of U-NET based generator and the Xu model discriminator for performing the steganalysis of the images. The generator is reconstructed by combining the multiple feature maps and the reconstructed generator is used for proper information embedding.

\begin{table*}[] 
\centering
\caption{Deep Adversial based Image Steganography techniques}\label{T:GAN1}
\resizebox{\textwidth}{!}{%
\begin{tabular}{|l|l|l|l|l|l|l|}
\hline
\textbf{Authors} & \textbf{Year} & \textbf{Architecture} & \textbf{Networks} & \textbf{Advantage} & \textbf{Comment}\\ \hline
Denis et al. \cite{DenisVolkhonskiy2017} & 2017 & SGAN & 3  & The SGAN model reduced the steganalyzer detection accuracy &  DCGAN \cite{Radford2016} model also performed well to secure the cover images. \\ \hline
Haichao et al. \cite{Shi2017} & 2017 & SSGAN & 3 & SSGAN performed better than SGAN & Visual quality is better and relaistic \\ \hline

Chu2017 et al. \cite{Chu2017} & 2017 & CycleGAN & 4 & Image hiding in another image & CycleGAN adversial attacks vulnerability helps to hide information     \\ \hline

Tang et al. \cite{Tang2017} & 2017 & ASDL-GAN & 2 & The automatic learning of the
embedding change probabilites at the pixel level & The proposed model is only evaluated for spatial domain \\ \hline

Denis et al. \cite{Volkhonskiy2017}  & 2017 & SEGAN & 3 & Secure and adaptive steganographic image generation & Symmetric key is used for encryption prior to hiding \\ \hline

Hayes2017 et al \cite{Hayes2017}  & 2017 & HayesGAN & 3 & First encoder and decoder based adversial learning for steganography  & The secured performance still has room for improvement  \\ \hline 
Zhu et al. \cite{Zhu20184} & 2018 & Hidden &  3 & Secured performance better than HUGO, S-UNIWARD, WOW  &  The secured performance can still be improved. \\ \hline
Zhang et al. \cite{Zhang2019}  & 2019 & GRDH, CycleGAN & 4 & First data hiding without cover image modification & Embedded Capacity is limited compared to traditional methods \\ \hline
Zhang et al. \cite{Zhang201998} & 2019 & SteganoGAN & 3 & SteganoGAN can accomdate higher payloads while evading the detection & Three different models are proposed to generate steganoGAN image \\ \hline
Zhang et al. \cite{Zhang20191}  & 2019 & ISGAN & 3 & Can hide the gray scale image in the color image & Model robustness need to be improved \\ \hline
Zhuo et al. \cite{Zhang2020}  & 2018 & Stego-ACGAN  & 3 &  Cover image modification is not required for data hiding & Only set of messages stored in dictionary can be embedded.\\ \hline

Zhangjie et al. \cite{Fu2020} & 2020 & HIGAN  & 3 & Higher visual quality and stronger security & Encoder network uses upsample, downsample and residual layer\\
\hline


Wang et al. \cite{Wang2020} & 2020 & GAN with feature maps & 2 & Better distortion measurement and secured steganographic scheme & Reconstructed U-NET for Generator and Xu model for Discriminator \\ \hline

\end{tabular}%
}
\end{table*}
The performance evaluation of the GAN based stego images are illustrated in Table \ref{T:GANP1}. The Table \ref{T:GANP1} categorized as the adversial model, performances, metric used, dataset used for models comparison. The SGAN generated stego images in \cite{DenisVolkhonskiy2017} are applied to HUGO steganography analyzers for assessing the security performance. The detection accuracy of the SGAN generated images reduced by 30\% in comparison with $\pm$1-embedding algorithm when HUGO steganalyser used. The authors \cite{Shi2017} compared the performance of the SSGAN with the SGAN \cite{DenisVolkhonskiy2017}. We can clearly see that the time to run 7 epochs took 227.5 seconds for SSGAN, whereas SGAN completed the  epochs in 240.3 seconds. Additionally, the steganalyser detection accuracy for SSGAN 0.72 is much lower than the SGAN accuracy 0.90 for the GAN generated images. The authors in \cite{Tang2017} showed that the ADSLGAN performed better than S-UNIWARD in terms of security performance. For 0.1bpp embedding rate, ASDLGAN achieved 26.92\%, whereas S-UNIWARD obtained 42.53\% accuracy when applied the Xu's steganalysis model. \cite{Volkhonskiy2017} presented data encryption based GAN model SEGAN has obtained the quality of encryption/decryption 99.98 and 99.96 for the MNIST and CIFAR-10 when the percentage of reconstruction bits is 16. Furthermore, the paper also demonstrated that SGAN and DCGAN models decrease the detection accuracy to an extent of similar to random classifier, which is impressive. Hayes2017 et al. \cite{Hayes2017} tested the performance of the HayesGAN on the datasets Boss, Celeba. Even though HayesGAN is not performed better than the state of the art steganalysis techniques like WOW, HUGO, the model showed that the image steganography can be performed using HayesGAN. For Boss datasets, HayesGAN achieved 79\% accuracy whereas for celeba datasets, it obtained 90\% accuracy. The Hidden model proposed in \cite{Zhu20184} obtained 50\% detection rate for the COCO dataset. The performance achieved by Hidden model is better than the HUGO, WOW and S-UNIWARD. However, the comparison of the Hidden with other GAN models is not available. The GRDH scheme proposed in \cite{Zhang2019} able to perform well to securely hide the data in the images.  On the dataset Celeba, the PSNR value 22.665 is obtained when random noise is given as an input for image generation case model BEGAN. But, the embedded capacity of the proposed method is very limited. The authors \cite{Zhang201998} proposed SteganoGAN for embedding data in the cover images is tested on two datasets Div2K, COCO and the performance metrics such as accuracy, Reed Solomon Bits Per Pixel, Peak signal-to-noise ratio are used for comparison. SteganoGAN performed better to securely hide the data. The payload capacity in Div2K dataset rather than the COCO dataset when performing the steganalysis 

The ISGAN model \cite{Zhang20191} performance on three different datasets LFW, PASCAL-VOC12 and ImageNet shows that the CNN based steganalysis network able to detect 78.5\%,74.3\% and 73.6\% images accurately. Furthermore, the authors showed that ISGAN achieved the state-of-the-art steganography performance in terms of the structure similarity index (SSIM) and PSNR. The author's proposed model in \cite{Zhang2020} achieved 100 percent prediction accuracy of auxiliary classifier when the training step is 10. The HiGAN model \cite{Fu2020} performance on the Imagenet2012 showed that the structural similarity between stego image and original images is 94\% and the peak signal-to-noise ratio is 30.95. These performance results indicate that HiGAN did performed well to hide the color image on another color image. The performance of the U-NET based Generator and Yu's model Discriminator \cite{Wang2020} on BOSSBase dataset shows that error rate 34.80\%  is obtained when the spatial rich model with ensemble classifier used. The authors reported that their proposed model achieved better performed than S-UNIWARD, ASDL-GAN.

\begin{table*}[] 
\centering
\caption{Deep Adversial  based based Image Steganography performance}\label{T:GANP1}
\resizebox{\textwidth}{!}{%
\begin{tabular}{|l|l|l|l|l|l|}
\hline
\textbf{Authors} & \textbf{Year} & \textbf{Architecture} & \textbf{Dataset} & \textbf{Performance} & \textbf{Metrics}\\ \hline
Denis et al. \cite{DenisVolkhonskiy2017} & 2017 & SGAN & Celebrities & Accuracy reduction 0.624 to 0.499 & Detection accuracy \\ \hline
Haichao et al. \cite{Shi2017} & 2017 & SSGAN  & CelebA  & Epochs:7, run time: 227.5 s, Accuracy: 0.90 to 0.72 & Run time, Accuracy  \\ \hline
Chu2017 et al. \cite{Chu2017} & 2017 & CycleGAN  & - & - & -\\ \hline
Tang et al. \cite{Tang2017} & 2017 & ASDL-GAN & BossBase & 0.1bpp, Accuracy: 26.92\% & Accuracy\\ \hline
Denis et al. \cite{Volkhonskiy2017}  & 2017 & SEGAN  & MNIST, CIFAR\-10 & MNIST: 99.98, CIFAR\-10: 99.96 & Quality of encryption \\ \hline
Hayes2017 et al \cite{Hayes2017}  & 2017 & HayesGAN  & Boss, Celeba & Boss:79\%, Celeba:90\% & Detection accuracy\\ \hline 
Zhu et al. \cite{Zhu20184} & 2018 & Hidden  &  COCO  & 0.203bpp, bit error  $ < 10^{-5}$, detection rate: 50\% &  Detection rate, bit error\\  \hline
Zhang et al. \cite{Zhang2019}  & 2019 & GRDH & CelebA & BEGAN, PSNR:22.665, StyleGAN, PSNR:28.333 & PSNR\\ \hline
Zhang et al. \cite{Zhang201998} & 2019 & SteganoGAN & Div2K, COCO & Depth=5,COCO Accuracy:0.84, RS-BPP:3.43 PSNR:29.73 & Accuracy, RS-BPP, PSNR  \\ \hline
Zhang et al. \cite{Zhang20191}  & 2019 & ISGAN  & LFW, PASCAL,ImageNet &  LFW: 0.785 PASCAL-VOC12: 0.743, ImageNet:0.736 & Detection Accuracy  \\ \hline
Zhuo et al. \cite{Zhang2020}  & 2018 & Stego-ACGAN  & MNIST & Prediction accuracy:100\% at training step 10 & Prediction accuracy \\ \hline
Zhangjie et al. \cite{Fu2020} & 2020 & HIGAN  & ImageNet2012 & SSIM:0.94, PSNR:30.95 & SSIM, PSNR\\\hline
Wang et al. \cite{Wang2020} & 2020 & U-NET  & BOSSBase  & 0.20bpp, Error rate: 34.80  & Error rate \\ \hline

\end{tabular}%
}
\end{table*}
\textbf{Lesson Learned}:
The deep adversial based steganographic algorithms has potential to generate maximum embedding capacity and less detectable stego images. Overall, based on the performance analysis of the state-of-the-art GAN based solutions, it is difficult to compare deep adversial stego model using a unique approach because the researchers use different datasets, dataset count, steganalyzer, performance metrics for evaluation. However, we can conclude that GAN based steganography solutions performed well compared to the traditional steganography solutions such as WOW, HUGO, UNIWARD in terms of embedding capacity, secured data hiding. We also found that there are no existing deep adversial stego solutions focused on hiding the malware in images. It is interesting to generate stegomalware using deep adversial solutions and evaluate the performance of those solutions to evade the advanced next generation solutions.

%% file: S4Steganalysis.tex
\section{Steganalysis}
\label{sec:steganalysis}

Steganalysis is the act of determining if the cover medium is hidden data or not and identifying the steganographic algorithm; estimating the hidden data if the cover medium is hiding the data and extracting the hidden data from the stego medium. Steganalysis can be performed in many ways. Signature based steganalysis is the simplest way and straight forward process. Steganalyzer may take the advantage of the steganography tool signature when embed the data to confirm if the given digital medium is stego or cover medium. For instance, "hiderman" steganography tool add the letters “CDN” at the end of the cover file when embedding the data. So, extracting the bytes of the file and comparing with known signature or specific patterns maybe helpful to identify the stego file. The file can be an image, audio, video or other format. We discuss the "file" herein in the context of images, since we determined that stegomalware is leveraging the cover image for hiding seen in section \ref{sec:history}. Later on, the statistical properties of the images in different domains are considered as a feature vector to capture the stego content and ML techniques are applied to classify stego and cover images. The spatial and JPEG rich models along with ensemble classifiers shown good performances in state-of-the-art ML category but they are not optimal models. The detailed review and categorization of the various feature extraction and classification using ML for steganalysis is described here \cite{Karampidis2018}. Recently, the deep learning based steganalysis received major attention and reported best stego detection performances compared to the rich model based detection techniques \cite{Evsutin2020}. In this section, our aim is to discuss and categorize the best state-of-art steganalysis techniques existed for the last two decades in chronological order highlighting the research progress and also investigate if the existing steganalysis methods are used for stegomalware detection.

\subsection{Image domain based feature and rich models steganalysis solutions}

Johnson et al. \cite{Johnson1998} discussed different ways to perform the steganalysis on various stego tools such as Hide4PGP, Mandelsteg, Syscop, hideandseek, S\-tools for uncovering the hidden data. Although there were no experiments designed in the paper, it opened up new ideas for approaching the steganalysis. The authors in \cite{Lyu2002} proposed QMF based  high order statistics feature extraction method and then applied SVM on the extracted features for steganalysis. The higher order statistics could resist counter based attacks. Lyu et al. \cite{Lyu2002} considered image quality measures (IQM) for capturing image features and applied multivariate regression to discriminate between cover and stego images. The IQM features are obtained using ANOVA technique. The authors in \cite{Fridrich2004} performed feature extraction in JPEG domain using DCT coefficients, which are first order and second order statistics and then applied linear classifier for the steganalysis of the images. The steganographic algorithms Outguess, F5 and Model based steganography with and without deblocking are tested for evaluating the proposed method. Also, the co-occurrence matrix of DCT coefficients is identified as one of the essential feature for JPEG image steganalysis. Kenneth et al. \cite{Sullivan2005} described the markov chain based interpixel dependencies as a feature sets for detecting the stego images, where the data is hidden using spread spectrum. The ML technique SVM is employed on the extracted features to classify the images. However, the proposed inter pixel dependencies based model is shown to be applicable only spread spectrum based steganography image detection. The authors \cite{Ismail2005} proposed BSM based features for detecting the stego images. SVM is considered for classifying the images. The correlation between the bit planes and the binary texture characteristics within the places can be different for cover and stego image to distinguish them. Zou et al. \cite{Zou2006} used two-dimensional markov model of thresholded prediction-error of an image for stego detection. The transition matrix of markov along the chain in vertical, horizontal and diagonal direction is considered as a feature and applied to the SVM for classifying the images. The state-of-the -art steganographic techniques spread spectrum(SS), non-blind SS, Quantization index modulation (QIM), LSB are considered for testing the proposed model. Jan et al. \cite{Jan2010} used different steganalysis features such as SPAM, markov process, cartesian calibration doubles (CC-PEV), cross-domain Feature (CDF) to detect the YASS steganographic images. The SVM with a Gaussian kernel is used as image classifier by applying the four feature sets on different YASS settings from 1 to 12. The combined feature set CDF achieved good performance to classify the images compared to the three individual feature sets.

The low dimensional feature selection along with linear classifier or SVM tend to be finding difficulty to classify content adaptive stego algorithms like HUGO proposed in 2010. So, researchers considered high dimensional feature extraction by performing the low feature selection from multiple submodels also called base learners. Then, ensemble classifiers are used for selecting the optimal base learners and obtaining the better detection performances. The recent works in this direction are discussed now onwards. Fridrich et al. \cite{Fridrich2011} proposed higher-order local model estimators of steganographic changes (HOLMES) to consider the combination of the MINMAX residual and the co-occurrence matrices with markov chain features to maximize the detection capabilities of the content adaptive algorithms like HUGO. Additionally, ensemble classifier with voting method for fusion is considered to classify the images. Fridrich et al. \cite{Fridrich2012} also presented the spatial rich models and ensemble classifiers for stego detection. Multiple residual classes such as first, second, third order, edge and square are described for feature (neighboring pixels cooccurrence matrices) generation of the submodel. In ensemble classifier, the feature selection of the submodels is done using different classes like CLASS-q, ITERATIVE-BEST-q, BEST-q-CLASS, BEST-q etc. The out of box (OOB) error can be used as a parameter to select the best performed classes for ensemble classification. The author's work prompted a new research direction towards using rich model and ensemble classifiers for steganalysis.

Kodovsky et al. \cite{Kodovsky20126} performed rich model and ensemble classifier stego detection in JPEG domain and proposed cartesian-calibrated JRM (CC-JRM). A detailed investigation of individual subsets performance is performed on the 6 well known JPEG steganographic algorithms. Adding Cartesian calibration increases the number of features and also improved the performance of steganalysis. Furthermore, the paper determined that BCHOpt is the most secure, and MBS and YASS are by far less secure stenographic algorithms when tested on the union of CC-JRM and the SRM. The detailed comparison of the various other feature sets on the JPEG steganographic algorithms is discussed in the paper \cite{Kodovsky20126}. Holub et al. \cite{Holub2013} proposed PSRM to retrieve the statistically significant features from the residual samples rather than the co\-occurrence matrices representation of residual samples in SRM. In PSRM, the neighboring residual samples are projected onto a random space and the first order statistics of the projections are considered as a feature set. The authors showed that the PSRM can be applied in multiple domains with few changes and tested against various secured steganographic algorithms for detection. Holub et al. \cite{Holub201498} also proposed low complexity feature set generation method DCTR to use in JPEG domain. These features are obtained from the decompressed JPEG using 64 kernel DCT. The authors \cite{Holub20159} presented PHARM feature set for stego detection in JPEG images. In contrast to the PSRM, PHARM only uses few number of residual from support kernels and represent them in first order statistics of their random projections. The computation cost and number of dimensions needed are much lower than the JRM and PSRM models. 

Song et al. \cite{Song2015} proposed 2 dimensional gabor filters to capture the image texture and edges from different scales and orientations and the histogram features are extracted from the filtered images so that adaptive steganographic images can be detected accurately than the state-of-the-art models. The gabor filter is applied on the JPEG decompressed images. The extracted feature set is applied to the ensemble classifier to classify the images. The experimental results shown that gabor filter performed better than the state-of-the-art feature set techniques like CC-JRM and DCTR. The authors in \cite{Abdulrahman2016} proposed SGF for generating new features and then adding to the color rich model features to enhance the feature set. The Gaussian filter features can identify the minor changes done to the image during the embedding process. These new features enhance the detection capability against the three stego algorithms S-UNIWARD, WOW, and Synch-HILL. Xia et al. \cite{XiaChao22017} improved the GFR \cite{Song2015} by proposing the symmetric merging of different Gabor filters and weighted histograms by considering the position of residuals. The combination of the symmetric merging and weighted histograms along with GFR is defined as GFR-Gabor symmetric merging and weighted histograms(GFR-GW) and the symmetric merging with GFR is denoted as GFR-Gabor Symmetric Merging (GFR-GSM). The inclusion of these two methods improved the detection performance against the J-UNIWARD and UED. The authors in \cite{Xia2019} improved the PHARM for enhancing the stego detection capabilities. Three changes have been done on PHARM to improve the performance. The authors reduce the maximum projection matrix size, selection of more than one phase pair per projection and considering the transposition symmetry to improve the PHARM performance. Additionally, the proposed improved PHARM detection accuracy better than DCTR feature sets. Xia et al.\cite{Xia2020} proposed improvements to DCTR and GFR feature sets for efficient stego detection. The different residual images, multiple filter sizes and different symmetrization rules by considering the filter type, filter size are utilized to improve the performance. The improvement models DCTRD, DCTR-W and SCA-DCTR, GFRD, GFR-W and SCA-GFR are presented in \cite{Xia2020}. The detection performance improved for SCA-DCTR compared to DCTR-W and the performance of DCTR-W improved compared to DCTRD. Similarly, the GFR improved version also follow the same trend. However, the feature extraction time increases for SCA-DCTR compared to DCTR-W  and also increases DCTR-W compared to DCTRD. Overall, the detection performances increased for improved versions of DCTR and GFR. Fend et al. \cite{Feng2020}  presented maximum diversity cascade filter residual (MD-CFR) feature set for steganalysis of images. The cascade filters are formed by combining the base filters and maximum diversity is considered for cascade filter selection. These filters are convolved with JPEG decompressed images to obtain the maximum diversity cascade filter residuals. The steganalysis is performed on four steganographic algorithms to test the proposed feature set with the state-of-the-art feature sets.

\textbf{Lessons Learned}:
Image steganalysis has advanced over the last decades with initially research progress focused on feature and ML based detections followed by rich model and ensemble classifier based solutions. Some of the steganalysis methods like CDF, SRM, JRM, PSRM, DCTR, PHARM, GFR, SGF stand out to be effective detection solutions in the pre deep learning era. However, we believe that the conventional image steganalysis research seems to be slow down with more contributions towards improving the state-of-the-art solutions GFR, DCTR, PHARM roughly for the last five years. The reason could be the availability of the deep learning technology with computation capabilities in the last few years and rapid growth of using deep learning in various application include steganalysis. The fact that deep learning models could provide good performance results compared to the conventional steganalysis is another reason towards this shift. Interestingly, we have not found any works evaluating the performance of the existing techniques to detect the stegomalware hidden in images or proposing new steganalysis techniques for stegomalware detection.

\begin{table*}[h!] 
\centering
\caption{Image domain based feature and rich models steganalysis solution}\label{T:ImagerichSA4}
\resizebox{\textwidth}{!}{%
\begin{tabular}{|l|l|l|l|l|l|l|}
\hline
\textbf{Authors} & \textbf{year} & \textbf{Technique} & \textbf{Machine Learning} & \textbf{Embedding Algorithms}  & \textbf{Advantages} & \textbf{Comment} \\ \hline
Johnson et al. \cite{Johnson1998} & 1998 & Generic  & - & Hide4PGP,Mandelsteg,Syscop,hideandseek, S\-tools &  One of the first works on steganalaysis  & Discussed various methods for existing tool steganalysis \\ \hline
Lyu et al. \cite{Lyu2002} & 2002 & QMF & SVM & Jsteg3, OutGuess(+-), EzStego, LSB,  F5  & higher order statistic is not vulnerable to counter-attacks & countermeasures on higher order statistic are possible \\ \hline
Ismail et al. \cite{Avcibas2003} & 2003 & ANOVA &  Regression  & Digimarc, PGS, steganos, stools and Jsteg & Anove improved steganalysis performance   & Image quality based features are used first time\\ \hline
Fridrich et al. \cite{Fridrich2004} & 2004 & higher order DCT & linear classifier & Outguess, F5, Model based  & Accurately detect the stego images generated by OutGuess  & Only applicable to JPEG domain images \\ \hline
Kenneth et al.  \cite{Sullivan2005} & 2005 & Markov Chain  & SVM & - & 95\% stego images are detected correctly &  Only detects spread spectrum based hidden data \\ \hline
Ismail et al.\cite{Ismail2005}  & 2005 & BSM & SVM & LSB, LSB +/−, Outguess, Outguess+, and F5 & BSM outperformed high order statistic for LSB & Can accurately detect high embedding payload images  \\ \hline
Zou et al. \cite{Zou2006} & 2006 & 2-D Markov chain &  SVM &  SS, non-blind SS, QIM and LSB & More than 90\% detection rate for SS, QIM, non-blind SS & Markov chain may need to be tested against commercial tools F5, Outguess \\ \hline


Jan et al. \cite{Jan2010} & 2010 &  SPAM, MP, CC-PEV, CDF & SVM & YASS  & fusion set CDF detect the YASS & feature fusion improve the detection even for  MME3 and nsF5 \\ \hline

Fridrich et al. \cite{Fridrich2011} & 2011 & HOLMES & EC & HUGO & HOLMES with EC classify HUGO stego images & CDF with SVM Guassian performed well compared to CDF with EC \\ \hline

Fridrich et al. \cite{Fridrich2012} & 2012 & SRM & EC & HUGO, EA, LSB matching & Rich models  handled high dim feature inputs & Automated steganalysis is discussed as one of the future direction \\ \hline
Kodovsky et al. \cite{Kodovsky20126} &  2012 & CC\-JRM & EC & nsF5, model based, YASS, MME, BCH, BCHopt  &  CC\-JRM improved the detection performance compared to JRM & Various feature sets in JPEG domain also compared to use in JPEG rich models \\ \hline

Holub et al. \cite{Holub2013} & 2013 & PSRM & EC & HUGO, WOW, nsF5,UED, NPQ, UNIWARD & PSRM improved detection performance compared to SRM models & PSRM works on spatial, JPEG, SI-JPEG with few changes \\ \hline

Holub et al. \cite{Holub201498} & 2014 & DCTR & EC & J\-UNIWARD & DCTR require few features and performed better than SRM, PSRM & DCTR feature sets has low computational complexity, lower dimensionality  \\ \hline

Holub et al. \cite{Holub20159} & 2015 & PHARM & EC & J\-UNIWARD, UED, SI-UNIWARD & PHARM outperformed SRM, PSRM for the J\-UNIWARD and UED stego methods & PHARM feature set is computationally efficient \\ \hline

Song et al. \cite{Song2015} & 2015 & 2D Gabor filters & EC & UED, JUNIWARD, SI-UNIWARD & Gabor filters feature sets performed better than DCTR, CC-JRM  & Gabor filters is not tested aganist adaptive JPEG steganography \\ \hline

Abdul et al. \cite{Abdulrahman2016} & 2016 & SGF & EC  & S-UNIWARD, WOW, Synch-HILL  & higher detection rates than CRM, CFARM, and GCRM & Gaussian filters new features added to CRM features for improving performance \\ \hline  

Xia et al. \cite{XiaChao22017} & 2017 & GFR-GW, GFR-GSM & EC & UED-JC and J-UNIWARD & Symmetric merging and weighted histograms improves the performance of GFR & Merging with the DCTR features may obtain good results \\ \hline

Xia et al. \cite{Xia2019} & 2019 & Improved PHARM & EC & J-UNIWARD, UED-JC & Less computation time and high detection accuracy than PHARM & Adding weighted histogram scheme to the model may even perform well \\ \hline

Xia et al.\cite{Xia2020}& 2020 & Improved DCTR and GFR & EC & J-UNIWARD, UED-JC &  Detection perforance of DCTR and GFR is improved &  The feature extraction time increases for the improved DCTR and GFR model  \\ \hline

Feng et al. \cite{Feng2020} & 2020 & MD-CFR & EC & JC-UED, UERD, J-UNIWARD, SI-UNIWARD & cascade filter residual obtained better performance than GFR,DCTR & Adding selective channel awareness may even enhance the performance more \\ \hline 

\end{tabular}%
}
\end{table*}

\subsection{Image domain based feature and rich models steganalysis performance}
The higher order statistic and SVM based method in \cite{Lyu2002} achieved 98.5\% classification accuracy for the stego images generated by the jsteg tools, when the natural images are considered for the evaluation. The authors \cite{Avcibas2003} performance evaluation on ANOVA based feature set steganalysis achieved in the best case 85\% detection rate for PGS technique. The steganalysis performance in \cite{Avcibas2003} still had a room for improvement. \cite{Fridrich2004} evaluated the blind feature based steganalysis on algorithms F5, Outguess, Model based (MB1) and Model based with deblocking (Mb2). The results presented in the paper show that the model based techniques MB1 and MB2 performed well with low detection reliability values 0.16, 0.21. Further, the Outguess based stego images are almost  detected using the blind features with linear classifier when the bpnzac is set to 0.05. Kenneth et al. \cite{Sullivan2005} evaluated their Markov chain and SVM based detection model in both the locally adaptive and globally adaptive hiding cases with spread spectrum method. The results showed that globally adaptive hiding performed well to classify the images compared to the locally adaptive hiding when considered the image datasets from diverse sources. The BSM performance evaluation in \cite{Ismail2005} on stego algorithms Outguess, F5 and LSB$\pm$ showed that the accuracy increased as the embedding payload bpp increases. When the bpp is 15, the proposed method obtained 92.17\% for LSB$\pm$ detection. The Outguess and F5 could not accommodate 15bpp for hiding the data in images. Zou et al. \cite{Zou2006} obtained good performance results for stego image classification for four stego algorithms. When the experiments are performed with non-linear kernel, their model obtained around 90\% accuracy for SS, non-blind SS, QIM and around 85\% accuracy for the LSB steganography image datasets. The article also showed that the non-linear kernel SVM performed better than linear kernel in stego image classification. 

The authors \cite{Jan2010} tested the performance of different feature sets on YASS algorithm and showed that the best performed CDF detected YASS with $P_{e}$ $<$ 15\% when bpnzac is 0.003. Additionally, for the MME3 and nsF5, the CDF obtained $<$ 10\% $P_{e}$ when bpnzac is 0.003. This shows that feature fusion methods help to achieve good performance even in steganalysis. The performance evaluation of the HOLMES in \cite{Fridrich2011} showed that the HOLMES methodology for stego detection is effective even for secure algorithm HUGO. For embedding capacity of 0.5bpp, the detection error for HOLMES is 12\% for HUGO compared to 7.3\% with LSB$\pm$ embedding. This shows that HUGO is much more secured than $\pm$ embedding for HOLMES detection. HOLMES also performed well compared to CDF \cite{Jan2010} when tested on HUGO.  The SRM with EC proposed in \cite{Fridrich2012} is evaluated on Bossbase datasets. The top39 submodels with 12,753 dimensional are selected for the evaluation of SRM+EC on LSB matching, HUGO, EA algorithms. For embedding payload bpp is 0.05, the SRM+EC is outperformed the Gaussian-SVM in terms of detection error and running time. The detection error for TOP39 rich model is 0.42 for HUGO model, 0.3255 for EA and 0.274 for LSB matching. The authors concluded that HUGO is secured in comparison to the LSB matching and EA. Overall, the rich models with EC are proved to be an efficient future direction to detect HUGO algorithms. The proposed model CC-JRM in  \cite{Kodovsky20126} obtained detection error 0.422 for BCH and 0.448 for BCHOpt when the embedding payload is 0.1. On the other hand, the JPEG techniques nsF5, MBS and MME obtained detection error 0.3298, 0.037 and 0.4307 respectively. This clearly shows that MBS is the least secured and BCHOpt is the highest secure algorithms out of the 6 JPEG steganography algorithms for CC-JRM. In \cite{Holub2013}, the PSRM performance is evaluated on Bossbase and Leticia datasets. As shown in the Table \ref{T:ImagerichSA4}, the PSRM performance for the secure algorithms in each domain is presented for comparison. UNIWARD clearly performed better than other algorithm in each domain. When the embedded payload is 0.1, S-UNIWARD achieved $P_{e}$ 0.3564 in spatial domain, which is higher than HUGO, WOW. The PSRM also showed more resilience for J-UNIWARD and SI-UNIWARD detection. The feature set DCTR performance in JPEG domain is evaluated in the paper \cite{Holub201498}.The authors mentioned that DCTR feature sets had low OOB error compared to the PSRM, SRM methods for the dataset BossBase when tested on J-UNIWARD algorithm, which means DCTR can detect hidden data better than classic SRM, PSRM.

Holub et al. \cite{Holub20159} proposed PHARM feature set performed better than the JRM, PSRM, DCTR for the J-UNIWARD, UED algorithms. For the quality factor 75 and embedding payload 0.1 in JPEG, The PHARM obtained detection error 0.31 for the J-UNIWARD and 0.18 for the UED models, which are better performances than JRM, PSRM. Abdul et al. \cite{Abdulrahman2016} proposed Gaussian filters feature sets to enhance the stego detection. For the embedded payload bpp 0.1, the detection rate for S-UNIWARD, WOW and Sync-HILL is 70.16\%, 69.09\% and  70.54\% respectively, which is reported to be better than color rich models. Song et al. \cite{Song2015} evaluated the detection performance of the GFR feature set on UED, J-UNIWARD, SI- UNIWARD detection. When the Quality factor is set to 75 and embedded payload is 0.2, the detection errors for UED, J-UNIWARD and SI-UNIWARD are 0.18, 0.3 and 0.47 respectively. Furthermore, the detection error for GFR is lower than the CC-JRM and DCTR, which is good. Xia et al. \cite{XiaChao22017} evaluated the performance of the improved GFR version GFR-GW, GFR-GSM on the UED and J-UNIWARD algorithms. The improved GFR version is better performed on UED compared to the J-UNIWARD. Additionally, the improved version GFR-GW(0.2943) performed better than GFR-GSM(0.3071) for UED. The improved PHARM in \cite{Xia2019} has obtained detection error 0.2911 for UED-JC and 0.4023 for J-UNIWARD algorithms when the selected parameters were QF 75 and bpnzac is 0.1. Improved DCTR and GFR performance in \cite{Xia2020} shows that detection error for DCTRD(0.2859) and DCTRD-W(0.2789) for UED is improved compared to J-UNIWARD detection. The detailed performance description of the 6 variations of improved
DCTR, GFR feature sets is given in detail \cite{Xia2020} and it is clear that improved versions significantly improved the detection performance. For 0.1bpnzac and QF 75, MD-CFR \cite{Feng2020} performed better for JC-UED with detection error 0.277 in comparison to UERD(0.373) J-UNIWARD(0.412) and SI-UNIWARD(0.494) detection.

\textbf{Lessons Learned}:
The steganalysis solution performances are evaluated and compared with other solutions using detection error and detection accuracy metrics. The researchers also used the standard stego algorithms such as HUGO, UNIWARD, HILL, WOW, UERD to assess the effectiveness of steganalysis solutions. The steganalysis solutions mentioned in chronological order in Table \ref{T:ImagerichSAP} showed that detection performances improved year by year. The notable steganalysis best performed techniques are CDF in early 2009 followed by the Rich models around 2011 and then PHARM, GFR, DCTR since 2015. But, none of these works evaluated performance on stegomalware detection and it is interesting to see how effective these methods to detect the hidden malware in cover images.

\begin{table*}[h!] 
\centering
\caption{Image domain based feature and rich models steganalysis performance}\label{T:ImagerichSAP}
\resizebox{\textwidth}{!}{%
\begin{tabular}{|l|l|l|l|l|l|}
\hline
\textbf{Authors} & \textbf{Year} & \textbf{Technique} & \textbf{Dataset} & \textbf{Performance} & \textbf{metrics} \\ \hline
Johnson et al. \cite{Johnson1998} & 1998 & Generic  & - & -  & -  \\ \hline
Lyu et al. \cite{Lyu2002} & 2002 & QMF & Natural images & jsteg: 98.5\%  & classification accuracy \\ \hline
Ismail et al. \cite{Avcibas2003} & 2003 & ANOVA & fapp2 & PGS:85\% Stools:75\% Jsteg:70\% & Detection rate \\ \hline
Fridrich et al. \cite{Fridrich2004} & 2004 & higher order DCT  & Greenspun & 0.05bpnzac, F5:0.24 Outguess:0.87 MB1:0.21 MB2:0.16 & Detection reliability \\ \hline
Kenneth et al. \cite{Sullivan2005} & 2005 & markov chain & Diverse sources &  Spatial: Local adaptive: 0.985, 0.893, globally adaptive: 0.982, 0974 & Recall, Precision \\ \hline
Ismail et al.\cite{Ismail2005}  & 2005 & BSM &  greenspun & 15bpp: LSB$\pm$:91.06 & Detection accuracy \\ \hline
Zou et al. \cite{Zou2006} & 2006 & 2-D markov chain & Multiple sources & 0.1bpp, SS:89.15\% non-blind SS:94.10\% QIM:97.03\% LSB:86.30\% & Accuracy \\ \hline

Jan et al. \cite{Jan2010} & 2010 & SPAM, MP, CC-PEV, CDF  & mother image db & 0.003\% bpnzac, CDF: $P_{e} <$ 15\% & probability of error \\ \hline

Fridrich et al. \cite{Fridrich2011} & 2011 & HOLMES & BossBase &  0.5bpp, HUGO, Holmes:12\%  CDF:28.4\%; LSB$\pm$, Holmes:7.3\%  CDF:13.4\% & Detection error  \\ \hline

Fridrich et al. \cite{Fridrich2012} & 2012 & SRM & BossBase &  0.05bpp, HUGO: TOP39:    0.424; EA: TOP39: 0.3255; LSB: TOP39: 0.274 & Running time, Detection error \\ \hline


Kodovsky et al. \cite{Kodovsky20126} &  2011 &  CC-JRM & CAMERA db & 0.05bpnzac,nsF5:0.3298,MBS:0.0373,MME: 0.4307;0.1bpnzac,BCH:0.422,BCHopt:0.448;0.077bpnzac,YASS:0.303 &  detection error   \\ \hline

Holub et al. \cite{Holub2013} & 2013 & PSRM & BossBase, Leica & 0.1bpp, S-UNIWARD:0.3564; QF75, 0.1bpnzac, J-UNIWARD: 0.4319; QF75,0.1bpnzAC, SI-UNIWARD:0.4952 & Detection error \\ \hline

Holub et al. \cite{Holub201498} & 2014 & DCTR & BoSSBase & DCTR:0.1523, SRM:0.2127, PSRM:0.148 & Outof the box(OOB) error\\ \hline

Holub et al. \cite{Holub20159} & 2015 & PHARM  & BOSSbase & QF75,0.2bnpzac, J-UNIWARD:0.31; UED: 0.18; SI-UNIWARD:0.47  & Detection error \\ \hline

Song et al. \cite{Song2015} & 2015 & 2D Gabor filters & BossBase & QF75, 0.2bpnzac UED:0.18, J-UNIWARD:0.3, SI-UNIWARD 0.47 & Detection error \\ \hline

Abdul et al. \cite{Abdulrahman2016} & 2016 & SGF & BossBase & 0.1bpp, S-UNIWARD:70.16\% WOW:69.09\% Sync-HILL:70.54\% & Detection rate \\ \hline

Xia et al. \cite{XiaChao22017} & 2017 & GFR-GW, GFR-GSM & BossBase & 0.1bpnzac, QF75, J-UNIWARD: GFR-GSM:0.4058  GFR-GW:0.3994; UED-JC: GFR-GSM: 0.3071  GFR-GW:0.2943  & Detection error \\ \hline

Xia et al. \cite{Xia2019} & 2019 & Improved PHARM & BossBase & 0.1bpnzac QF75, J-UNIWARD:0.4023, UED 0.2911 & detection error \\ \hline
Xia et al.\cite{Xia2020} & 2020 & Improved DCTR and GFR &  BossBase & 0.1bpnzac QF75, J-UNIWARD: DCTRD:0.4120, DCTRD-W:0.4082; UED: DCTRD:0.2859, DCTRD-W:0.2789 & detection error \\ \hline
Fend et al. \cite{Feng2020} & 2020 & MD-CFR & BossBase & QF75, 0.1bpnzac, J-UED:0.277, UERD:0.373, J-UNIWARD:0.412, SI-UNIWARD:0.494 & detection error  \\ \hline
\end{tabular}%
}
\end{table*}

`


\subsection{Deep learning models for Image steganalysis}

Even though neural networks proved to process the high dimensional data and reduce to one dimension \cite{Holden2006}, the neural network struggle to achieve optimal learning time and thought to be less effective than machine learning algorithms until 2010 \cite{Hinton2006}. But, the advancements in hardware GPU capabilities and innate feature learning capabilities has made deep learning a first choice to addressing the complex classification problems in different applications. 

Determining the feature sets for image steganalysis required domain knowledge and deep understanding of the image pixel level operations. Additionally, the performance need to be improved for the accurate stego image detection using the combination of feature sets and machine learning EC methods. Furthermore, the average running time for high dimensional feature images using ML classifiers is higher. So, researchers explored the application of deep learning models in steganalysis. CNN is a well known to be used for image processing and classification in deep learning. The modification of CNN in accordance with classifying the hidden data images and cover images may be helpful for obtaining optimal detection performance in steganalysis. We discuss various deep learning models present in the prior art addressing the image steganalysis and Table \ref{T:DLSA} illustrates the different deep learning steganalysis solutions proposed in the prior art. The research works are listed in chronological order for ease of solution  comparison and research progress analysis. 

Qian et al. in\cite{Qian2015} proposed GNCNN steganalysis model for stego image detection. The GNCNN architecture contains an image processing layer, five convolutional layers and three fully connected layers. The uniqueness of the GNCNN is that the Gaussian function is considered as non-linear activation function instead of the $Relu$ to add at the output of the convolution layers. The authors described that Gaussian function is better to distinguish the stego and cover image. The GNCNN is tested against the WOW, HUGO and S-UNIWARD steganography algorithms and also compared with the SRM and SPAM feature sets with SVM classifier. Xu et al. \cite{Xu2016} presented a CNN based steganalysis architecture Xu-net to detect the residual based stego detection. The architecture contains a high pass filter for generating residual from the image, the convolution module and the linear classification module. The convolution module comprises 5 groups of convolution blocks including the convolution layers, average pooling, activation function like $Relu$, $TanH$ and batch normalization. The linear classification module contains fully connected and soft max for image classification. The novelty of the Xu-net is the design of the CNN layers for stego detection. The Xu-net is evaluated against the residual based algorithms S-UNIWARD and HILL and reported that Xu-net provides comparable detection performance with SRM feature sets. Ye et al. \cite{Ye2017} proposed SCA-TLU-CNN architecture for image steganalysis. The SCA-TLU-CNN consist of the 9 convolutional layers, 1 fully connected layer and 1 softmax for binary classification. Each convolution layer includes the author's proposed activation function $TLU$ instead of $ReLU$. The selection channel awareness is incorporated in the first layer of the proposed model. The model is tested against WOW, S-UNIWARD and HILL and compared with SRM and maxSRMd2 feature sets. 

Chen et al. \cite{Chen20171} presented CNN based payload estimator for stego detectors. The CNN model softmax module is replaced with mean square error function and the payload values are considered as classification labels. The binary classifier output is fed into 3 fully connected layers with $Relu$ activation function to obtain payload estimate. The estimators have been tested with spatial domain and JPEG domain stego algorithms. The authors \cite{Xu2017} proposed modified CNN solution for detecting the J-UNIWARD stego images. The proposed CNN architecture includes a preprocessing layer to convert JPEG into spatial domain and then applying filter banks to retrieve the magnitude of DCT subbands followed by truncation to limit the input data. After the preprocessing, the data is processed through 20 convolution layers while pooling is achieved with convolution with stride 2 rather than max or average pool. The linear classification contains a fully connected layer followed by a softmax layer for assigning the class labels to the feature vectors. Inspired by Xunet \cite{Xu2016}, the authors \cite{Chen2017} proposed phase aware CNN architecture for JPEG phase awareness detection. The phase split layer is introduced at the output of Group 2 feature maps in Xunet to subsample each feature map into 64 sublattices. Based on how the phases processed later in phase aware CNN, two architectures such as Pnet and Vnet are proposed. In Pnet architecture, the 64 phases are channeled independently, and the net becomes wider towards the last layers. This model requires more memory and the computation complexity increases. On the other hand, all the 64 channels are concatenated into a group and fully convolved into the convolution layer of the next group in Vnet architecture. The Vnet run faster and takes advantage of the correlation between different phases in the subsequent layers of the architecture. Yedroudj et al. \cite{Yedroudj2018} presented Yedroudj-net CNN architecture for spatial steganalysis. the authors designed the Yedroudj-net by considering the best layer components, modules used in ye-net, Xu-net for stego classification. The Yedroudj-net comprises a preprocessing layer, five convolution layers similar to Xu-net and typical CNN classification module. The preprocessing layer contains predefined high pass filter for learning the robust signals in images. The convolution layer also followed by the batch normalization, $Relu$ non-linear activation function, absolute value activation in convolution block 1, average pooling with stride 2 and global average pooling. The classification module contains three fully connected layer followed by softmax activation function for image classification. The proposed model is tested against WOW and S-UNIWARD spatial algorithms.

The authors in \cite{Li20182} designed a training model ReST-Net concatenating the prior art \cite{Xu2016} Xu-CNN architectures while choosing different activation functions. The ReST-Net consists of three parallel subnets and concatenated with the classification module to perform the image classification. Each subnet is the modified Xu-CNN architecture with the convolution group 2 and 4 are replaced with dynamic activation modules (DAM). The DAM comprises three parallel convolution layers, in which one of the $ReLU$, $Sigmoid$ and $TanH$ activation function is  applied to each convolution layer to learn the steganography artifacts. The concatenated feature maps are passed through the next group. Average pooling, batch normalization and activation functions are used in all the layers of the convolution groups and the last layer contain global average pooling. The classification module includes the fully connected layer and softmax function for classification. Tsang et al. proposed \cite{Tsang2018} CNN steganalysis architecture adapted from Ye-net \cite{Ye2017} for stego detection of any image size. The batch normalization is added to each $Relu$ of the Ye-net. Additionally, the 9th convolution stride changed to 1. The moments extraction module is added to capture moments such as maximum, minimum, average and variance of feature maps to identify the image size and other characteristics. For a larger image, the network with moments is trained with smaller size crop images first. Then, the larger image is used to extract the  moments and trained two Inner product layers on the larger image moments to obtain larger image detector. The authors in \cite{Boroumand2019} presented deep residual architecture universal steganalyser "SR-net" for spatial and JPEG stego detection while minimizing heuristics and externally enforced elements in the model. The SR-net is composed of 12 convolution layers and the layers can be any one of the defined 4 layers types. The 4 layer types are defined based on the existence of residual shortcuts and pooling. The two layers of type 1 don't contain the residual shortcuts or pooling. type 2 of layer 2 to 7 contain residual shortscuts and no pooling. The layers 8 to 11 has type 3, in which both pooling and residual shortcuts exist. type 4 has one last convolution layer in the SR-net, which contain average pooling layer and no residual shortcuts. 

Deng et al. \cite{Deng2019} proposed global covariance pooling based CNN steganalysis architecture to improve the training time and also improve the detection performance compared to the state-of-the art SR-net architecture. The proposed model comprises the preprocessing layer with HPF and truncation, 4 groups containing the convolution layers and the linear classifier for image classification. The novel global covariance pooling layer is incorporated in the Group 4, which already contain 2 convolution layers. The group 1 includes 4 convolution layers followed by average pooling with stride 2. The group 2 and 3 contains 2 convolution layers followed by average pooling with stride 2. The model is tested against the SRnet, the proposed model with average pooling instead of  Global covariance pooling. The authors in \cite{Yousfi2020} presented OneHot CNN architecture to effectively detect the stego images in JPEG domain. The OneHot encodes the DCT coefficients of the images into binary volumetric representation of the DCT plane. The encoded DCT values are fed to two convolution modules followed by Global average pooling layer.  The classification module consists of fully connected layer for binary image classification. The proposed model is tested against the nsF5 and J-UNIWARD detection and showed that OneHot CNN is better than JRM. Li et al. \cite{Li20205} performed steganalysis based on the feature fusion of SRNet base learners and used ensemble classifiers to obtain the better performance. Several decision combinations such as majority voting, product combination are used to combine the SRNet base learners. Furthermore, the ensemble classifiers are applied to the base learner's serial and parallel feature fusion to obtain the results. The authors in \cite{Zhang20205} proposed CNN based architecture Zhu-net for efficient detection of image steganography. Various changes in preprocessing, convolution layers and pooling layer has been proposed to model the Zhu-net. The 3x3 kerenl sizes are recommended instead of 5x5 in the preprocessing layer to reduce the number of parameters and capture the local region features. Additionally, the depth wise convolution layer is proposed to improve the signal-to-noise ratio, utilizing the channel correlation of the residual in the convolution modules. Finally, the spatial pyramid pooling (SPP) is also presented as the last pooling layer prior to the classification module to represent the features with multi level pooling. Further, additional datasets are considered to further boost the detection performance. The authors in  \cite{Yedroudj2020} performed data enrichment by improving the datasets so that the detection performance increases. The data enrichment method "pixels-off" removes few pixels from the image to enrich the datasets. As the number of images in the dataset increases, it helps to increase the overall detection performance. The pixels-off is tested against the Yedroudj-net and covariance pooling  CNN for performance evaluation.

The authors in \cite{Ahn2020} presented LSER network to improve the detection performance. The LSER architecture contains 3x3 convolution layer followed by local resource group layer and two local source downsample group layer. Then, second order global pooling with iterative matrix square root normalization, fully connected layer and softmax to classify the stego and cover images. LSER mainly contains two characteristics such as ensemble residual and local source skip connection. The residual block with no batch normalization is considered in the LSER architecture. Local source skip connection allows bypassing features from different levels so that precise features are used for representation. The LSER is tested against against both spatial and JPEG algorithms. Jang et al. \cite{Jang2020} proposed feature aggregation networks (FANet) and leveraged $Relu6$ as an activation function for image stego detection. The feature aggregation module contains two down sampling, two up sampling and one residual block to aggregate the feature maps at various level and resolutions. The feature aggregation is performed near the input data to expand the number of channels of convolutions blocks. Overall, the FANet architecture contains 16 blocks varying the block types from 1 to 7 followed by linear classifier for classification. The authors in \cite{Xu2021} presented Reparameterization Vgg (Repvgg) \cite{Ding202109} block and Squeeze-and-excitation (SE) based feature fusion steganalysis architecture SFRNet. The Repvgg block contains a stack of 3x3 convolution layers and $Relu$ activation, while having multibranch topology. It will improve the inference and efficient memory utilization. The Squeeze-and-excitation block is used to improve the detection accuracy rate. The SFRNet architecture comprises the feature extraction and fusion layer followed by the Repvgg blocks with $TLU$, $ReLU$ and average pooling spreading over five stages and three SE blocks are incorporated between the stages. The linear classifier includes three fully connected layer and softmax function for stego and cover image classification. 

The DFSE-Net proposed in \cite{Liu2021} composed of the diverse filter module (DFM)  and squeeze-and-excitation (SE) modules for effective detection against the content adaptive steganographic methods. The diverse filter modules combine three different scale convolution filters to obtain the diverse information and squeeze-and-excitation module to strengthen the key channels. The DFSE architecture contain preprocessing layer with $TLU$ and High pass filter, convolution with batch normalization, three DFSE modules including DFM and SE and then followed conventional linear classifier with fully connected and softmax function. The authors in \cite{Reinel2021} presented GBRAS-Net architecture, which comprises the preprocessing layer with filter banks, multiple depth wise and separable convolution layers for improving the performance and skip connections to speed up the learning. The authors in \cite{Singh2021}  proposed deep fractal network based  architecture SFNet for image steganalysis. The SFNet comprises multiple CABR(Convolution, Absolute, Batch, $RelU$) and CBR(Convolution, Batch, $RelU$ blocks with balanced width and depth in the network followed by the global averaging, fully connected and softmax function.The SFNet does not require preprocessing steps and can achieve good performance using depth and wide coverage of the network. Mondal et al. \cite{Mondal2021} presented H-Stegonet hybrid deep learning technique by combining the MHSRMNet and StegoRUNet. StegoRUNet is the modified version of Residual U-net \cite{Zhang2018}. MHSRMNet process the feature vector into multiple bins to reduce the dimension and then concatenate those bins to get the final decision output. The two nets are combined and used dense and dropout layers to perform the classification. The authors tested the  model against HUGO and WOW S-UNIWARD detection. 

\textbf{Lessons learned}:
The image steganalysis research has shifted towards deep learning steganalysis than the conventional steganalysis is lately. There are number of deep learning steganalysis solutions proposed in the literature to address stego detection. The solutions mainly focus on proposing changes in preprocessing layer, activation function, convolution layers arrangement, blocks fusion, linear classifier to capture the stego elements. As seen in conventional steganalysis review, the stego algorithms WOW, HILL, UNIWARD, UED are mainly used for performance comparison. As the deep learning field is actively progressing in image processing applications, there are more research opportunities in future to utilize the deep learning models for image steganalysis and improve the performances even more. We have not found any literature works performing stegomalware detection using deep learning models and pose to be a security malware research direction in future to test the existing deep learning models and proposing new models for stegomalware detection.

\begin{table*}[h!] 
\centering
\caption{Deep learning steganalysis}\label{T:DLSA}
\resizebox{\textwidth}{!}{%
\begin{tabular}{|l|l|l|l|l|l|}
\hline
\textbf{Authors} & \textbf{year} & \textbf{Technique} & \textbf{Embedding Algorithms} & \textbf{Advantages} &  \textbf{Comment} \\ \hline
Qian et al. \cite{Qian2015} & 2015 & GNCNN & HUGO, WOW, and S-UNIWARD & GNCNN achieved comparable performance to SRM & GNCNN still has room for detection improvement  \\ \hline
Xu et al. \cite{Xu2016} & 2016 & Xu-Net or Xu-CNN & S-UNIWARD and HILL & Xu-net obtained comparable detection performance to SRM  & The Xu-net only learns from the noise residual.  \\ \hline
Ye et al. \cite{Ye2017} & 2017 & SCA-TLU-CNN or Ye-Net & HUGO, WOW, and S-UNIWARD & superior performance compared to SRM, maxSRMd2 &  TLU and selection channel knowledge improved the performance  \\ \hline
Chen et al. \cite{Chen20171} & 2017 & CNN Payload estimator & WOW and S-Uniward, J-UNIWARD and UED-JC &  Estimated the size of payload using CNN & softmax is replaced with MSE \\ \hline
Xu et al. \cite{Xu2017} & 2017 & CNN-J-UNIWARD & J-UNIWARD & Outperformed SCA-GFR & Only applicable to J-UNIWARD  \\ \hline
Chen et al. \cite{Chen2017} & 2017 & Pnet,  Vnet & J-UNIWARD, UED-JC & JPEG Phase awareness incorporation in the CNN & SCA-GFR still performs better than individual Vnet for J-UNIWARD detection \\ \hline
Yedroudj et al. \cite{Yedroudj2018} & 2018 & Yedroudj-net & S-UNIWARD, WOW & Yedroudj-net outperformed  Xu-net, Ye-net, Rich models+ EC & Only applicable for spatial steganalysis \\ \hline
Li et al. \cite{Li20182} & 2018 & ReST-Net & S-UNIWARD, HILL, CMD-HILL & ReST-Net performed better than Xu\-CNN \cite{Xu2016} and TLU\-CNN  & training time can be much longer than Xu-CNN \\ \hline
Tsang et al. \cite{Tsang2018} & 2018 & SID & LSBM and WOW & Stego detection on arbitrary image size &  feature maps statistical moments are the key to preserve image size \\ \hline
  
Boroumand et al. \cite{Boroumand2019} & 2019 & SRNet & S-UNIWARD, HILL, WOW, J-UNIWARD, UED-JC & SRnet improved  performance significantly in JPEG domain & Enforced elements in the architecture which are universal and minimize the heuristics \\ \hline
\cite{Deng2019} & 2019 & Covariance pooling  CNN & S-UNIWARD, HILL, WOW & Improved training time and detection performace compared to SRnet & Selection channel awareness may improve the performance even more\\ \hline

Yousfi et al. \cite{Yousfi2020} & 2020 & OneHot CNN & nsF5, J-UNIWARD & Onehot CNN performed better than JPEG rich models &  Onehot along SRNET combination can obtain promising results \\ \hline

Li et al. \cite{Li20205}& 2020 & SRnet Ensemble Classifier & WOW and J-UNIWARD & The feature fusion with EC SRNet performed better than SRNet alone &  The training sets carefully selected for multiple SRNet base learners \\ \hline

Zhang et al. \cite{Zhang20205} & 2020 & Zhu-Net & WOW, S-UNIWARD and HILL &   improved performance compared to SRM, \cite{Ye2017}, \cite{Xu2016}, \cite{Yedroudj2018} and \cite{Boroumand2019} & SPP module may be used for stego detection of any image size \\ \hline
Yedroudj et al. \cite{Yedroudj2020} & 2020 & pixels-off & S-UNIWARD,WOW & Improved detection performance when use data enrichment & The data enrichment seems to be one of the future aspect to improve the stego detection\\ \hline
Ahn et al. \cite{Ahn2020} & 2020 & LSER  & WOW, S-UNIWARD, J-UNIWARD, UED-JC & LSER performed better than SRNet and Zhu-net & LSER may have running time overhead.  \\ \hline

Jang et al. \cite{Jang2020} &  2020 & FANet &  J-UNIWARD, UED & FANet obtained better performance compared to SRNet & ReLU6 as a activation function for better generalization.\\ \hline

Xu et al. \cite{Xu2021} & 2021 & SFRNet & HUGO, WOW, S-UNIWARD, and MiPOD & SFRNet performed better than SRNET and Zhu-net & The combination of RepVgg block and Squeeze and excitation module is used in SFRNet\\ \hline

Liu et al. \cite{Liu2021} & 2021 & DFSE-Net & WOW, S-UNIWARD & performed better than Xe-net, Ye-net and Yedroudj-Net & the model is only deal with images with same size \\ \hline

Reinel et al. \cite{Reinel2021} & 2021 & GBRAS-Net &  WOW, S-UNIWARD, MiPOD, HILL and HUGO  & Performed better than Zhu-net, SR-Net & depthwise and separable convolutional layers, and skip connections \\ \hline

Soumik et al. \cite{Mondal2021} & 2021 & H-Stegonet & S-UNIWARD, WOW & H-stegonet outperformed Zhu-net, SRNet, Ye-Net & \\ \hline

Brijesh et al. \cite{Singh2021} &  2021 & SFNet & WOW,S-UNIWARD, HILL & Outperformed SRnet and SCA-SRNET & The fractal network can be applied in JPEG domain too\\ \hline

\end{tabular}%
}
\end{table*}

 \subsection{Deep learning Steganlysis performance}

The performance evaluation of GNCNN \cite{Qian2015} illustrated that GNCNN has obtained comparable performance with SRM feature sets and better performance compared with SPAM feature sets. When the payload capacity 0.3bpp, GNCNN achieved detection error 0.338 on HUGO, 0.343 on WOW and 0.359 on S-UNIWARD stego detection. These results showed that S-UNIWARD is slightly more secured than HUGO, WOW for GNCNN steganalysis. The Xu-net \cite{Xu2016} achieved detection accuracy 80.24\% for S-UNIWARD and 79.24\% for HILL when the embedded capacity is selected as 0.4bpp. Xu-net architecture showed similar performances for both  Hill and S-UNIWARD. Additionally, for S-UNIWARD, SRM obtained detection accuracy of 79.53\%, which is comparable performance to Xu-net. Ye et al. \cite{Ye2017} model SCA-TLU-CNN performance evaluation depicts that the detection error for WOW, S-UNIWARD and HILL are 0.1691,0.2224 and  0.2538 respectively for resampled images when the payload is 0.2bpp. HILL is more secure against SCA-TLU-CNN when compared to WOW and S-UNIWARD. But, we can clearly see that the SCA-TLU-CNN performed well compared to GNCNN and Xu-net to detect the stego images, although the dataset sample size used for evaluation may not be the same. In \cite{Chen20171}, the detector used for payload estimator has the following detector error when applied to spatial and JPEG algorithms. The detection error for WOW and S-UNIWARD are 0.2796 and 0.3452, whereas the detection error for J-UNIWARD and UED-JC is 0.4040 and 0.2450 when quality factor is 75. The authors in \cite{Xu2017} tested the modified CNN for J-UNIWARD detection using BoSSBase and CLS-LOC image dataset. With the parameters QF75 and embedding capacity of 0.2bpnzac, the modified CNN achieved 0.1947 detection error for J-UNIWARD detection and performed better than SCA-GFR. The phase aware CNN proposed in \cite{Chen2017} performance shows that phase aware CNN can detect UED-JC better than J-UNIWARD with almost 50\% reduction in detection error, as shown in Table \ref{T:DLperformanceSA}. Further, the Pnet, Vnet performed better than SCA-GFR  detection in UED-JC. However, SCA-GFR still performed slightly better than Pnet and Vnet in J-UNIWARD detection. The authors \cite{Yedroudj2018} presented CNN based spatial steganalyzer "Yedroudj-net" for stego detection. The Yedroudj-net able to detect WOW better than S-UNIWARD stego images. For embedding rate 0.2bpp, the $P_{e}$ for S-UNIWARD is 36.7\%, which is higher than the WOW 27.8\%. Additionally, the Yedroudj-net outperformed the prior art Xu-net(32.4\%), Ye-net(33.1\%) and SRM+EC(36.5\%) for the WOW detection. Similar performance for Yedroudj-net is achieved for J-UNIWARD detection when compared to the existing models. So far. we could see that SCA-TLU-CNN reported best performances compared to all the other models in all the image domains.

The ReST-Net model in \cite{Li20182} is evaluated for S-UNIWARD, HILL and CMD-HILL detection. The model is efficiently detecting S-UNIWARD than the HILL and CMD-HILL stego images. The detection accuracy for S-UNIWARD is 71.35, whereas the HILL and CMD-HILL produced 70.64 and 65.14 when embedding rate is 0.2bpp. Additionally, the ReST-Net performed better than the Xu-CNN, TLU-CNN models for the three stego algorithms. This performance improvement is due to the feature maps concatenation from parallel subnets as well as layers in convolution groups. The size independent detector (SID) in \cite{Tsang2018} is tested against with RTRIP detector for proving that the SID works for any image size. For embedded rate 0.12bpp and the image size 256x256, the SID achieved detection error 0.259 on WOW, which is comparable to RTRIP detection error 0.261. The same trend followed for 1024x1024 image size with detection error 0.1391 and 0.1445 for the SID and RTRIP respectively. The SR-net architecture \cite{Boroumand2019} performance is evaluated on spatial and JPEG stego detection to compare the performance against the state-of-the-art methods. For embedding rate 0.2bpp, the detection error of the SRnet spatial models S-UNIWARD, HILL and WOW are 0.20, 0.23 and 0.16 respectively. This shows that SRnet can detect WOW  slightly better than  HILL and S-UNIWARD. Additionally, the SRNet is also performed better than the SCA-YeNet for all the three stego algorithms. In JPEG domain, the model is tested against the J-UNIWARD and UED-JC. With QF 75, 0.2bpp, the detector errors for J-UNIWARD and UED-JC are 0.1889 and 0.568. It shows that J-UNIWARD is more secured against SRNet compared to the UED-JC. The model obtained superior performance in JPEG domain compared to the prior arts. At this point, SR-net and SCA-TLU-CNN models has given the best performances for deep learning based stego detection.

The global covariance layer based CNN \cite{Deng2019} performance shows that the detection accuracy 80.05, 77.11 and 84.33 obtained for S-UNIWARD, HILL and WOW detection when the embedding rate 0.2bpp. The WOW stego images are more likely to be detected with the covariance CNN in comparison with the S-UNIWARD and HILL. Under the same experimental setting, the SRnet obtained 84.05 detection accuracy  and the model with average pooling obtained 83.78 detection accuracy for the WOW detection. This shows that the author's model slightly improved the detection performance compared to SR-net. Additionally, the average time taken to complete one iteration during training is significantly reduced for the covariance CNN (65ms) compared to the SRNet (261ms). The peformance of OneHot CNN in \cite{Yousfi2020} showed that the detection error of OneHot CNN is 3.49 for nsF5 and 7.36 for J-UNIWARD, when the quality factor and embedding rate is set to 100 and 0.2bpp. Furthermore, the authors mentioned that OneHot CNN performed better than JRM with detection error 4.17 for the same parameter settings. They also showed that combination of SRnet with OneHot CNN fusion improves the performance significantly compared to the SRnet for nsF5, J-UNIWARD, UED-JC. In \cite{Li20205}, the authors evaluation of SRNet base learners serial and parallel feature fusion combinations with ensemble classifier showed that both feature fusion methods performed better than SRnet for WOW and J-UNIWARD detection. When the embedding rate 0.2 and serial feature fusion is used for classification, the $P_{e}$ is 0.1872 for WOW and 0.2367 for J-UNIWARD detection. The similar performance 0.1878 for WOW and 0.2308 for J-UNIWARD is achieved when used parallel feature fusion and EC. For embedding rate 0.2bpp, the Zhu-net performance in \cite{Zhang20205} obtained the detection error 0.233 and 0.285 for WOW and S-UNIWARD detection. The authors reported that zhu-net performed  better than Xu-net, Ye-net, SRM+EC, Yedroudj-net, SRnet. Furthermore, the addition of the more datasets, the zhu-net detection error reduced from 0.233 to 0.131, which is significant performance improvement. The testing performance of the "pixels-off" in \cite{Yedroudj2020} shows that Yedroudj-Net with 400 pixels off Boss dataset achieved 23.5 detection error for WOW and 26.5 for S-UNIWARD in comparison with 27.71 for WOW and 35.42 for S-UNIWARD for the original datasets. This clearly showed that "pixels off" data enrichment techniques improves the decision performance for Yedroudj-Net. The similar phenomenon is also true for covariance CNN. The LSER \cite{Ahn2020} performance is evaluated using BoSSBase and BOWS2 datasets. For embedding rate 0.2bpp, the LSER achieved 0.2375 and 0.2846 for WOW and S-UNIWARD detection in spatial domain. It shows that WOW is more secured than S-UNIWARD against the LSER detection. For embedding rate 0.2bpnzac and QF 75, LSER obtained 0.1176 and 0.3115 detection error for UED-JC and J-UNIWARD detection in JPEG domain. Overall, the J-UNIWARD is more secure than other three algorithms for stego detection using LSER. Additionally, the authors mentioned that LSER performed well compared to SRNet and Zhu-net in both spatial and JPEG domain for all the four stego techniques.

The FANet performance in \cite{Jang2020} is tested using ALASKA-V2 image datasets. The FANet obtained 71.22 and 84.24 detection accuracy for J-UNIWARD and UED when the QF 75 and embedding rate 0.2bpnzac is selected. The J-UNIWARD stego images are more secured compared to the UED when FANet is used for detection. Further, the FANet performed better than the SRNet (70.14 and 79.32 for J-UNIWARD and UED) when the embedding rate is 0.2bpnzac and QF 75.  The performance of DFSE-NET  \cite{Liu2021}  reveals that detection error for WOW and S-UNIWARD is 0.247 and 0.341 respectively when the embedding rate is chosen to be 0.2bpp. The DFSE-NET is also performed better than Xu-Net, Ye-Net and Yedroudj-Net with detection errors 0.345, 0.306 and 0.332 respectively. The SFRNet proposed in \cite{Xu2021} performance showed that Mi-POD is slightly secured than other spatial techniques against the SFRNet steganalysis. Additionally, the article reported that SFRNet outperformed prior art models SRNet, Zhu-net, DFSE-Net in terms of detection accuracy and testing time. SFRNet took 9 sec whereas other model consume more than 25 sec to test the images.  Mondal et al. \cite{Mondal2021} architecture H-Stegonet obtained the classification error 35.5 and 41.4 for the WoW and S-UNIWARD, when the embedding rate is 0.2bpp. When the embedding rate 0.2bpp, GBRAS-Net \cite{Reinel2021} obtained the best detection accuracy 80.3 for WOW and least detection accuracy 68.5 for HILL. In comparison with popular prior art solutions like Zhu-net (76.9), SR-net (75.5) and Ye-net (66.9) for WOW detection, GBRAS-Net performed much better. The SFNet \cite{Singh2021} performance indicated that HILL is more secured compared to WOW and S-UNIWARD for stego detection, as shown in \ref{T:DLperformanceSA}, when the embedding rate is 0.2bpp. Furthermore, the authors reported that SFNet performed better than SRNet and SCA-YeNet. 

\textbf{Lessons learned}:
The literature deep learning model performances are mostly compared using detection accuracy, detection error and BoSSBase is considered as a standard benchmark database for evaluation. Based on the performance evaluations of deep learning models listed in Table \ref{T:DLSA}, the models SCA-TLU-CNN, SR-Net, Zhu-net, LSER, SFR-Net and  GBRAS-Net showed notable and improved detection performances. These models may be used as a reference for performance comparison in DL steganalysis future contributions. One of our future work is to evaluate the performances of the existing DL solutions in stegomalware conceal in images detection.

\begin{table*}[h!] 
\centering
\caption{Deep learning models for Image steganalysis performance}\label{T:DLperformanceSA}
\resizebox{\textwidth}{!}{%
\begin{tabular}{|l|l|l|l|l|l|}
\hline
\textbf{Authors} & \textbf{year} & \textbf{Technique} & \textbf{dataset} & \textbf{Performance} & \textbf{metric} \\ \hline

Qian et al. \cite{Qian2015} & 2015 & GNCNN & BOSSbase 1.01, ImageNet & 0.3bpp, HUGO: 0.338, WOW: 0.343, S-UNIWARD: 0.359 & Detection error \\ \hline
Xu et al. \cite{Xu2016} & 2016 & Xu-Net & BOSSbase 1.01 & 0.4bpp, S-UNIWARD:79.53 and HILL:75.47 & Accuracy  \\ \hline
Ye et al. \cite{Ye2017} & 2017 & SCA-TLU-CNN  & BOSSbase 1.01, BOWS2 & 0.2bpp, WOW: 0.1691, S-UNIWARD: 0.2224, HILL: 0.2538 & Detection error \\ \hline
Chen et al. \cite{Chen20171} & 2017 & CNN Payload estimator & BoSSBase & $\alpha$ 0.1 WOW:0.2796, S-UNIWARD:0.3452; QF75, JUNI75:0.4040, UED-JC75:0.2450 & Detection error \\ \hline
Xu et al. \cite{Xu2017} & 2017 & CNN based J-UNIWARD detection &  Bossbase v1.01,CLS-LOC  & QF-75, bpnzAC:0.2 CNN-J-UNIWARD: 0.1947 & Detection error \\ \hline
Chen et al. \cite{Chen2017} & 2017 & Pnet, Vnet & BossBase & QF75, 0.2bpnzac, J-UNIWARD(Pnet: 23.50 Vnet:24.57); UED-JC(Pnet: 9.55 Vnet:10.07) & Detection error\\ \hline
Yedroudj et al. \cite{Yedroudj2018} & 2018 & Yedroudj-net & BOSSBase v.1.01 & 0.2bpp, WOW: 27.8\% S-UNIWARD: 36.7\% & Probability  error \\ \hline
Li et al. \cite{Li20182} & 2018 & ReST-Net  & BOSSBase v1.01 & 0.2bpp, S-UNIWARD:71.35, HILL:70.64, CMD-HILL:65.14 & Detection accuracy \\ \hline
Tsang et al. \cite{Tsang2018} & 2018 & SID  & BOSSbase 1.01  & 0.12bpp WOW; 0.01 change rate LSBM; 256x256 LSBM SID:0.243 WOW:0.259; 1024x1024 LSBM:0.0856 WOW:0.1390  & detection error \\ \hline
Mehdi et al. \cite{Boroumand2019} & 2019 & SRNet &  BOSSbase and BOWS2 & 0.2bpp, Spatial(S-UNI: 0.2090, HILL: 0.2353, WOW: 0.1676); QF75, JPEG(J-UNIWARD-0.1889, UED-JC -.0568) & Detection Error\\ \hline
Deng et al. \cite{Deng2019} & 2019 & Covariance pooling based CNN & BOSSBase and BOWS2  & 0.2bpp, S-UNIWARD:80.05, HILL:77.11, WOW:84.33 & Detection accuracy \\ \hline
Yousfi et al. \cite{Yousfi2020} & 2020 & Onehot CNN & union of BOSSbase 1.01 and BOWS2 & QF:100, 0.2bpp nsF5: 3.49; 0.4bpp J-UNIWARD:7.36 & Detection error \\ \hline
Li et al. \cite{Li20205} & 2020 & SRNet ensemble classifier & BOSSBase v1.01 & SF-EC, 0.2bpp WOW:0.1872, 0.2bpnzac J-UNIWARD:0.2367; PF-EC, 0.2bpp WOW:0.1878, 0.2bpnzac J-UNIWARD:0.2308  & Prob error \\ \hline
Zhang et al. \cite{Zhang20205} & 2020 & Zhu-Net & BOSSBase v1.01, BOWS2 & 0.2bpp, WOW: 0.233 S-UNIWARD: 0.285 & detection  error\\ \hline

Yedroudj et al. \cite{Yedroudj2020} & 2020 & pixels-off & BOSS, Alaska &  0.2bpp, Yedroudj-Net: WOW: 23.5, S-UNIWARD: 29.3; CovPool-Net: WOW: 23.34, S-UNIWARD: 26.64 & detection error \\ \hline
Ahn et al. \cite{Ahn2020} & 2020 & LSER  & BOSSbase 1.01, BOWS2 & 0.2bpp, WOW:0.2375; QF75, 0.2bpnzac, S-UNIWARD:0.2846, UED-JC:0.1176, J-UNIWARD:0.3115 &  detection error \\ \hline

Jang et al. \cite{Jang2020} &  2020 & FANet &  ALASKA-V2  & 0.2bpnzac, QF75, J-UNIWARD:71.22 UED:75.09 & Detection accuracy \\ \hline

XU et al. \cite{Xu2021} & 2021 & SFRNet & BOSSBase 1.01 & 0.2bpp, HUGO:75.4, WOW: 76.8 S-UNIWARD:72.5, and MiPOD: 75.2; test time:9 sec & Detection accuracy, test time\\ \hline

Liu et al. \cite{Liu2021} & 2021 & DFSE-Net & BoSSBase & 0.2bpp, WOW:0.247, S-UNIWARD:0.341 & Detection error \\ \hline

Reinel et al. \cite{Reinel2021} & 2021 & GBRAS-Net & BoSSBase, BOWS & 0.2bpp, WOW: 80.3 S-UNIWARD:73.6, Mi-POD:68.3, HILL:68.5, HUGO:74.6 & Detection accuracy \\ \hline

Mondal et al. \cite{Mondal2021} & 2021 & H-Stegonet & BoSSBase, BOWS & BoSSBase, 0.2bpp, WOW:41.4, S-UNIWARD:35.5 & classification error\\ \hline

Singh et al. \cite{Singh2021} &  2021 & SFNet & BOSSBase & 0.2bpp, WOW:0.1579, S-UNIWARD:0.1964, HILL: 0.2438 & Detection error\\ \hline

\end{tabular}%
}
\end{table*}

%% file: S5Stegoframework.tex
\section{Stegomalware Detection framework}
\label{sec:Stegoframework}

In this section, we describe the stegomalware creation process,  multimedia malware analysis framework to identify the enterprise organization network targeting stegomalware, and the different network architectures in datacenter, cloud or multicloud environments used to deploy the malware analysis framework.

\subsection{Stegomalware Creation Process}
\begin{figure}[!h]	
\centering
\includegraphics[width=8.3cm,height=8cm]{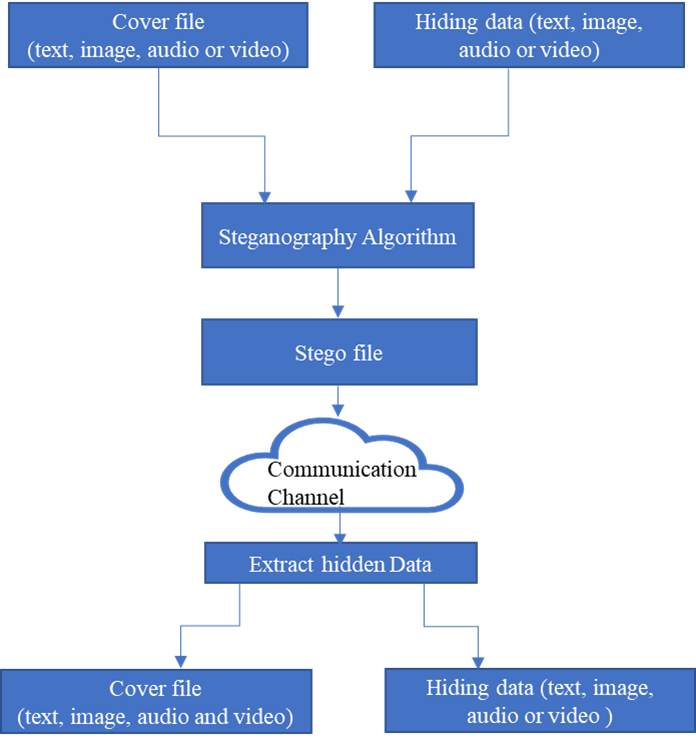}
\caption{Generic Stegomalware creation process} 
\label{f.stegprocess}
\end{figure}

The stegomalware creation process involves selecting the cover or carrier medium like image, audio, or video files as an input file; the intended hiding data may be C\&C server IP address, URL, malware payload, shellcode commands, other malicious intent Linux or windows commands to run on the compromised victim machine when triggered, and inputting the cover medium and hiding data to the chosen steganography tool or algorithm for generating the stegomalware file. The Figure \ref{f.stegprocess} illustrates the top-down representation of the stegomalware creation process.

Steganography can be performed using the tools mentioned in Table \ref{T:tools} or GAN based stego generation or adversary customized  tools like WOW, UNIWARD, GFR, HILL for generating more sophistication stegomalware to evade the victim security defense tools. There are different ways to hide the content in the cover medium so that the content embedded carrier file looks like the original carrier. The steganography tools may use the steganography algorithms like classical LSB, PVD or transform domain techniques like DCT, DWT to hide the malicious content. In addition, the hidden content is firstly encrypted with private key using symmetric encryption algorithms like RC4, DES or AES to have another layer of protection along with steganography. Consequently, the stegomalware is delivered to the victim machine through communication channel. The most generic communication channel in most of the stegomalware cases is internet. Subsequently, the receiver collects the stegomalware file through phishing emails or another form of weaknesses in the victim machine. An exploit kit running on the victim machine instruct the malicious code to extract the hidden content from the stego image. If the hidden content is the malicious attacker server IP address or URL, then the exploit kit uses those artifacts for connecting to the remote server without being caught by the security tools. Then, it may download another malicious code script from the remote server to perform the exfiltration of the confidential data or encryption keys to encrypt the victim data. It is also possible that the malicious code hidden in the stego malware and then decode the content to run on the victim machine. The victim can only see the cover file and may not be able to know the hidden malicious activity happening behind.  

\begin{figure}[!h]	
\centering
\includegraphics[width=7.3cm,height=7.5cm]{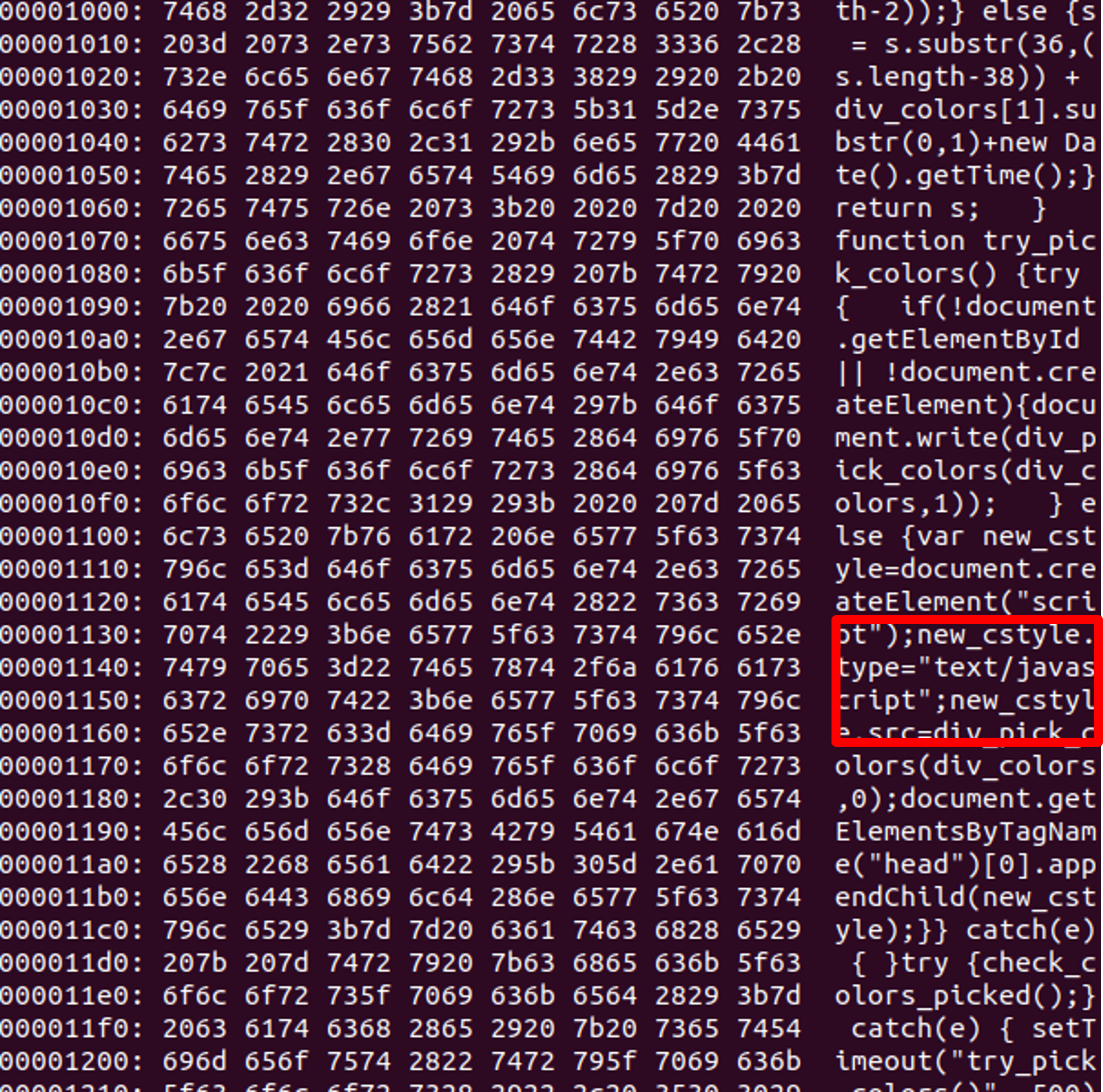}
\caption{Malicious JavaScript hidden in Image} 
\label{f.stegsample}
\end{figure}

We have collected the stegomalware samples from virushare.com and performed the string analysis using “xxd” command in Ubuntu machine. The Figure \ref{f.stegsample} shows the excerpt of the stegomalware representing the HTML and JavaScript code in the cover image and their hexadecimal representation of the values. The basic string analysis tools like “strings”, “exiftool”, “binwalk”, “foremost”, “pngcheck”, “identify” and “ffmpeg” in Unix based operating system are enough sometimes to identify the stego image. For instance, the adversary uses the Exif header, which normally contain with camera hardware or other source of image captured relevant information to hide the content. So, running “exiftool” on the suspicious command may show the exif header and an investigator can visually validate the hidden content in the header for verification during the forensic process.

\subsection{Stegomalware Analysis Framework}
The proposed stegomalware analysis framework may be semi-automated or fully automated based on the size of organization, frequency of the steganography images seen in the organization, number of inbound multimedia files received by the organization, and the organization business operations focus either providing security services or product for the customers. The stegomalware analysis process is as follows:

In general, an employee may report a suspicious image file seen in the received phishing email or security team may receive a suspicious alert regarding the malicious outbound communication to previously known malicious IP address when an employee accesses an image or video file. So, the employee reported suspicious images will be instantly uploaded to the standard storage location, which is isolated from the rest of the application infrastructure to stop the accidental infection while storing the file. Additionally, the security team member may also upload the suspicious file for analysis. The received file can be scanned for the detection of malware behavior and determine maliciousness of the file using the signature-based detection tools. If the file is identified as malicious with matching hash values of the known malware signature or other behavioral characteristics, we may red flag the image and perform the preventive actions. The preventive actions can be isolating the infected machine from other network machines or updating the malware hash signature of the image in end point security policies to block the image malware and stopping the infection in the network. 

\begin{figure}[!h]	
\centering
\includegraphics[width=8.3cm,height=9cm]{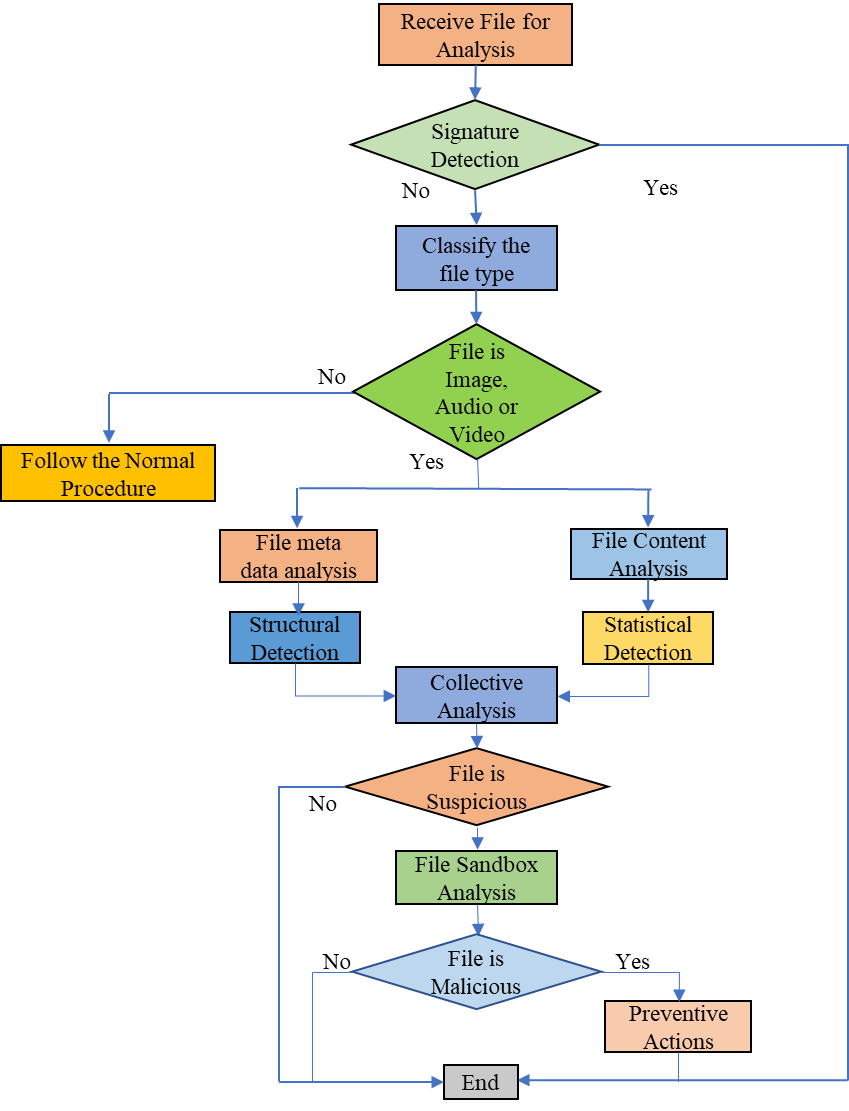}
\caption{The Proposed Framework} 
\label{f.s56}
\end{figure}

If the file is not identified as malicious in the preliminary signature detection as shown in Figure \ref{f.s56}, we forward the image to the next step in the pipeline to determine the file format and type of the file. The file type can image, audio or video and the file format may be JPEG, PNG, GIF for images, WAV for audio format and video format may be MP4, MOV, AVI but not limited. One of the challenging tasks is to identify the stegomalware when multimedia file can be any format. The camera hardware used to capture the original cover medium can be different and hence can have different Exif header for each file. So, the accurate detection using structural changes and statistical features is needed. If the file is not an audio, image, or video file, we may use the existing procedure to forward it to the anti-malware tool for analysis. The antimalware tool can be vendor offered for purchase or the homegrown tool for further checking the maliciousness of the file. If the file is image, audio or video, we perform the structural and statistical analysis on the file for malware detection. The structural analysis includes changes in timestamp and dates, unusual file properties such as file size, checksum and content modification, anomalies in the Exif header content. We use the open-source tools mentioned in  \cite{DominicBreuker2017} and StegSpy for structural analysis of the file. An anomalous structural property would flag the file is suspicious for further analysis. 

In addition, the statistical analysis of the files is performed to find more evidence on the maliciousness of the file. The statistical properties may include byte and n-gram histogram of the files, the pattern changes in the pixels of the image or video frames and least significant bit changes in the images. The existing steganography detection tools such as StegExpose, Stegdetect can be used for detecting the statistical anomalies in the multimedia files as part of the proposed malware framework. The collective response from the structural and statistical anomaly scores are combined and evaluate the maliciousness or malware suspicion of the file. If the file is indicated as suspicious or malicious, then the file is forwarded to  cuckoo sandbox environment for dynamic malware analysis and identify the behavioral characteristics of the file. There is highly likely that the file hidden malicious content can be extracted and could
perform the malicious activity as per the embedded code instructions. For instance, the shell code embedded in the image file may be executed and tried to connect to the remote server for executing malicious commands and may exfiltrate the data. So, based on the behavior of the stegomalware, we may have to take the preventive in the environment if it is malicious. The preventive actions again can be updating the malware signature for the indicators of compromise like IP address, domains, and other hexadecimal code signatures for detection of the malware in the infrastructure environment. If the file behaves normally during the dynamic analysis, we may ignore the file for further actions and may track these files for avoiding the false positives in the future. The Figure \ref{f.s56} shows the workflow of the proposed stegomalware analysis framework.

\subsection{Enterprise Architectures for The Proposed Framework}
\subsubsection{Enterprise Datacenter stegomalware detection Architecture}
\begin{figure}[!h]	
\centering
\includegraphics[width=8.3cm,height=6cm]{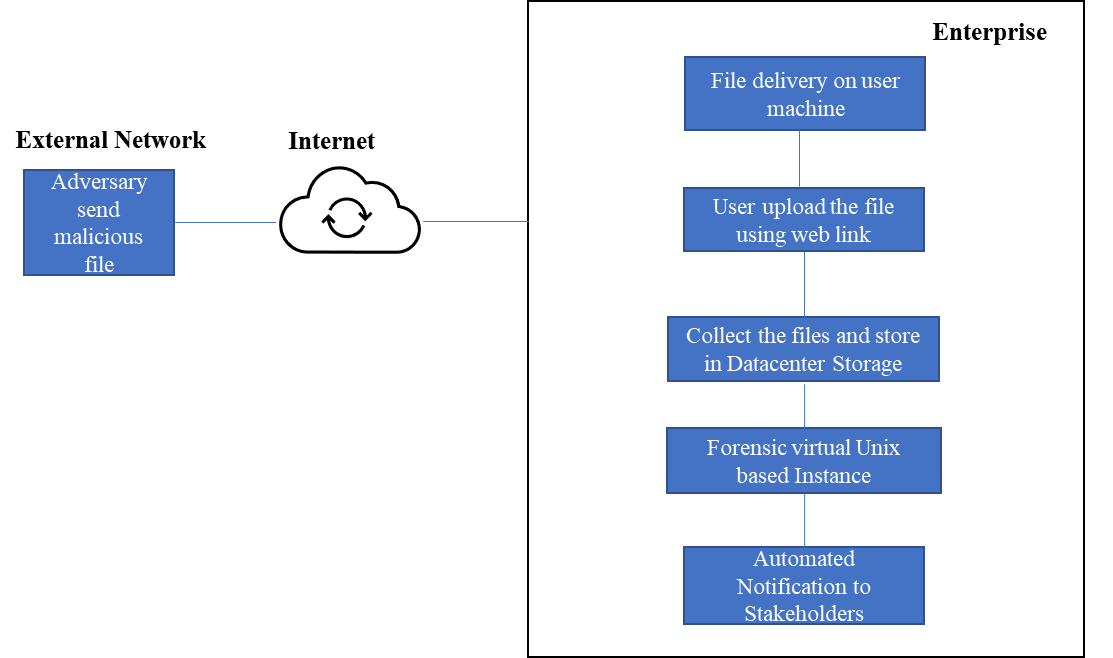}
\caption{The Framework Deployment in Enterprise Datacenter Network} 
\label{f.s51}
\end{figure}

Our proposed steganography process related to malware analysis framework deployment in a typical data center  is described here, as shown in Figure \ref{f.s51}. Let us suppose, an adversary may find the target email address from open-source threat intelligent platforms or dark web and send a phishing email attached with malicious multimedia files from external network. The delivered email is stored in an email exchange server in the target organization data center supported by the security vendor. The targeted user may upload the suspicious file for analysis using secured link enabling the file with password protection if they find it as suspicious. A security team member may also identify the file for malware analysis using the security tool alerts. These files are stored in the direct attached storage area in a separate Virtual Local Area Network (VLAN) environment. The direct attached storage is supported by backup storage devices for cold storage. When the new file is arrived in the direct attached storage, the file is submitted or submitted set of files in the batch form after certain time interval to the forensic virtual instance from the direct attached storage instance. Our proposed malware analysis pipeline can be implemented in the forensic virtual machine with connection to the remote attached storage.

If the file is identified as malicious, then the security team members are notified using email service for taking next preventive actions. We may expect three main scenarios if the detected file is malicious. The first scenario would be the file is hiding the C\&C server IP address or domain address. Then, the victim machine needs to be contained by disconnecting from the internet and perform the forensic analysis on the root cause. The malware artifacts in association with the identified malicious file  need to be  updated in the network or end device security tool set and search for any other victims being compromised with the malware. The containment process needs to be repeated from all the infected machines. In particular, the network tools like IPS and Firewall policies may need to be updated to block the C\&C IP address and domain name temporarily. The second scenario would be the file contains malicious executable. In this case, the machine still need to be contained and perform the forensic analysis to determine the behavior of the executable. Based on the behavioral results, further actions need to be taken like updating the file hashes in security policies for detection and blocking. The third scenario would be the file contain the shell code. In this case, the shell code snippet artifacts need to be analyzed in an isolated environment and determine the properties of the shell code and their capabilities. Obviously, the first action would be blocking the remote server IP and domain in the victim network environment.

\subsubsection{Stegomalware Framework in Cloud Architecture}
The enterprise organizations may also opt to implement the proposed approach in cloud environment, as the cloud environment offers many benefits such as save infrastructure and operations cost, security, flexibility, and quick deployment. The top three cloud services based on the revenue in the industry are Amazon Web Services (AWS) \cite{AWS}, Google Cloud Platform (GCP) \cite{GoogleCloud} and Microsoft Azure \cite{MicrosoftAzure}. These cloud services offer storage solutions for less cost, running instances and containers in multitenant environment, Internet of Things (IoT) applications, ML and AI solutions. We consider the Amazon AWS to implement the proposed model and choose different AWS resources for designing the solution in this work. It is assumed that the adversary targeted employee user machine is controlled by the Microsoft active directory service so that the employee machines fall under the enterprise environment in one umbrella. So, when the user received a malicious or suspicious multimedia file or someone would like to report the suspicious file for malware analysis, the file is submitted to the cloud environment through internet.

\begin{figure}[!h]	
\centering
\includegraphics[width=8.3cm,height=6cm]{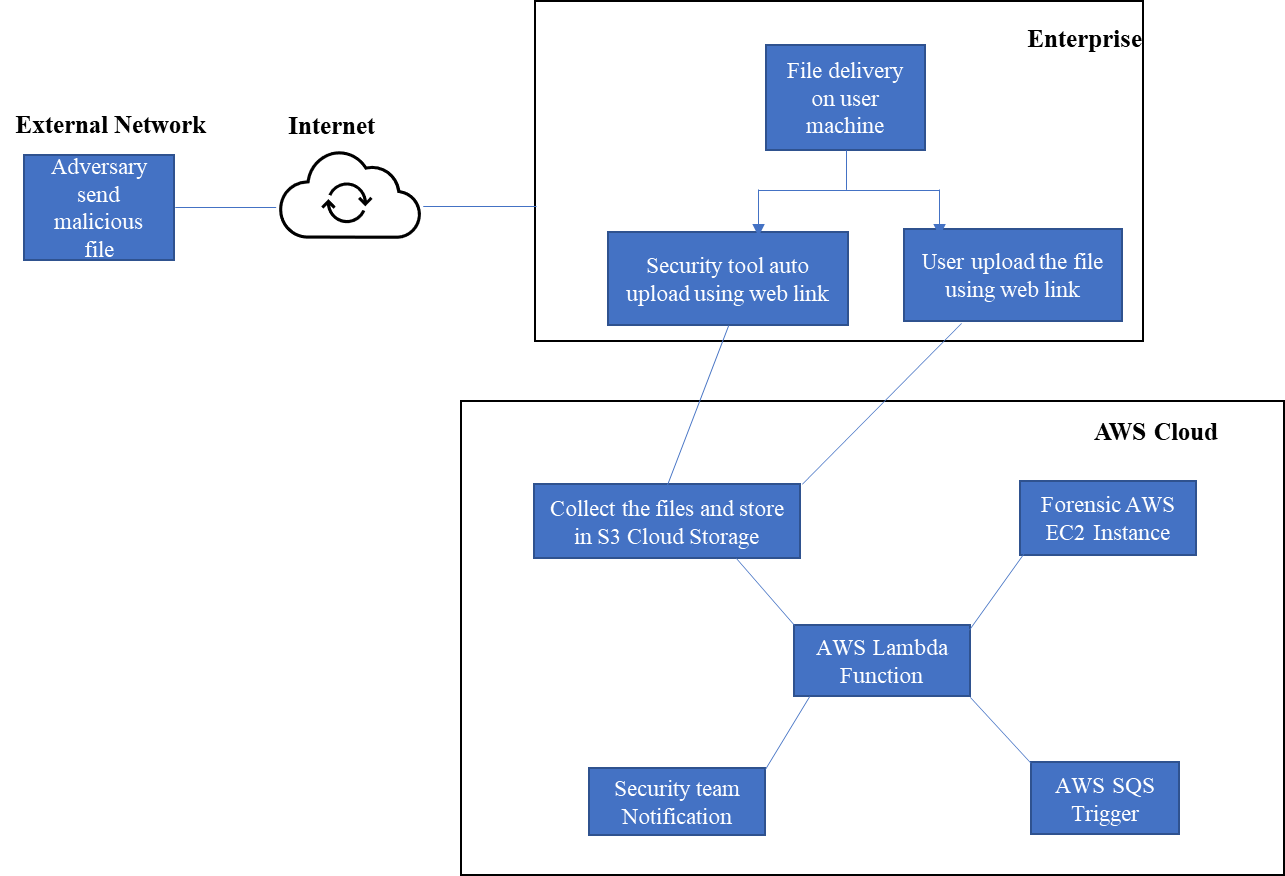}
\caption{The Framework Deployment in AWS Cloud} 
\label{f.s52}
\end{figure}

These files are stored in Amazon simple storage service (S3) location and has been set the file access level to private for not disclosing to the public. Whenever a new file is uploaded to the S3 bucket, AWS lambda function triggers to instruct the file to be submitted to the forensic AWS Elastic Computing Cloud (EC2) instance. The lambda function has given the resource level access to monitor the S3 buckets for new files and submit to the EC2 instance. At this point, our proposed framework stegomalware detection process starts in the EC2 instance. It first submits the file for possible signature-based detection using open-source tools. Subsequently, the submitted file is analyzed for the possible hidden malware content and the final decision on the file is extracted for further actions as described in the previous framework description. If the file is determined as malicious after structural, statistical, and dynamic malware analysis, lambda function retrieves the file output and triggers the AWS simple queuing service (SQS) and simple email service (SES) to notify the users, security team and other stakeholders to take further actions as needed. The lambda can also be used to extract the IoC from the file and may leverage the security tool set API functionality to update the IoC’s in the tools for detecting the future attack attempts in the target environment. The Figure \ref{f.s52} shows the different AWS cloud components involved to implement the proposed solution and the connectivity between the targeted user network as well as the AWS cloud environment.

\subsubsection{Stegomalware Framework in Multi-Cloud Architecture}
\begin{figure}[!h]	
\centering
\includegraphics[width=8.3cm,height=6cm]{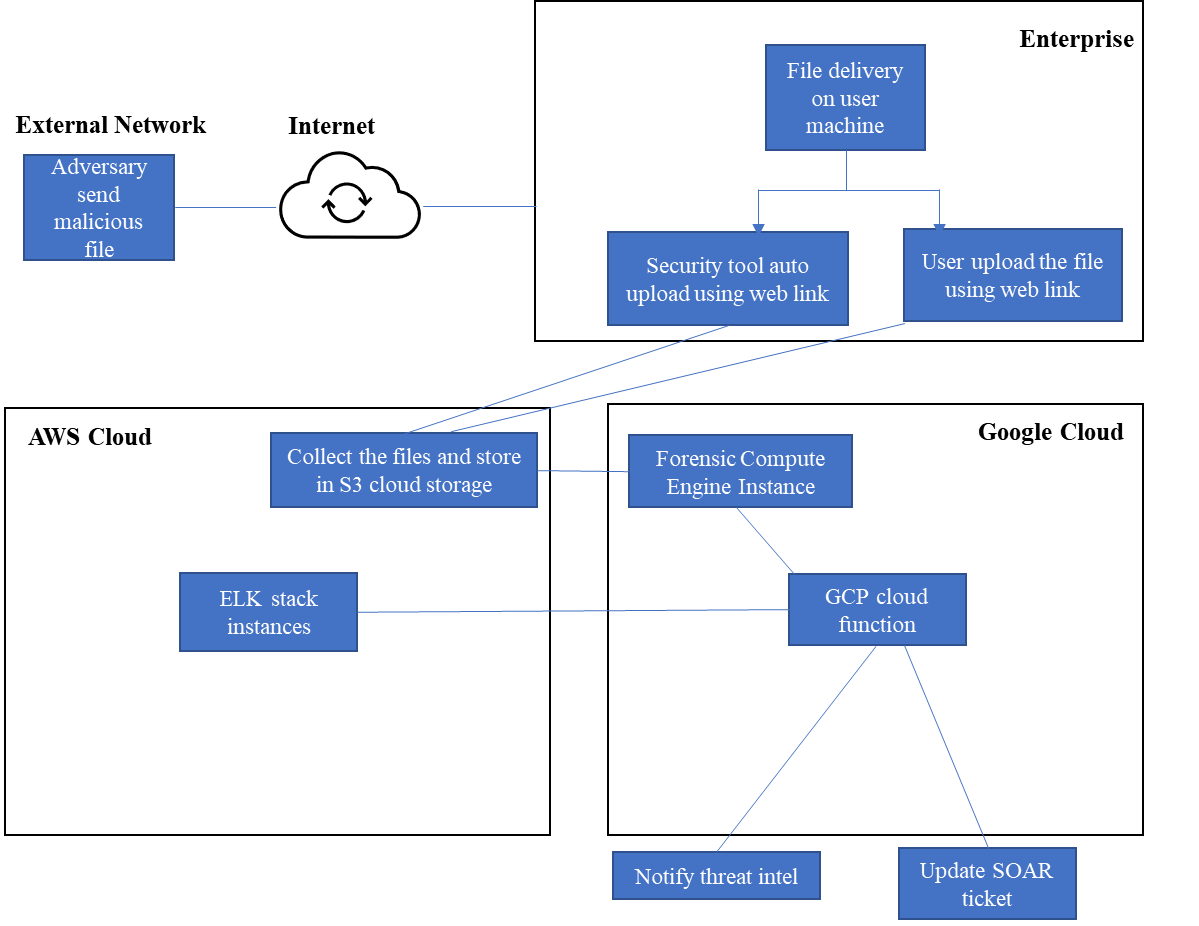}
\caption{The Framework Deployment in Multicloud} 
\label{f.s53}
\end{figure}

As the diverse number of applications running in parallel in most of the enterprises, it is very likely that the enterprises may use more than one cloud service for their businesses. We have provided the AWS and GCP cloud usage scenario as a multi cloud for implementing our stegomalware analysis framework. The multicloud may isolate the resources and may improve the overall security. When someone submits the file for analysis, the file is still stored in the AWS S3 bucket like discussed in the AWS cloud scenario. But, the proposed framework is deployed in the Google cloud environment, as shown in the Figure \ref{f.s53}. Like AWS lambda functions, we may use Google cloud function to retrieve the file from S3 and submit the files to the Google computer engine instance for analysis. The framework running in the computer engine instance will determine the file disposition using malware analysis techniques particularly steganalysis techniques. If the file is identified as malicious, the cloud function performs the API calls to threat intelligence and security ticketing tools to  notifying the stakeholders so that preventing actions can be taken. The other functionalities can be implemented using cloud function like interact with other tools and update the status of the file outcome. For instance, the cloud function can also update the Elastic Logstash Kibana (ELK) instances for saving the file output records and use for event correlation with other security events in the enterprise.

%% file: S51Dataset.tex
\section{Datasets}
\label{sec:dataset}

The quality of the dataset clearly influence the outcome of the ML/DL applications. The same holds true for solving the steganalysis/steganography with ML/DL in two ways. One way is proposing new steganographic algorithms and performing steganalysis on it for validation. Other way is proposing new steganalysis detection technique and validating the performance of the proposed detection method. In either way, the proper selection of datasets plays a major role for obtaining good performance results. We have reviewed number of papers contributing to the steganography or steganalysis state of the art and the most widely used datasets to solve the information hiding problems are tabulated in the Table \ref{T:dat} with detail overview. Images are extensively used as a cover for steganography/steganalysis. So, most of the datasets used in the prior art are image datasets and we only included image based datasets in the reported datasets.

Although there were couple of image based datasets existed for image segmentation, object classification in images, image resolution, image edge detection for the last few decades, the first notable information hiding dataset was available on public is in 2007 as part of the watermarking breaking contest "Break Our Watermarking System (BOWS)" in images held by ECRYPT network. Later, The BOWS verion2 of the contest image dataset released with three challenges to encourage the researchers to participate in the competition. The watermarking algorithm "Broken Arrows" also used in those challenges. In the following year 2010, The Break Our Steganography  System (BOSSbase) contest was conducted with a dataset of 10000 JPEG images in PGM format. The embedded algorithm "HUGO" is applied to generate the stego images in the dataset. This dataset was the first standard dataset available to use by researchers for steganography validation and performing steganalysis. The BoSSBase is considered as a base dataset in image steganography and steganalysis for validation and performance comparison of various techniques proposed by the researchers. Various embedding algorithms like HILL, S-UNIWARD are also applied on the Bossbase dataset images to test deep learning based steganalysis techniques detection performance, as illustrated in the table \ref{T:DLSA}.

There were other standard image datasets like Dresden, Erlangen, Coco, Raise and  DIV2K mainly build for the image forensics applications also used for steganography. Essentially, GAN based steganography solutions proposed by researchers used these datasets to hide the data in the images. Unfortunately, as there is no standard dataset is being considered for GAN based steganography, the performance comparison of various GAN based steganography solutions is complicated. It is highly desired to have a standard dataset for testing GAN based solutions, instead of considering image processing datasets.

All the above-mentioned datasets are generated focusing on the image resolutions used to view in computer device. But, the authors in \cite{Newman2019} created image generation application, which can generated mobile device supported image datasets. The required parameters like image format, resolution need to submitted to the application and then application create image datasets using the image database running on the application backend. This application is a great resource when someone would like to perform research on the mobile device based image steganography. 

Recently, Hugo et al. LSSD \cite{Hugo2021} created 2 million huge image steganalaysis dataset, which contain gray and color images, including the J-UNIWARD based steganographic images. For large scale steganalysis experiments, This open source image database can be helpful for research community.

Overall, majority of the datasets used in the state of the art are mainly focused on the data hiding in images and not clearly made for the purpose of the stegomalware detection. This may be the fact that the malware analysis is completely different technology discipline compared to the information hiding. But, In 2019, The authors in \cite{Cohen2020} proposed MalJPEG solution to detect the stegomalware hidden in JPEG image format, which is the only work we have come across focused on detecting stegomalware with datasets and using machine learning techniques. 

\begin{table*}[h!] 
\centering
\caption{Image datasets used in the state of the art for steganography and steganalysis}\label{T:dat}
\resizebox{\textwidth}{!}{%
\begin{tabular}{|l|l|l|l|l|l|l|l|}
\hline
\textbf{Dataset}  & \textbf{Year} & \textbf{Sample Size} & \textbf{File Format} & \textbf{File Sizes} & \textbf{Purpose} & \textbf{Steganography Algorithm} \\  \hline

USC-SIPI \cite{SIPI}  & 1977-2021 &	291  &	TIFF &	256,512,1024 & Generic Image Research & - \\ \hline

BOWS \cite{BOWS2} &	2007 &	 &	JPEG, PGM &	512	& Watermarking & Broken Arrows \\ \hline

BOWSv2 \cite{BOWS2}	& 2008	&	& JPEG,PGM & 512	&	Watermarking & Broken Arrows  \\ \hline

BOSSbase \cite{BossBase}  &	2010 &	10000  &	JPEG &	512	& Steganalysis contest & Hugo \\ \hline

Dresden	\cite{Dresden} & 2010 &	17000 & - & -	& Image Forensics & - \\ \hline

ImageNet \cite{Russakovsky2015}  & 2010 & 14197122  & JPEG & Avg 469x387 & Image Classification & - \\ \hline

Erlangen \cite{Manipulationimage}  &	2012 &	135  & JPEG & varied & copy-move forgery analysis & -\\ \hline

Coco  \cite{COCO}  \cite{Lin2014}  &	2015 &	328K  & - & - & Image research & Image Object classification\\ \hline				

Raise  \cite{RAISE} \cite{DangNguyen2015}  &	2015 &	8156  & JPEG, TIFF & Varied &	Image Forensics & -\\ \hline

DIV2K \cite{DIV2K}  &	2017 &	1000  &	PNG	& varied & Image resolution based research & - \\ \hline

Stegoappdb \cite{Newman2019}  &	2019 &	810000  &	JPG, PNG &	varied &	Steganography Research &  variants of F5 using PixelKnot app \\ \hline

Alaska \cite{Alaska}  & 2019  &	50000 &	JPEG & 512, 640, 720, 1024 & Image source classification &	Naive LSB, nsF5, UED, EBS, J-Univward \\ \hline

IStego100K \cite{Yang2020867}  &	2019 &	208,104	 &	JPEG &	1024 & Steganography Research & J-uniward, nsF5, UERD \\ \hline

Steganograpghy dataset \cite{LSBIEEE} &	2019 &	70000  & - & - &	Steganography Research &	LSB  \\ \hline

MalJPEG \cite{Cohen2020} & 2019 & 156,818  & JPEG & Varied & \textbf{Stegomalware Research}& \\ \hline

AlaskaV2 \cite{Cogranne2019} & 2020 &	80005  &	JPEG &	256,512	& Image source classification & JMiPOD, J-UNIWARD, UERD   \\ \hline

LSSD \cite{Hugo2021} \cite{LSSD} & 2021   & 2 million & JPEG  & 256  & Steganalysis  &  J-UNIWARD\\ \hline
\end{tabular}%
}
\end{table*}

\textbf{Lessons Learned}:
Our steganography and steganalysis review on datasets used for research purpose shows that there are no public stegomalware datasets available for contributing to the stegomalware research. Recent work on JPEG dataset used \cite{Cohen2020} for stegomalware is not available in public. However, there are number of datasets available to test the generic data hiding in images. For instance, the BoSSBase, BOWS are extensively used for evaluating the prior art steganography and steganalysis solutions.

%% file: S52Performancemetric.tex
\section{Evaluation Metrics}
\label{sec:metrics}
The secret hiding capability of a given steganographic algorithm and secret detection abilities of a steganalaysis techniques can be assessed using various performance metrics. These metrics may vary based on the cover medium used for concealing or revealing the secrets. We will describe various performance metrics used for image cover medium and their essence for evaluation of image stego techniques here.

The three factors such as the effectiveness of hiding the information without being identified by the steganalysis techniques (secrecy), the difference between the cover image and the stego image (distortion) and number of bits can be hidden in a pixel of a cover image (capacity) are the assessment indicators of a steganography algorithm for a cover image. Ideally, a good steganography algorithm should have high secrecy, high distortion and  high capacity to hide the secrets. The detail description of the performance metrics are mentioned in the following paragraphs.

\textbf{Peak Signal to Noise Ratio (PSNR)}:
The PSNR metric is helpful to assess the quality of the image. It can be used to measure the distortion between stego and cover images. It is defined as a function of Mean Square Error (MSE). For a given two monochrome images A and B with width W and height H, the MSE is defined as \cite{Zhang201998}

\begin{equation}
   MSE =  \frac{1}{WH} \sum_{i=0}^{W-1}\sum_{j=0}^{H-1} [A(i,j) - B(i,j)]^2 
\end{equation}
  where the i, j are the location of  $i^{th}$ row and  $j^{th}$  column pixel value.
  
The PSNR is defined as 
\begin{equation}
   PSNR = 20.\log_{10}(\frac{N}{\sqrt{MSE}})
\end{equation}
"N" is the maximum difference between pixels in an image

If the mean square error for a steganography algorithm is higher, it is comparatively difficult to perform the steganalysis compared to lower MSE steganography algorithm. In contrast, the PSNR values should be high to make the stego image looks the same as the cover image so that it will be hard to crack using steganalysis. Normally, when the PSNR is higher than 30 dB, it is very difficult for human eyes to distinguish the stego image from the cover image. The PSNR comes under the factor "distortion" for hidden data assessment.

\textbf{Structural Similarity Index (SSIM) }: The PSNR may not be an ideal metric to assess all the steganographic algorithms and it's just one of the metric for measuring the image quality. So, we may use  SSIM for image quality measurement and is often used in broadcast industry. 

Given two images A and B and their respective mean values $\mu_X$ and $\mu_Y$ as well as the variance values $\alpha_X^2$ and $\alpha_Y^2$ , covariance $\alpha^2_{X_Y}$
The SSIM is represented as 
\begin{equation}
    SSIM = \frac{(2\mu_X \mu_Y + K_1R)(2\sigma^2_{X_Y} +K_2R)}{(\mu_X^2 +\mu_Y^2 + K_1R)(\sigma_X^2 +\sigma_Y^2 + K_2R)}
\end{equation}

The SSIM values ideally lies in the range between -1 to 1. The lower value of SSIM indicates that the cover image and stego image are difficult to distinguish. In general, the k1 and k2 values will be 0.01 and 0.03 respectively \cite{Zhang201998}. For a good steganographic algorithm, the SSIM  should be as low as possible.

\textbf{Embedded capacity (EC)}: 
Embedded capacity is the ratio of the total number of embedded bits in an image to the total size of the image. The embedded capacity is also represented as the bits per pixel (bpp). For a given stego image with Width W and Height H, and the number of embedding bits are E, then the Embedded capacity is denoted as 

\begin{equation}
Embedded capacity = \frac{E}{WH} 
\end{equation}

\textbf{Bits Per non-zero DCT Coefficient (bpnzac)}:
The embedded capacity for the JPEG domain images is the number of bits embedded in the DCT coefficients of an image. This parameter is selected to choose the proportion of the bits embedded and used to evaluate the performance. 

\textbf{Quality factor (QF)}:
The JPEG compression is measured using quality factor. The quality factor is the quality of an image after JPEG compression. It is usually represented in percentages and typically in between 75\%-100\% QF is used for evaluating the steganography and steganalaysis of JPEG domain image. The images can be categorized as low, medium and high quality images based on the quality factor in the range between 70\%-80\%, 80\%-90\% and greater than 90\%. 

Steganalysis:

\textbf{Probability of error($P_E$):} The probability of error of an image steganalysis method is the average  total number of incorrectly identified as stego images and incorrectly identified as cover images i.e. the average of the false positives and false negatives. In other words, the average error in detection of the cover and stego images.

The probability of error or detection error is denoted as
\begin{equation}
P_E = \frac{1}{2}(P_{F_A} + P_{M_D})
\end{equation}
PFA is probability of false alarm which gives the probability of cover images being classified as stego images and PMD is probability of missed detection which gives the probability of misclassified stego images as cover. The Probability of error or detection error rate or classification error is the main performance metric used for evaluation of the ML/DL based image steganalysis techniques for information hiding detection.

\textbf{Detection Accuracy}
The detection accuracy is another metric widely used in steganography and steganalysis performance evaluation. The detection accuracy is measured as the ratio of the total number of correction classification of the cover images and stego images divided by the total number of correct and incorrect classification of both the cover and stego images. Let TP is the correct classification of the stego images, TN is the correct classification of the cover images, FP is the incorrect classification of the cover images and FN is the misclassification of the stego images.

\begin{equation}
Detection Accuracy (DA) = \frac{TP+TN}{TP + TN + FP + FN}
\end{equation}

The summation of the Probability of error or detection error and detection accuracy is always 1.

\begin{equation}
Probability of Error (P_E)  = 1 - DA
\end{equation}

\textbf{Bit error rate (BER)}:
Bit error rate quantify the robustness of embedding data in the cover medium. For a steganography algorithm with B bits embedded in the image and $B_E$ is the number of errors occurred while extracting the embedded data, the Bit error rate is denoted as 

\begin{equation}
BER  = \frac{B_E}{B}
\end{equation}

\textbf{Mean Absolute Error (MAE)}

Mean absolute error is determined as the average of the absolute value of errors. The absolute error is the absolute value of the difference between the predicted and target values. For a given two monochrome images A and B with width W and height H, the MAE is defined as 

\begin{equation}
MAE  = \frac{1}{WH} \sum_{i=0}^{W-1}\sum_{j=0}^{H-1} |A(i,j) - B(i,j)|
\end{equation}

The MAE can be used to measure for measuring the stego image medium quality compared to cover medium.

\textbf{Image Quality Index (Q Index):} 
The measurement of the image distortion using the factors such as loss
of correlation, luminance distortion, and contrast distortion signifies the image quality index \cite{Wang2002}. Let x and y are the cover and stego images. The mean and variance of the cover and stego image pixels values are denoted as $\overline{x}$, $\sigma^2_x$ and $\overline{y}$, $\sigma^2_y$. 

The quality index is represented as
\begin{equation}
    Quality index = \frac{4\sigma_{x_y} \overline{x} \overline{y}}{\sigma^2_x+\sigma^2_y[ (\overline{x})^2+(\overline{y})^2]}
\end{equation}



The quality of stego images generated by GAN models are also measured with different performance metrics. The mostly used quantitative indicators are  Frechet inception distance (FID) \cite{Heusel2017}, inception score (IS) \cite{Salimans2016}, Wasserstein distance for GAN model evaluation. However, for data hiding using GAN, the conventional steganalysis metrics like detection error and detection accuracy are used in performance evaluation.

%% file: S6ResearchChallenges.tex
\section{Research Challenges}
\label{sec:challenges}

\subsection{Advanced Stegomalware detection}
The existing anti-malware and end point security solutions tend to be ineffective to analyze the concealed malicious content in multimedia files, and thorough byte level analysis in image, audio or video using those tools may end up with performance issues. As the stegomalware is rarely seen for sophisticated attack campaigns in the wild, the signature-based tools may not be updated. The update  sometimes may be delayed because researchers perform the in-depth analysis and distribute the analysis to the security community. By that time, the attacker might do few modifications in the cover medium so that previous Indicators of Compromise (IOC) like hashes might be invalid. So, signature-based detection of the antimalware solution may be ineffective for the detection. The statistical based detection may experience false positives, and considering how common the multimedia files seen in the enterprise, the base rate fallacy tendency may be difficult to overcome. This leads to sophisticated stegomalware detection techniques and methods are required for accurate, robust and efficient detection of the stegomalware.

\subsection{Lack of the multimedia stegomalware datasets}
As the multimedia cover medium is encapsulated in a malware module during the propagation of the malware, the stegomalware hiding in the cover medium may evade the detection tools.  In addition, the multimedia malware is mostly used in advanced persistent threat attack stages. So, it's even more likely that stegomalware not getting noticed by the security tools to flag them as malicious. So, the detection and collection of stegomalware samples is challenging. Notably, the authors in \cite{Cohen2020} proposed “MalJPEG” for hidden malware detection in JPEG images and collected a sample set of JPEG files for analysis. Although the dataset was used for performing the experiments, the JPEG image sample size is very small. Moreover, the datasets used for performing JPEG stegomalware evaluation are not available in public for research purpose. So, the standard multimedia stegomalware dataset is highly desired to perform the academic research and detect the future attacks using data analytics models such as ML and DL. Additionally, the paper  \cite{Cohen2020} only focused on the JPEG image dataset for evaluation. But, there are different file formats for image, audio and video. These file formats also can be used to hide the malicious content and evade the traditional malware solutions. So, multi format standard steganography datasets are required to use for stegomalware classification. One of our future work is to build datasets comprising multiple multimedia file format stegomalware images for research purpose. 

\subsection{Synthetic stegomalware datasets}
The deep learning generative networks such as GAN, VAC have gained major attention for using image processing applications. Despite GAN improves the performance of the image processing applications without having a dataset, GAN applications are also being used for malicious purpose. For instance, an adversary may generate the Deepfakes \cite{Yang20195} \cite{Matern2019}  to mimic the celebrities or targeted individual to defame them or spreading the fake news or even performing the social engineering attacks in enterprise \cite{Wilcox2015}. Similarly, GAN can also be leveraged to generate stego images hiding secret data \cite{Zhang201998}. We may use this GAN data hiding capability to solve one of the major challenges  for stegomalware detection i.e. lack of datasets. So, we envision the future contributions of generating synthetic stegomalware datasets so that ML/DL based stegomalware detection models can be implemented and tested for accurate stegomalware detection. 

\subsection{Deep learning based Stegomalware detection}
The existing malware steganalysis tools \cite{Openstego2015} \cite{StefanoDeVuono2003} mainly rely on structural and statistical properties to detect the malware hiding in images. The detection performance of these tools still need to be improved. Additionally, an adversary may use advanced steganography techniques \cite{Pan2016} \cite{Holub2014} to evade the detection. As discussed in this paper, the DL models improved the detection performance compared to conventional steganalysis \cite{Xu2021} \cite{Singh2021} for hiding data. However, the stegomalware may also use encryption capabilities to hide the malware in images to evade the detection. Furthermore. malware artifacts like IP addresses or URL requires very low embedding payload. So, the stegomalware detection using Deep learning is more complex than stego data detection in images. The current state-of-the-art also show that the DL detection performance need further improvement for stego data detection \cite{Singh2021} \cite{Xu2016}. We believe that the research towards DL based stegomalware detection is one of the fruitful direction to be followed.

\subsection{Universal Stegomalware Detection}
The existing structural and statistical analysis solutions were proposed mainly focusing on the specific file format of images. For instance, the state-of-the-art steganography and steganalysis solutions are focused on the JPEG images \cite{Tabares-Soto2019} \cite{Karampidis2018}. These solutions may be compatible to detect stegomalware hidden in the JPEG images. But, an adversary can use PNG or GIF images to store the malware payload and the existing solutions may not work well for stegomalware detection, as the steganalysis techniques target certain characteristics of the file structure and file content presentation to identify the stego content in an image. So, there is a strong need to propose universal steganography detection solutions for detecting the hidden malicious content in images of various formats.

\subsection{Attack centric stegomalware detection models rather than structural and statistical models}
The anomaly stegomalware detection solutions based on structural and statistical models can be easily evaded if an intelligent adversary can use advanced steganography techniques. For instance, instead of using the least significant bit for storing the hidden content, the adversary may use the highly secured alternative methods to evade the anomaly detection solutions. So, we envision that the attack centric based solutions like unique solution for hiding the shell code or malicious EXE file or  malicious IP or domain hiding detection are helpful for robust and accurate stegomalware detection rather than relying on the state-of-the-art solution focused on detection of the generic data hidden in the images.

\subsection{Machine Learning based stegomalware detection} 
Although ML techniques are extensively used for malware classification in security field, there has been little work done on utilizing the machine learning techniques for stegomalware attack detection and classification. For instance, we see that Maljpeg applied Adaboost algorithm to detect the malicious JPEG images based on the features constructed from the JPEG image file structure \cite{Cohen2020}. Apart from that, there are no known prior works applied machine learning to detect the stegomalware. The main reason could be the lack of the datasets available for classifying the malware. We can see that the stegomalware detection using various machine learning models still need to be explored, particularly, proposing stego feature extraction methods from various image file formats and evaluating the effectiveness of the suitable machine learning models to image stego malware detection.  

\subsection{Audio and Video Stegomalware Detection}
Adversaries utilizing advanced malware hiding techniques to deceive the antimalware tools. Recently, the audio media files in wav format is used as a cover medium \cite{Cimpanu2019} to conceal the malicious DLL files. These are difficult to detect using existing security solutions because lack of malware samples for ML/DL based detection and signature techniques are ineffective. Additionally, the state of the art mainly focused on the image steganography and steganalysis. We believe that stegomalware based audio steganography and steganalysis research need to advanced for proposing solutions to effectively detect the stegomalware. The research opportunities include creating stegomalware benchmark audio and video datasets, using advanced malware hiding techniques like GAN and proposing novel detection solutions like Deep learning solutions to accurately detect the stegomalware.

\subsection{Hiding Malware in Neural Network models}
In recent times, deep neural networks are widely used in real time applications in recent times, as the DNN provide better performances with little domain knowledge and especially the existence of DL models as a service business model makes it even easier to use the pretrained models provided by those services. But, these models adapted from service providers can be dangerous if the service provider has malicious intent. Similar to the malware hidden in package repository libraries and installing the malware whoever downloads from the repository, the neural network models can be used as a cover to hide the malware and install the malware in the victim machines when a particular trigger occurred. The stegomalware hiding in neural network is feasible as the neural network contains number of parameters, which are insensitive to the minor changes and the result will not be impacted. Furthermore, the existing antimalware solutions are not capable of detecting stegomalware to leverage neural network models. Liu et al. \cite{Liu2020} proposed stegonet to hide the malware in deep neural networks. They used resilience training, value mapping and sign-mapping techniques to inject the malware payload in to the neural network model. To trigger the malware install from the neural network model, the logits trigger, rank trigger and fine-tuned rank triggers are proposed in the paper. The logits triggers can be considered as matching the key-lock pair and the trigger event as the key is supplied to match the pair.  An adversary may leverage the existing vulnerabilities in the DNN software like TensorFlow, Caffe to install the malware payloads. The authors in \cite{Wang2021} also showed that the malware can be hidden in neural network "Alexnet" with minimal accuracy loss, maximum payload embedding and the existing antimalware tools are unable to classify the model as malware. For instance, the authors could hide 36.9 MB malware file in 178 MB Alexnet model with 1\% accuracy loss. Overall, it is clear that neural networks can be the sweet spot for hiding the stegomalware and execute advanced malware attacks on the targeted organization. So, there is a huge research potential to contribute to the detection of stegomalware hiding in neural networks and proposing advanced techniques to hide the malware as well.


%% file: S7Conclusion.tex
\section{Conclusion}
\label{sec:conclusion}


In this paper, we performed a detail review of the stegomalware targeting the enterprise as part of the cyberattacks, and the state-of-the-art academic research image steganography and steganalysis techniques including the recent GAN stego image generation and DL based steganalysis for stego image detection. The detail description of the stegomalware history, tools and used file format specification are presented to comprehend how difficult to generate image stegomalware in the past. Additionally, we presented the existing stegomalware generation and detection techniques in  prior art in accordance with the image steganography and steganalysis. We have also provided a detailed comparison of the GAN based stego image generation models and DL based image steganalysis methods. Additionally, we have proposed anomaly based stegomalware detection framework for enterprise to detect the malware payload hidden in the images and discussed the components needed to deploy the in different network environments. Overall, based on our findings, we believe that there are a good deal of research opportunities to be pursed in the stegomalware generation and detection domain including stegomalware datasets generation, advanced stegomalware detection, robust and accurate DL based detection models but not limited to.